\documentclass[12pt]{article}

\title{ \vspace*{-0.3in}  \textbf{\Large Occupational Retirement and Pension Reform: The Roles of Physical and Cognitive Health}   }

   \author{\href{https://wjyecon.weebly.com/}{\textbf{Jiayi Wen} } \thanks{E-mail: wjyecon@gmail.com. School of Economics and Wang Yanan Institute of Studies in Economics (WISE), Xiamen University. This paper has been circulated under the name \textit{Occupational Retirement and Social Security Reform: the Roles of Physical and Cognitive Health}, from which significant revisions have been made. I am grateful to Pedro Mira, Manueal Arellano and Josep Pijoan-Mas for their advice throughout. I would also like to thank Mariacristina De Nardi, Nezih Guner,  Seonghoon Kim, Sagiri Kitao, Donhoon Lee, 
Ponpoje Porapakkarm, Georgoa Topas, Wilbert van der Klaauw as well as seminar and conference attendants at EALE/SOLE/AASLE World Conference, AASLE Annual Meeting, Econometric Society European Meeting, Asian Meeting of Econometric Society, CEMFI, Xiamen University, New York Fed, Singapore Management University, University of Tokyo, GRIPS, Peking University, Renmin University, Melbourne University, Vanguard for their constructive comments. I acknowledge the grant from the National Natural Science Foundation of China (Grant No. 71903167) and the financial support from the Spanish Ministry of Economics and Competitiveness (Grant no. BES-2015-072739). All remaining errors are mine.}}

\date{\small{Wang Yanan Institute for Studies in Economics (WISE)\\ and School of Economics, Xiamen University}  \\ \quad \\
	This version:  May 28, 2023 \\ \vspace{0.1cm} Please visit \href{https://wjyecon.weebly.com/}{\tt here} for the latest version}

   \usepackage{natbib}
   \usepackage{amsmath}
   \usepackage{indentfirst}
   \usepackage{graphicx} 
   \usepackage{float}
   \usepackage{caption}
\usepackage[margin=1in]{geometry}
   \usepackage[justification=centering]{caption}
   \usepackage[normalem]{ulem}
   \usepackage[toc,page,title,titletoc,header]{appendix}
   \usepackage{multirow}
   \usepackage{epstopdf}

   \usepackage{booktabs}
   \usepackage{mathtools}
   \usepackage{rotating}
   \usepackage[colorlinks=true,linkcolor=blue,urlcolor=black,bookmarksopen=true]{hyperref}
   \usepackage{bookmark}
   \usepackage[titletoc]{appendix}   
   \usepackage{soul}    
   \usepackage{setspace}
   \usepackage{numprint}

\usepackage[table,xcdraw]{xcolor}   

\usepackage{array}
\usepackage{multirow}

\npdecimalsign{.}
\nprounddigits{3}

\hypersetup{
  colorlinks,
  citecolor=[RGB]{0,102,204},
  linkcolor=[RGB]{0,0,0},
  urlcolor=black}

\makeatletter

\newcommand{\Rmnum}[1]{\expandafter\@slowromancap\romannumeral #1@}
\makeatother

\begin{document}

\newpage

  \maketitle

\begin{abstract}
   \begin{singlespacing}

Despite increasing cognitive demands of jobs, knowledge about the role of health in retirement has centered on its physical dimensions. This paper estimates a dynamic programming model of retirement that incorporates multiple health dimensions, allowing differential effects on labor supply across occupations. Results show that the effect of cognitive health surges exponentially after age 65, and it explains a notable share of employment declines in cognitively demanding occupations. Under pension reforms, physical constraint mainly impedes manual workers from delaying retirement, whereas cognitive constraint dampens the response of clerical and professional workers. Multidimensional health thus unevenly exacerbate welfare losses across occupations.



   \end{singlespacing}
\end{abstract}


  \setstretch{1.25}

\newpage

\section{Introduction}
Many governments aim to delay retirement through reforms to ensure pension sustainability.\footnote{Among others, the French government raises the retirement age from 62 to 64 in 2023. Policy Option Projections released in 2021 by Social Security Administration in the US examines options to increase FRA to age 68. The UK government was looking at bringing forward plans to increase the pension age to 68 between 2044 and 2046. The retirement age of Japan’s civil servants is raised to 61 in 2023 and will reach 65 by 2031.  The 14th Five-Year Plan of Chinese Government in 2021 reemphasized the plan to increase retirement age. } As health declines rapidly with age, do workers have enough capacity to respond to these reforms?  Understanding the role of health in retirement is essential for designing effective pension policies. However, while there is vast literature on this topic, our knowledge has mainly centered on the physical dimensions.\footnote{See \cite{french2017health}, \cite{blundell2016retirement}, \cite{coile2015economic}, \cite{lumsdaine1999new} for surveys.}

To gain a comprehensive understanding of the role of health in retirement, it is critical to consider the multidimensional nature of health and the heterogeneous requirements of jobs.  Firstly, the impacts of health on retirement should depend on how much it is required. Stephen Hawking achieved 138 thousand Google citations on a wheelchair, while Forrest Gump excelled in American football and ping-pong despite being slow-witted. The consequences of poor physical and cognitive health would be reversed if their careers were to switch. Secondly, changes such as skill-biased technical change, structural transformation,  automation, and the Covid-19 pandemic,  have substantially transformed the  jobs.\footnote{ See discussions such as \cite{katz1992changes}, \cite{acemoglu2011skills}, \cite{deming2017growing}.} Since 1970s, the share of manual employment has shrunk by 23\% whereas professional employment has increased by almost 50\%. On average, jobs are now less physically demanding but require more cognitive abilities.  Thirdly, policy reforms are targeting individuals at advanced ages. Poor cognitive health may become increasingly relevant given the rapid growth of dementia prevalence with age (\cite{van2005epidemiology})


This paper examines the impacts of physical and cognitive health on retirement, with a specific focus on how they interact with the heterogeneous ability requirements across occupations.  Using a dynamic structural model, I find that cognitive health has a relatively minor effect on retirement at younger ages but becomes increasingly important after age 65.  Moreover, physical and cognitive health disproportionately explain employment declines across occupations. When simulating retirement behaviors under pension reforms, I observe that poor physical and cognitive health distinctly hinder workers from delaying retirement across occupations, which unevenly exacerbates the welfare losses caused by the reform. 
To compensate for the losses magnified by poor physical health, the subsidy for a professional worker is 4,408 dollars less than a manual worker, but  this gap narrows to 2,951 dollars by 33\% once the cognitive dimension is also considered.  A back-of-the-envelop calculation shows that, considering the physical dimension alone, the subsidy for poor health would be 505 dollars less for a typical worker in 2015 compared to 1968 as jobs become less physically demanding. However, these amounts are overestimated by 55\% compared to the 325 dollars when taking into account the increasing cognitive demands.

To shed light on the link between health and job requirements, I start by merging the restricted version of the Health and Retirement Study (HRS) data with occupation information from the Occupational Information Network (O*NET).  While the public HRS only has 25 masked categories, the restricted version provides each worker's three-digit occupation code, enabling me to exploit the fine variations of ability requirement information for over 900 occupations from O*NET. I find that older workers are more likely to retire with physical decline in occupations with higher requirements of physical abilities, whereas retirement tends to occur with cognitive decline in more cognitively-demanding occupations. Do these patterns also exist in common occupation classifications? By the manual, clerical and professional categories, I show that these ability-requirement gradients translate into the occupation gradients. While the gradients are still present when looking at education, they become more modest. The gradients are also weaker when using subjective health measures and between-individual variations.  Moreover, I find that people experience a faster decline in cognitive health during their 60s than their 50s, a phenomenon unique to cognitive health and absent in either physical or self-reported health. These facts emphasize the importance of focusing on occupation, which directly captures the heterogeneous health requirements, and being careful about the measures and variations used for identification.

Motivated by these facts, this paper develops and estimates a dynamic programming model of retirement and saving decisions for male workers across the manual, clerical and professional occupations in the United States.  The key feature of the model is that it incorporates multiple health dimensions and allows them to affect retirement differentially across occupations through various channels.  Literature indicates five main channels through which health affects retirement: disutility of work, productivity, life expectancy, medical expense, and disability insurance. For all these channels, the model allows different effects by physical and cognitive health.  The disutility of work and productivity channels are allowed to be occupation-dependent, characterizing their heterogeneous ability requirements. The differential retirement effects of health across occupations are also likely to arise from socioeconomic gaps. The model captures rich heterogeneity  at the individual level: wage, assets, Social Security benefit, private pension, employer-provided health insurance, Social Security Disability Insurance, physical and cognitive health, life expectancy, education, and so on.\footnote{The state space includes 153.12 million state points in total. Meanwhile, assets and AIME are both continuous state variables, which are discretized but require interpolation. Decision rules have to be numerically solved at each of these state points. The resulting computation burden is very heavy. } The joint distribution of these variables differs across occupations empirically, which is important for capturing heterogeneous retirement behaviors. The model also embeds detailed rules of Social Security according to the Social Security Administration (SSA). These rules are essential for calculating public pension benefits and capturing policy incentives to retirement.

I use indirect inference to estimate the model on a sample of older males in the United States, combining data from the HRS 1996-2012 with the administrative data on earnings history  from the SSA. Consistent with the literature, physical health is measured by an index constructed from objective and detailed health variables about physical dimensions, while the measure of cognitive health is obtained from a standard memory test. The reduced form evidence suggests objective measures and variations over time are important for pinpointing the occupation gradients.   By indirect inference, I make use of the flexibility of auxiliary models to exploit specific data variations to identify the structural parameters.  The spirit is in line with a few but burgeoning structural studies that are more careful in choosing variations for identification (e.g. \cite{fu2019estimation}).\footnote{With the exception of \cite{french05} and \cite{french2011effects}, much of the existing work essentially uses a mix of within- and between-individual variation to identify the effect of health on retirement in structural models. Reduced form evidences have revealed the different results identified by cross-sectional and panel data variation. Discussions can be found in \cite{lumsdaine1999new} and \cite{blundell2021impact}.  } 

Based on the estimates, the  positive  part of this paper seeks to understand the quantitative significance of physical and cognitive health in retirement from a dynamic perspective. Both the history and the expectation of health have impacts on determinants of retirement  that are endogenous, such as savings.  Using the model, I am able to  shut down the channels of poor health and simulate labor supply changes throughout. The findings reveal that muting cognitive health leads to a smaller increase in labor supply than muting physical health at younger ages. However, the relative effect of cognitive health starts to rise exponentially after the age of 66, reaching as high as 200 percent as of age 75.  
 Moreover, the importance of physical and cognitive health varies considerably across occupations. Overall health, encompassing both dimensions, explains a share of employment decline between age 51 and 70 that is comparable to existing studies.  However, I find cognitive health explains a notably larger share of the employment decline in professional and clerical occupations, whereas physical health shows the opposite effect.

To underscore the significance of diverse job requirements, I conduct simulations of labor supply under hypothetical ability requirements. In reality, professional workers retire, on average, 2.4 years later than manual workers. However, I find that this gap would decrease to 1.3 years if manual workers faced physical requirements as lenient as those of professional workers. Conversely, the gap would widen to 3.1 years if professional occupations demanded cognitive abilities equivalent to manual occupations.

These variations in retirement effects of health are found under the current pension regulations. However, how important is health under pension reforms? 
In practice, policymakers tend to focus on the ability to work in physically demanding occupations while overlooking the cognitive constraints faced by office workers.\footnote{For instance, the Switzerland government reduced the retirement age of construction workers from 65 to 60 in 2003.} In the normative analysis, I simulate the changes in labor supply and welfare that would occur if the Full Retirement Age (FRA) were increased to 70, with a specific focus on the distinct constraints imposed by poor health across occupations. To understand behavioral responses to this policy change,  I enhance the policy simulation with analysis under the framework of heterogeneous potential outcomes. The theoretical framework demonstrates that poor health always leads to ``never-takers'' who are unable to respond to the reform. Whether poor health reduces the share of ``compliers'', i.e. workers responsive to the reform, appears to be indeterminate. ``Defiers'' to the reform may also exist because of the dynamic features of pension rules. Using the structural model, I am able to observe potential outcomes for the same individual, thereby defining ``complier'', ``defier'', ``always-taker'' and ``never-taker'' directly at the individual level.

Simulation results reveal that poor physical health  increases the share of ``never-takers'' and reduces the proportion of ``compliers'', particularly among manual occupations. Meanwhile, poor cognitive health contributes to ``never-takers'' notably in clerical and professional occupations.  To assess the extent to which labor supply responses to the reform are discounted by poor health, I compare changes in the average retirement age, with and without the constraint of poor health.   The findings indicate that, due to poor physical health, manual, clerical, and professional workers are less able to postpone their retirement by 13.7\%, 4.8\%, and 3.3\%, amplifying the welfare losses caused by the reform by 13\%, 6.7\% and 2.5\%. In contrast, poor cognitive health shows no impact on manual workers but reduces the labor supply responses of clerical and professional workers by 7.7\% and 5.0\%, leading to additional welfare loss by 0.5\%, 11.4\% and 10.6\% respectively.  To iron out the disparity in welfare losses due to poor physical health, the subsidy needs to compensate a manual worker 4,408 dollars more than a professional worker, while the subsidy for poor cognitive health turns out to be 2,341 dollars less.

Finally, I perform a back-of-the-envelope calculation to shed light on the impact of poor health, considering the evolving occupational composition since the 1970s. In order to offset the welfare losses caused by the reform and amplified by poor physical health, the subsidy for a typical worker has decreased by 505 dollars between 1968 and 2015, reflecting the declining physical demands of jobs. However, when accounting for the rising requirements for cognitive abilities, this reduction amounts to only 325 dollars, 36\% smaller.

The rest of this paper is structured as follows. The next section introduces the literature and Section 3 presents the empirical facts. Section 4 is devoted to the model, and Section 5 to the solution and estimation methods. Section 6 describes the data. Section 7 presents the estimates. Analyses using the model are shown in Section 8. Section 9 concludes.


\section{Related Literature}
This paper is mostly related to the literature that studies retirement behaviors and pension policies based on dynamic structural models.  Building on early works such as  \cite{rust1997social}, more recent studies enrich this type of model in various aspects.\footnote{ See \cite{french05}, \cite{vanderklaauwwolpin08}, \cite{blaugilleskie08}, \cite{french2011effects}, \cite{haanprowse14}, \cite{gustman2015effects}, \cite{iskhakov2020effects}, among others.} As an important factor in retirement, several studies share a particular focus on health. \cite{boundetal10} carefully deal with measurement bias in the self-reported health within a dynamic structural framework.  \cite{capatina2015life} quantifies the effects of health on life-cycle employment via channels of leisure, wage, medical expense, and life expectancy.  \cite{gustman2018role} model the detailed transition process of a number of specific health measures and incorporate it into a dynamic retirement model. 

The main contributions of the current paper come twofold: (1) Instead of taking health as an unitary concept, it considers the physical and cognitive dimensions separately,  and shows the rising importance of cognitive health with age. This finding is important for policies targeting advanced old ages. (2) It further takes into account the interactions of multiple health dimensions with job requirements. Much of the existing work focuses solely on factors of the labor supply side while overlooks ones of the demand side.\footnote{For instance, the survey by \cite{blundell2016retirement} suggests: “\textit{Our focus in both sections is on factors affecting the supply of labor among older workers. We do not devote much attention to factors potentially affecting the demand for older workers ..... Trends in employment of older workers have been rather similar across a large number of countries over recent decades... This suggests that supply-side factors
may be the most important.}”} Considering the interaction between health and occupation is  critical for policy evaluations concerned with the heterogeneous behavioral responses and the unequal welfare consequences. Moreover, it indicates the varying role of health in retirement over time due to the shifting occupation composition. It also provides new insights for understanding retirement across locations with different occupation composition. To the best of my knowledge, the work by \cite{jacobs2019health} is the only one that also focuses on occupational retirement, which explores the heterogeneity between blue- and white-collar occupations based on the self-reported health.\footnote{These two papers are developed independently at the same time. }

This paper also adds to the reduced form literature on the retirement effect of health.  First of all, this literature also rarely considers the interaction between multiple health dimensions and job requirements. The work by \cite{blundell2021impact} is the only one considers cognition, which finds a statistically significant but modest effect. The current paper shows that this small average effect masks strong gradients along occupations. 
Secondly, there is still no consensus on how large is the effect of health. The main debate centers on the proper health measures and the variations for identification. By providing thorough evidence,  the aforementioned paper by Blundell and the coauthors conclude that different measures yield similar estimates once a sufficient battery of objective measures are included. The current paper shows that the gradients are notably stronger using objective measures and within-individual variations, which suggests that measures and variations may still matter as far as the heterogeneous effects are concerned.

This paper also speaks indirectly to studies about the health capacity to work at older ages. While \cite{cutler2013health} and \cite{coile2016health} find ample capacity for older workers, this paper reveals considerable inequality in this capacity due to the interaction between health and occupation.

There are increasing studies about the consequence of cognitive decline, which mainly focus on older people's financial decisions, such as \cite{finke2017old} and \cite{widera2011finances}. This paper demonstrates the impacts of poor cognitive health on retirement, which is another important decision that has long-term implications for the older people.

Finally, by estimating the structural model carefully following the guidance of reduced form evidences, this paper adds to the burgeoning literature aims at a unification of structural and reduced form work (\cite{todd2023best}, \cite{li2023demographic}, \cite{fu2019estimation}, \cite{attanasio2012education}). While experimental or quasi-experimental variations are hardly available under this context, as mentioned, reduced form studies emphasize the different results based on different measures and variations. By indirect inference, the model targets these facts, upon which the structural parameters obtain identification.

\section{Empirical Facts}

This section documents empirical facts about physical and cognitive health, ability requirements, as well as their interaction with retirement.  I show that cognitive health declines much faster at advanced ages. Occupations that require substantial cognitive but little physical ability account for a large share of the employment. This section also shows that while the correlation between cognitive decline and labor force exit is small on average, there is a strong gradient by the extent to which cognitive health is required. The gradient is also found for physical health.  The education gradient exists but is smaller than the occupation gradient.

\subsection{Physical and Cognitive Health}
 Cognitive functioning is typically classified into two principal categories: crystallized and fluid cognition, building on the widely-accepted Gf-Gc theory in psychology (\cite{cattell1963theory}). Gf-Gc theory has been developed into the Cattell-Horn-Carroll theory which HRS has adopted to develop cognitive measures (\cite{wallace1995overview}.
 Crystallized cognition mainly reflects knowledge and the influence from education and experience, whereas fluid cognition represents the outcome of the influence of biological factors on intellectual development  (\cite{mcardle2002comparative}). Compared to crystallized cognition, fluid cognition experiences rapid decline at older ages.
This study thus chooses to focus on a crucial aspect of the fluid cognition, the (episodic) memory. This measure is widely used in literature on the financial consequence of cognitive decline (e.g. \cite{finke2017old}). As pointed out by   \cite{mcardle2011cognition}: ``episodic memory is a very general measure of an important aspect of fluid intelligence since access to memory is basic to any type of cognitive ability.'' Another advantage to focus on the memory is that it is directly measured in HRS by  the number of recalled words in a standard memory test. This word-recall measure is given top priority to represent fluid cognition, according to \cite{wallace1995overview}. Therefore, this paper uses this measure for cognitive health, which is at the individual level and longitudinal. 

Physical ability relates to body's capacity to perform activities that require strength and endurance.  A major focus of studies on the retirement effect of health is the reported bias in subjective health measures.  The justification bias rises if retirees tend to report worse health to justify their retirement, leading to upward bias in estimates of the retirement effect of health. On the other hand, classical measurement errors suggest the estimates will be underestimated because of the attenuation bias. More specific health-related variables are considered having less reported bias and are widely used as instruments for subjective measures (\cite{blundell2021impact}, \cite{disney2006ill} , \cite{bound1999dynamic} ).  In line with the literature, I measure physical health as the self-reported health instrumented by more objective measures. It is constructed as a health index, obtained as the fitted value from a regression of self-reported health on a bunch of more objective health variables. More details are provided in Appendix \ref{Appendix:PH}.

\begin{table}[H]
 	\centering
 	 		\footnotesize
 	\begin{minipage}{0.825\textwidth}
 	\caption{Variations of Cognitive Health by Education and Occupation}
		 \tabcolsep=0.12cm
		 \begin{tabular}{p{0.28\textwidth}>{\centering}p{0.10\textwidth}>{\centering}p{0.10\textwidth}>{\centering}p{0.10\textwidth}>{\centering}p{0.10\textwidth}>{\centering}p{0.10\textwidth}>{\centering\arraybackslash}p{0.10\textwidth}}
\hline
\hline
                     & \multicolumn{3}{c}{Education} &   \multicolumn{3}{c}{Occupation}  \\
 \cmidrule(lr){2-4}
 \cmidrule(lr){5-7}                     
\multicolumn{1}{c}{} & LHS                  & HS                            & SC\&C              & Man.               & Cler.                      & Prof.         \\ \hline
Age 51               & 9.20                 & 10.58                         & 11.00                & 10.43                & 11.87                          & 12.10                \\

Age 61                  & 8.42                 & 9.66                          & 10.16                & 9.67                 & 10.75                          & 11.64                \\
Age 71                  & 7.15                 & 8.29                          & 8.76                 & 8.31                 & 9.27                           & 10.26                \\
 Standard Deviation   & 3.18                 & 3.10                          & 3.18                 & 3.17                 & 3.21                           & 3.14                 \\ 
\hline
Drop from 51 to 61         & -0.78                & -0.92                         & -0.83                & -0.76                & -1.12                          & -0.46                \\
Drop from 61 to 71          & -1.27                & -1.37                         & -1.40                & -1.36                & -1.48                          & -1.38                \\
 \hline
\end{tabular}\\
 		\label{table:chtable}%
 		\scriptsize {This table presents variations of cognitive health by education and occupation. The values are predicted from the regressions with cubic ages to reduce noise. Individual fixed effects are controlled to obtain within-individual variations. LHS: less than high school; HS: high school; SC\&C: some college and college.  Standard deviation is calculated for ages 51-75. 
 			 \par}				
 	\end{minipage}
 	
\end{table}

 \begin{figure}[H]
 \centering
 	\caption{Age Profiles of Cognitive Health }
 	\begin{minipage}{0.85\textwidth}
 		\hspace{1cm}\includegraphics[height=5.6cm]{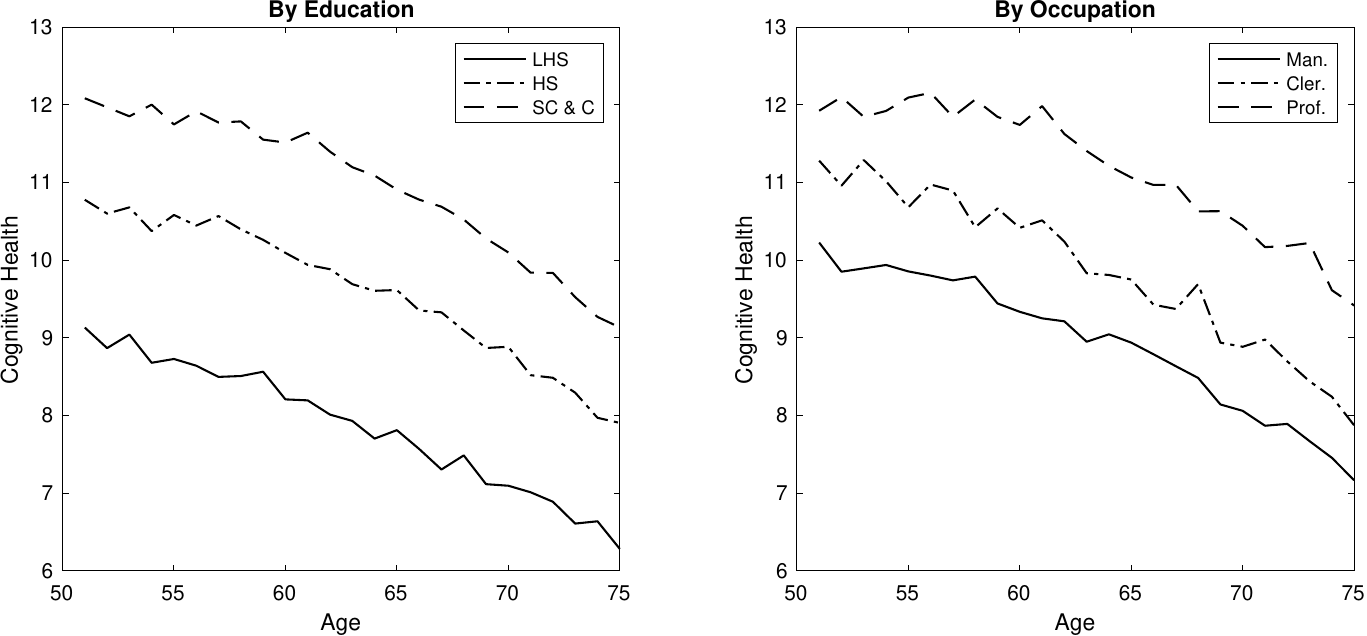} \\	    
 		\scriptsize{The figures present the variations of cognitive health from age 51 to 75. Results are obtained by regressing cognitive health on age dummies, controlling for individual fixed effects.  LHS: less than high school; HS: high school; SC\&C: some college and college. Man.: manual; Cler.: clerical; Prof.:professional. \par}	
 	\end{minipage}
 	\label{figure:chgraph}

 \end{figure}

  Table \ref{table:chtable} and Figure \ref{figure:chgraph} present the variations of cognitive health by education and  occupation. Occupations are based on the classification that is widely applied in other studies (e.g. \cite{katz1992changes}).  Broadly speaking, declining trends parallel between education and occupation categories.    
  While some of the studies in psychology argue that fluid cognition starts to decline as early as age 20-30 (e.g. \cite{salthouse2009does}),  conservative opinions suggest it starts around the middle of age 50s (e.g. \cite{ronnlund2005stability}). Similar patterns are found in the results.

  There are two crucial findings to be highlighted. First, the cognitive health declines much faster at more advanced ages. By education, cognitive declines between age 61 and 71 are around 50\%-70\% greater than the ones between age 51 and 61. By occupation, the declines are larger by 79\% and 32\% for the manual and clerical occupation. Most striking, the decline during ages 61-71 is three times as large as ages 51-61 for the professional occupation.  Second, the accelerating declines are only found for cognitive health. Results in Appendix B show that the declines of physical health, measured either by the health index or self-reported health, do not speed up at more advanced ages.

\subsection{Ability Requirements }

The previous subsection shows how physical and cognitive health vary with age. How are physical and cognitive health required by the demand side of labor market?

To shed light on this, I exploit the detailed information from O*NET data set.\footnote{The O*NET program, sponsored by the Employment and Training Administration under the Department of Labor, is the primary source of occupational information for the United States. This data set provides detailed occupation-specific descriptors, such as work styles and required knowledge, for more than 900 occupations. In particular, it provides occupation-specific ratings about how different abilities are required.}  O*NET defines nine  specific abilities under its physical category and twenty one specific abilities under its cognitive category.  Based on ratings of these specific abilities, I construct two indexes that respectively measure the extent of physical and cognitive ability requirements by the  principal component analysis.

\begin{figure}[H]
\centering
\caption{Distributions of Employment by Required-Abilities}
\begin{minipage}{0.98\textwidth}
		\hspace{1cm}\includegraphics[height=5.5cm]{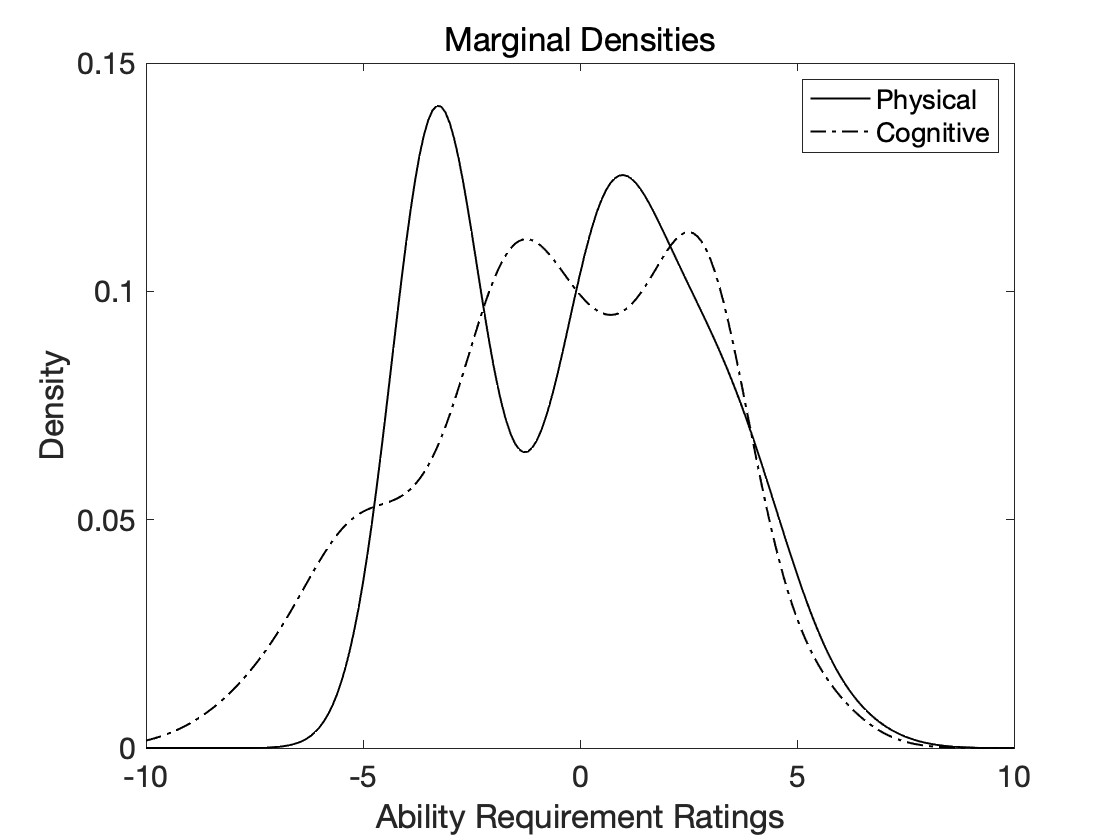} 
		\includegraphics[height=5.5cm]{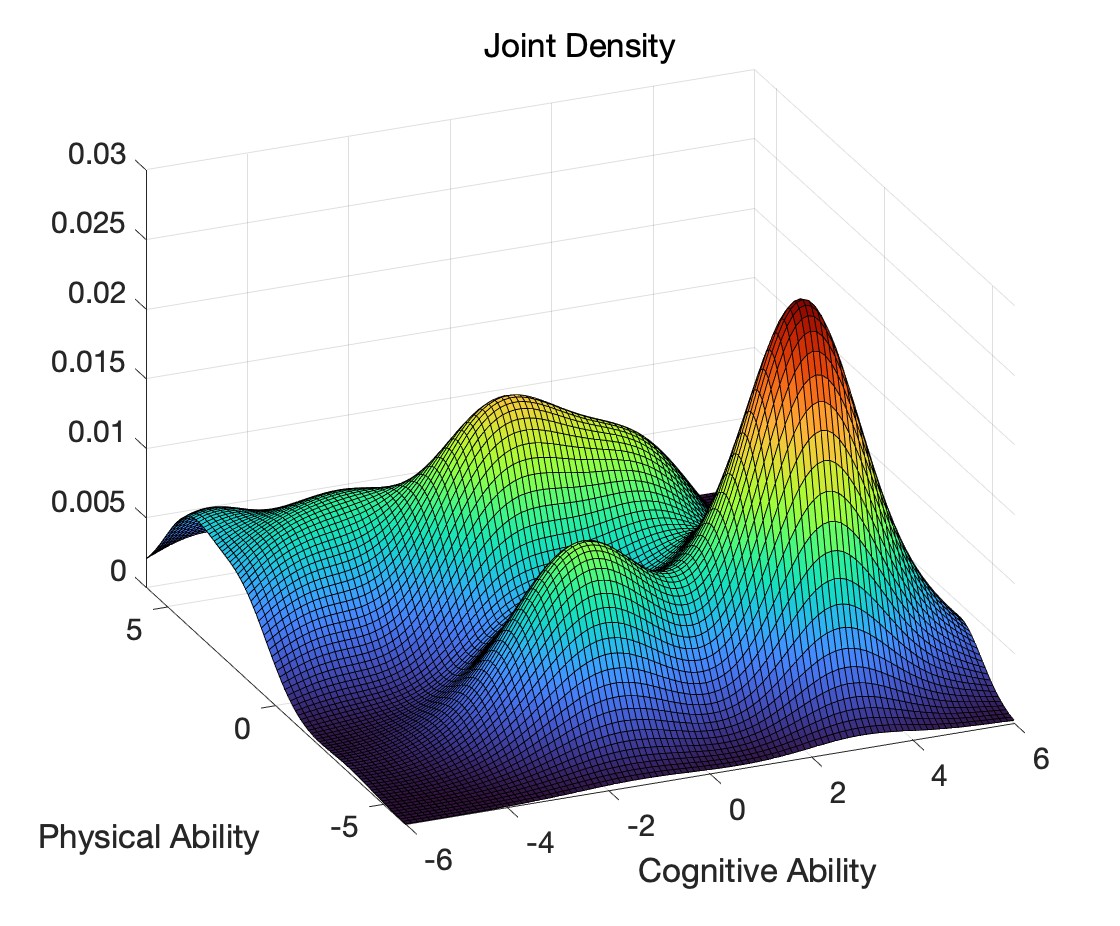}  \\
	\scriptsize{These figures show the kernel densities of employment by the ability requirements of occupation. The left figure presents the marginal distributions. The right figure presents the distribution by physical and cognitive ability requirements jointly. \par}	
\end{minipage}
\label{figure:phage}
\end{figure}

Figure 2 shows the distribution of employment by required abilities, estimated by the kernel density on the employment data of Census 2000. On the left figure, the  bimodal distribution for the physical ability shows that while many employments require the physical ability, the others demand little. To the contrary,  most employments share the requirement of cognitive ability centering around the median level.  In terms of the joint distribution, we can see that most employments require either cognitive or physical ability, showing a negative correlation between the requirements. The most prevalent employments require substantial cognitive but little physical ability.

\begin{figure}[H]   
	\centering
	\caption{Trends of Physical and Cognitive Ability Requirements of The Average US Job}
	\begin{minipage}{0.79\textwidth}
	\hspace{2cm} \includegraphics[width=9cm]{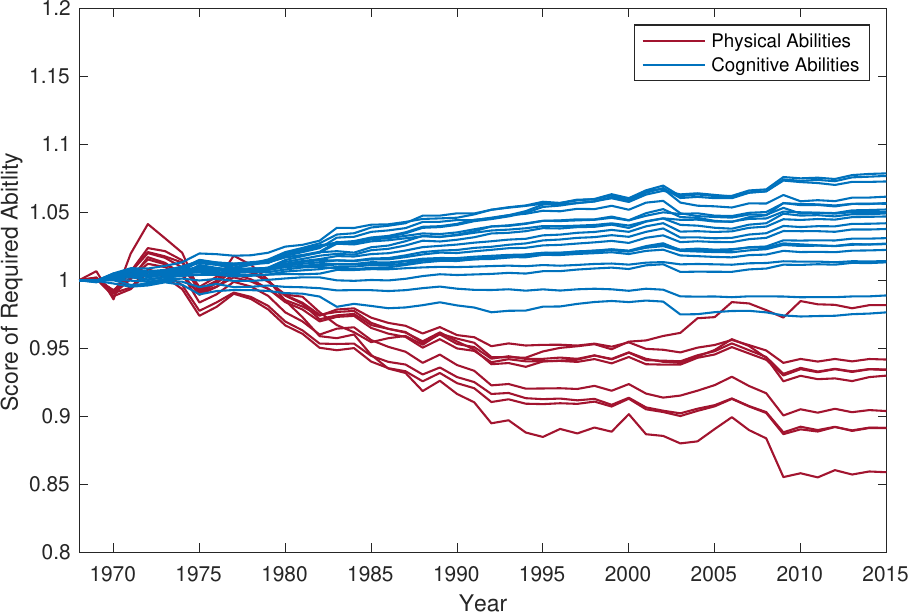} \\
		\scriptsize The figure presents the trends of scores measuring the extent of each ability is required by the average US job. Each line represents a specific ability under the physical or cognitive category, defined by O*NET.  The scores are normalized to 1 in 1968. The ability of spacial orientation, which is under the cognitive category, has experienced a significant decline since 1968 and has been excluded from the figure without undermining the main message.
	\end{minipage}
	\label{figure:skillchange}%
\end{figure}

How do current jobs require abilities different from decades ago?     Figure \ref{figure:skillchange} shows that the requirement for every physical ability has declined since 1968, whereas eighteen out of the twenty-one cognitive abilities have increased.\footnote{   These trends capture the compositional change and reflect how each physical or cognitive ability is required by the \textit{average} U.S. job from 1968 to 2015.   For each ability, O*NET only provides cross-sectional ratings. The data constraint does not allow us to keep track of the  longitudinal variation in ability requirement for each given occupation. To have a sense on the change over time, I  merge the ability requirement scores of each occupation by  O*NET, measured in 2016, with the longitudinal employment shares of each occupation from the Current Population Survey (CPS) 1968-2015.  }  
Given this systematic shift in the labor demand side, the cognitive dimension of health deserves more attention in studying older workers' retirement behavior.

\subsection{Ability Requirement and Occupation Gradients}
Given the significant heterogeneity in ability requirements, the same extent of health decline tends to have different impacts on retirement. For instance, the same cognitive decline could be more disturbing for a clerk than a construction worker. To shed light on this insight, I explore the relationship between different dimensions of health and labor supply based on the linear probability model. I regress the indicator of labor force participation (LFP) on the measures of physical and cognitive health, with and without interacting with variables measuring the extent that corresponding abilities are required.

This exercise requires information on individual's health and retirement status, as well as how much abilities are required by work. To this end, I merge the O*NET data with the restricted version of HRS data.  For each individual, the restricted version of HRS provides the three-digits occupation code instead of the grouped occupations in the public HRS. By merging with O*NET, I am able to exploit finer variations in the variables that measure ability requirements.\footnote{ The public HRS provides masked occupation codes for 17 occupations before 2004 and 25 occupations after 2004, whereas the restricted HRS provides unmasked codes for hundreds of occupations. }  The main focus of the literature on the retirement effect of health is whether to use subjective or objective health measures, as well as what variation to be used for identification. I explore the relationship between labor supply and health in the following exercises with particular care about these two issues.

\begin{table}[H]
	\centering
	\footnotesize
\caption{Ability-Requirement Gradients in Associations  between Labor Supply and Health}
	\begin{minipage}{0.92\textwidth}
		 \tabcolsep=0.35cm\begin{tabular}{p{0.32\textwidth}>{\centering}p{0.11\textwidth}>{\centering}p{0.11\textwidth}>{\centering}p{0.11\textwidth}>{\centering\arraybackslash}p{0.11\textwidth}}
\hline
\hline
           & (1)                  & (2)                  & (3)                  & (4)                               \\ \hline
    
Physical Health   & 0.025***            & 0.008*             & 0.007              & 0.059***            \\
           & (0.004)            & (0.005)            & (0.005)            & (0.008)            \\
Cognitive Health    & 0.009**            & 0.005              & 0.005             & 0.004              \\
           & (0.004)            & (0.004)            & (0.004)            & (0.004)            \\
Physical Health   $\times$ Ability Req.     &                      &                      & 0.016***            & 0.032***            \\
         &                      &                      & (0.005)            & (0.007)            \\
Cognitive Health   $\times$ Ability Req.      &                      &                      & 0.008**            & 0.009**            \\
        &                      &                      & (0.004)            & (0.004)            \\
\hline
Individual FE   &          No            &  Yes               & Yes                & Yes               \\ 
Physical Health Index  &          No            &    No              &   No               & Yes               \\ 
\hline
		\end{tabular} \\
		\label{table:gradients}
		\scriptsize{This table presents the standardized coefficients of the regression of labor force participation on different dimensions of health, and their interactions with corresponding ability requirements.  Standard deviations are reported in parenthesis. *** p$<$0.01, ** p$<$0.05, * p$<$0.1.\par }
	\end{minipage}
\end{table}%

Column (1) and (2) show the overall relationship between physical and cognitive health  and labor supply, based on the widely used self-reported health and the test memory. The self-reported health arguably mainly captures the physical domains of health, as respondents are unlikely to consider themselves as unhealthy simply because they have poor memory. Estimates based on the OLS with pooled variations in Column (1) show significant correlation between both dimensions of health and labor supply. The magnitude of the coefficient of physical health is threefold of the cognitive health.  For Column (2), whose identification is based on the within-individual variations, the estimated coefficients become notably smaller. Moreover, the coefficient of cognitive health becomes statistically insignificant. This is in line with the existing studies, which suggest panel data approach tends to yield smaller retirement effect of health (\cite{french2017health}).

However, once the heterogeneous ability requirements for physical and cognitive health are taken into account, I find remarkable gradients in the correlation between health and labor supply. For Column (3), while the correlation between health and LFP is insignificant at the average ability requirement level, the interactive terms of both the physical and cognitive health are statistically significant. It suggests that workers  with decline in physical or cognitive health are more likely to retire if they are from occupations with the higher ability requirements.  These gradients become further stronger when the objective physical health measure is adopted, as shown in Column (4). One doubt about the necessity of introducing cognitive health is that self-reported health may be comprehensive enough to reflect cognitive domains of health. The results  based on the self-reported health together with tested memory indicate the extra information uniquely captured by the cognitive measure.

 How important are physical and cognitive health in secular occupation classifications? Table \ref{table:onet1} and \ref{table:onet2} in Appendix \ref{Appendix:OCC} show that the manual, clerical and professional occupations have stark difference in ability requirements. Based on this classification, Table \ref{table:occgradients} shows that  the ability-requirement gradients are also translated to the occupation gradients. One standard deviation decline in physical health corresponds with 3.6 percentage points higher probability of retirement for professional workers, whereas the magnitude almost triples for manual workers with 9.7 percentage points. To the contrary, one standard deviation of cognitive decline associates with 2.3 percentage points higher probability of retirement for professional workers, two thirds of the one of physical health. At the same time, no such relationship is found for manual workers.

\begin{table}[H]
	\centering
	\footnotesize
\caption{Occupation and Education Gradients in \\ The Associations  between Labor Supply and Health}
	\begin{minipage}{0.90\textwidth}
		 \tabcolsep=0.22cm\begin{tabular}{p{0.18\textwidth}>{\centering}p{0.10\textwidth}>{\centering}p{0.10\textwidth}>{\centering}p{0.10\textwidth}>{\centering}p{0.10\textwidth}>{\centering}p{0.10\textwidth}>{\centering\arraybackslash}p{0.10\textwidth}}
\hline
\hline
VARIABLES    & LHS  & HS       & SC \& C     & Man.   & Cler. & Prof. \\ \hline
Physical Health     & 0.070*** & 0.073*** & 0.051*** & 0.097*** & 0.032**  & 0.036**      \\
             & (0.022)  & (0.013)  & (0.011)  & (0.012)  & (0.013)  & (0.016)      \\
Cognitive Health     & -0.015   & 0.003    & 0.009*   & -0.009   & 0.006    & 0.023***     \\
             & (0.012)  & (0.007)  & (0.005)  & (0.006)  & (0.006)  & (0.007)      \\
\hline
\end{tabular} \\
		\label{table:occgradients}
		\scriptsize{This table presents the coefficients of the regression of labor force participation on different dimensions of health, and their interactions with corresponding ability requirements.  Individual fixed effects are controlled and physical health index is used. Standard deviations are reported in parenthesis. *** p$<$0.01, ** p$<$0.05, * p$<$0.1.\par }
	\end{minipage}
\end{table}%

If we focus on education instead of occupation, the gradients remain but become notably  smaller. \footnote{\cite{blundell2021impact} do not find the education gradient for cognitive health, though the gradient for physical health also exists in their results. The difference may come from the different measures of cognition.}  These findings indicate the necessity to focus on the interaction of different dimensions of health with occupations, which directly reflect the heterogeneous health requirements of work.


\section{Model}

Motivated by the empirical facts, I develop a dynamic programming model of retirement and saving decisions for  males aged 51 and above in the United States. The main features of this model are the multiple dimensions of health and their retirement effects through occupation-dependent channels. 
 Individuals across occupations also differ in a number of aspects empirically, due to the rich heterogeneity captured by the model, such as physical health status, cognitive health status, education, labor earnings, history of earnings, Social Security benefits, private pension benefits, health insurance, disability insurance, life expectancy, unobserved individual type and so on.

\subsection{Choice Set}

  The model focuses on the labor supply and saving decisions of older workers in different occupations.  Occupations are predetermined by choices at younger ages, which is abstracted from the model, while the endogeneity of the initial distribution will be taken care of.
Individuals who have worked in the last period belong to one of the three occupation categories:  $j=1$ includes manual and service occupations;  $j=2$ consists of sales and clerical occupations, and  $j=3$ covers managerial and professional occupations. These three categories will be referred as the manual, clerical and professional occupation throughout.  This classification is consistent with the vast studies about the changes in labor markets. (e.g. \cite{katz1992changes}, \cite{david2013growth}  ), as I intend to highlight the implications of physical and cognitive health for these common occupations. I use the term ``occupation'' interchangeably with the occupation groups defined here.\footnote{Details about the classification are provided in Appendix \ref{Appendix:OCC}.  While the model abstracts from the shifts between the manual, clerical and professional categories, it implicitly allows for the change within each group. Conceptually, there are three behavioral responses to poor health: (1) changing occupation within the category; (2) changing occupation across the category; (3) exiting the labor force. The model allows for the first response and focuses on the last response, which should be the major one.  The incidence of between-category shift is low in the data. One reason could be that the categories are broad enough and occupation changes happen mainly within the category.}

 At each age, the worker chooses whether to participate in the labor force. The labor supply decision at age t is denoted by $d_t$, which equals to 1 if the individual chooses to work.  Being out of the labor force is assumed as an absorbing state. Besides the labor supply decision, the individual also chooses how much to consume. Consumption is a continuous decision variable. Consumption optimization is subject to an upper bound imposed by the borrowing constraint, and a lower bound which is the consumption floor insured by the government. The consumption floor captures public transfers provided by programs such as the Supplemental Insurance Income (SSI) for the poor people. Individuals make their labor supply decisions up to age $A^*=75$ and consumption decisions until age $A^{**}=90$.

This article refrains from modeling individual's Social Security application decision separately from their decision of leaving the labor force, \`{a} la \cite{blaugilleskie08}, \cite{vanderklaauwwolpin08} and \cite{boundetal10}. \cite{gustman2008projecting} report that the percentage of individuals who have fully retired and are receiving benefits ranges from 86.5\% to 95.2\%, on which \cite{coile2015economic} concludes: ``Claiming delays are fairly uncommon, relatively short, or both''. I follow \cite{blaugilleskie08} and \cite{boundetal10} to assume individuals start to collect Social Security benefits in their first year being out of the labor force after reaching the early retirement age 62.\footnote{Studies incorporating the joint decisions of labor supply and Social Security applications are referred to \cite{gustman2015effects}, for one example.  } 

\subsection{Utility Function}
  The utility function consists of both the pecuniary and non-pecuniary components. Pecuniary utility is derived from consumptions in the constant relative risk averse (CRRA) form, with the coefficient of risk aversion $\nu$. \footnote{The numerator of the pecuniary conponent includes  the constant -1, so that the utility function is continuous in $\nu$ around $\nu=1$. This guarantees the smooth parameter searching in estimation.} Non-pecuniary utility $L_t$ depends on individual's labor supply status $d_t$ and the interaction with his physical health  and cognitive health. $h_t^p$ and $h_t^c$ are indicators of poor physical and cognitive health which equal one if the status is poor and zero otherwise.  There is disutility $\lambda_{1j}^{type}$ when the individual chooses to work,  capturing the loss of leisure time. Moreover, there is extra disutility of work under poor physical or cognitive health, characterized by the parameters $\lambda_{2j}$ and $\lambda_{3j}$. These parameters are specific to occupation j,  reflecting the insight that workers with poor physical or cognitive health can suffer from working differently across occupations.

 There is also unobserved individual type based on the finite mixture structure as \cite{french2011effects} and \cite{vanderklaauwwolpin08}. Individuals with  unobserved types vary in their disutility of work $\lambda_{1j}^{type}$. Each individual's type is given by the type probability functions, which link individual's unobserved type to the observed state variables in the initial wave, including initial physical health, cognitive health, education, labor supply status and occupation, as well as initial assets. The endogeneity of the initial distributions are assumed being taken care by these unobserved individual heterogeneity. 
  \begin{align}
  	U(C_{it}, d_{t})=&\frac{1}{1-\nu} (C_{t}^{1-\nu}-1) +L_{t} \\
\text{where} \qquad
  	L_{t}= \lambda_{1j}^{{type}} & d_{t} + \big (\lambda_{2j} h_{t}^p+\lambda_{3j} h_{t}^c \big)  \cdot d_{t} +\varepsilon_{t}^{d}
  \end{align}
The non-pecuniary  utility is also subject to choice-specific idiosyncratic preference shocks $\varepsilon_{t}^{d_{t}}$. It is assumed following the identical and independent extreme value type one distribution. 
The structural interpretation of this shock is that it captures those state variables unobserved to researchers but observed to individuals in the model. Notice that the nonpecuniary utility is assumed to be additive to the pecuniary utility, so that consumption decisions are independent of this idiosyncratic shock once conditioned on the labor supply decision. Therefore, conditioning on the observed state variables in the model, the idiosyncratic preference shock $\varepsilon$ introduces heterogeneity only to the  labor supply decision. To further allow for  heterogeneous consumption decisions, the model introduces another idiosyncratic unobserved state variable directly affecting the budget constraint. This extra shock is assumed as a component attached to the total income.\\

\subsection{Budget Constraint} 

Assets accumulate according to Equation (3), where $A_t$ is the asset at the beginning of period t, $Y_t$ is the total income, $C_t$ is the consumption, $ME_t$ is the out-of-pocket medical expenditure, and $r$ is a fixed interest rate. $A_{t+1}$ is the decision variable, characterizing how much assets the individual leaves for the next period.
\begin{align}
(1+r)A_{t}+Y_{t}=C_{t}+A_{t+1}+ME_{t}
\end{align}

 The asset transition is subject to a borrowing constraint $A_{t+1}=(1+r)A_t+Y_t-ME_t-C_t\ge A_{min}$, where $A_{min}$ is the minimum asset required and it can be negative. The borrowing constraint caps the consumption in each period. There is also a consumption floor $C_{min}$, capturing government transfers such as the Supplement Security Income(SSI) and the Medicaid for individuals in deep poverty, in line with \cite{hubbard1995precautionary}. Therefore, in each period the individual can choose his consumption between the range $[C_{min}, C_{Max}]$, where $C_{max}=(1+r)A_t+Y_t-ME_t-A_{min}$ and $C_{min}$ is a constant. The government transfer takes place in the extreme case when individual's asset and income are too low, or the out-of-pocket medical expense is too high, leading to $C_{max}<=C_{min}$. The amount of government transfers thus equals to $max\{0,C_{min}-((1+r)A_t+Y_t-ME_t-A_{min})\}$.  \\


\subsection{Income}
An individual's total income consists of the labor earnings $W_t^j$, the Social Security benefits $SS_t$, the private pension benefits $P_t$, the spousal income $Y_t^s$, the income from Social Security Disability Insurance $SSDI_t$, and an unobserved idiosyncratic component $\zeta_t$:
\begin{align}
	Y_t=   W_t^j \cdot \mathbf{1}(d_t=1)+(SS_t+P_t) \cdot (1-\mathbf{1}(d_t=1))+Y_t^s+SSDI_t+\zeta_t 
\end{align}

\subsubsection{Labor Earnings}

 The labor income is a product of the skill rental price $r^j$ and the index of human capital. Embedding the  human capital model by \cite{grossman1972concept} into the earnings equation by \cite{mincer1974schooling}, the human capital index depends on individual's experience $X_t$, education $E$, as well as the status of physical and cognitive health. These human capital determinants are allowed to have different returns across occupations, as emphasized by \cite{kambourov2009occupational}. Poor physical and cognitive health are allowed to have penalty on the human capital specific to the occupation.
\begin{align}
W_{t}^j=r^j \cdot \exp \big( \kappa_1^j +\kappa_2^j X_t +\kappa_3^j X_t^2 +\kappa_4^j E + \kappa_5^j h_t^p+\kappa_6^j h_t^c \big)
\end{align}

\subsubsection{Social Security and Private Pension}
The model calculates Social Security retirement benefits $SS_t$ closely following the rule of Social Security Administration. Specifically, retirement benefits are calculated in following steps: First, individual's highest 35 years earnings are included to calculate the Average Indexed Monthly Earnings (AIME). The earnings before age 60 are adjusted by the national average wage index to reflect the real wage increase. In the second step, Primary Insurance Amounts(PIA) is computed as a piecewise linear function of AIME, with three separate percentages for different portions of AIME. It functions as a progressive taxation. Finally,  the PIA is multiplied by an adjustment factor dependent on the age at which the individual claimed benefits. For example, when the full retirement age (FRA) is 65, an individual who claimed his benefits at age 65 receives the amount as much as 100\% of the PIA, while another individual claimed at age 62 can only receive 80\%.  
 AIME serves as a state variable in the model. It will be constructed from the administrative earnings history data from the Social Security Administration (SSA). 


Based on the formula by which AIME is computed, the transition of AIME takes the following form.  $W_t$ denotes the current labor income and $\dot{W}_{t-1}$ denotes the earnings history until age t-1:
\begin{align*}
AIME_{t+1}=AIME_{t}+ \max\{0,W_t-\min(\dot{W}_{t-1})\}/35
\end{align*}

Notice that the AIME is updated only when the current labor income is higher than the lowest labor income among the existing 35 years included for previous AIME calculation. If the individual hasn't accumulated 35 working years, $\min(\dot{W}_{t-1})$ is 0 and working always contributes to increasing the AIME. Modelling the transition process precisely requires us to keep track of the whole earnings history of the individual to calculate the $min(\dot{W}_{t-1})$, which is intractable.  
Following \cite{french05}, this paper approximates it by $\alpha_t AIME_t$ with $\alpha_t=1$. More details about the modelling of Social Security are provided in Appendix \ref{Appendix:SS}.

Private pension is difficult to model because the plans vary with each individual. \cite{boundetal10} solves the dynamic programming model by each individual, at the expense of having a sample with only 196 individuals. \cite{vanderklaauwwolpin08} restricts the sample to low income individuals with private defined benefit pension from the previous jobs and drop those with defined contribution pensions and defined benefit pensions from the current employers. Alternatively, \cite{french05} and \cite{french2011effects} approximate the private pension benefits as a function of state variables existing in the model. The goal of this paper requires a representative sample of individuals from each occupation. To maintain reasonable sample size, this paper also approximates the private pension benefits. Appendix \ref{Appendix:PrivatePension} provides details on the modeling of private pension.

\subsubsection{Social Security Disability Insurance (SSDI)}
This paper also includes Social Security Disability Insurance (SSDI) by allowing its benefits to shift the budget constraint. The data suggests SSDI only accounts for a small portion of household's total income, 1.6\% on average. However, different from the Supplemental Security Income (SSI), which targets at individuals with limited income and can be captured by the consumption floor $C_{min}$ in our model, SSDI is accessible to individuals regardless of their current income as long as they meet certain medication conditions and requirements for work and tax history. Omitting SSDI may bias the estimated effects of different types of health on individuals' behaviors.    Tractability limites adding the SSDI application  as another decision variable. The SSDI status is therefore taken as exogenous.\footnote{Readers interested in structural work with endogenous DI application and retirement are referred to \cite{boundetal10} and \cite{french2012effect}, as well as \cite{low2015disability} under the life-cycle setting.  } I assume the eligibility for SSID is a function of age and both dimensions of health, which will be estimated with the data. Conditioned on being eligible, the amount of SSDI benefits is calculated precisely based on  the rule of SSA, as a function of individual's AIME.

\subsubsection{Other Income Sources}
The spousal income is predicted by a set of demographic variables similar to \cite{french05} and \cite{french2011effects}.\footnote{ To avoid adding extra state variables, I assume spousal income depends on individual's age and education. Instead, \cite{french05} assumes spouse income is a function of individual's own income and age.}  The random income component $\zeta_{t}$ is unobserved to researchers but observed to the individuals in the model. It follows a normal distribution with mean 0 and variance $\sigma_{\zeta}^2$. It introduces further heterogeneity to individual's consumption. Without this component, individual's consumption will be homogeneous conditioned on the observed state variables.     \\

\subsection{Health, Medical Expense and Health Insurance}

The physical and  cognitive health transit jointly with stochasticity. Transition of the joint health from period t to t+1 depends on individual's age $t$, education $E$, occupation $j$ and labor force participation $d_t$, as well as the joint health status in period t $h_t$. The descriptive facts show that the transition of health mainly differs in levels and maintains broadly parallel trends across occupations. Nevertheless, the model allows the health transition to depend on labor supply by occupations to accommodate the  effects of retirement on health, as revealed by a strand of literature on the effect of retirement on health (e.g. \cite{eibich2015understanding}, \cite{rohwedderwillis10}).  The transition is also subject to a stochastic component $u_t$ with normal distribution. The joint health status is a vector of the physical health $h_t^p$ and the cognitive health $h_t^c$ , which takes four states: both physical and cognitive health are in good status \{$h_t^p=0, h_t^c=0$\}, only physical or cognitive health is in good status \{$h_t^p=0, h_t^c=1$\} or \{$h_t^p=1, h_t^c=0$\}, and both are poor \{$h_t^p=1, h_t^c=1$\}.
\begin{align}
h_{t+1}=H(t, E, j, d_t, h_t, u_t  )
\end{align}

Similar to \cite{vanderklaauwwolpin08}, the out-of-pocket medical expenditure $ME_t$ is a function of the joint health status $h_t$ and the insurane type $H^*_t$. 
\begin{align}
ME_t=g(h_t,H^*_t)
\end{align}

\noindent The health insurance $H^*_t$ has four types. The first three types are employer-provided insurance, which are applicable when the individual is younger than 65. When the individual is above age 65, he receives the public health insurance Medicare. 
\begin{itemize}
 \setlength\itemsep{0.01cm}
\item $H^*_t=0$: No insurance;
\item $H^*_t=1$: Employer-provided health insurance without retiree coverage;
\item $H^*_t=2$: Employer-provided health with retiree coverage;
\item $H^*_t=3$: Public health insurance Medicare.
\end{itemize}
\vspace{2mm}



Medical expenditure may affect older workers's retirement dicision by interacting with the health insurance (\cite{rust1997social}, \cite{french2011effects}). The insurance without retiree coverage requires the individual working for the employer to be insured, whereas the one with retiree coverage provides insurance until age 65 regardless of the individual's subsequent employment. Individuals without reteree coverage therefore face extra incentive to work until 65. Conditioned on the states in period t, the insurance type in period t+1 is given by the following transition rules:
\begin{itemize}
 \setlength\itemsep{0.01cm}
  \item If $t\ge65$, $H^*_t=3$;
  \item If $H^*_t=2$,  $H^*_{t+1}=2$ until $t+1 \ge 65$;
  \item If $H^*_t=1$,  $H^*_{t+1}=1$ if $d_t=1$ and  $H^*_{t+1}=0$ if $d_t=0$;
  \item If $H^*_t=0$,  $H^*_{t+1}=0$
\end{itemize}

\subsection{Value Function}

 Taking all factors aforementioned into account, the individual makes his labor supply and consumption decisions at each age to maximize the current utility plus the present discounted utility from the future. The Bellman equaiton is given by Equaiton (8), where $\Omega_t$ represents the union of all state variables at age t, both observed and unobserved. 
\begin{align}
V_t(\Omega_{t})=\max_{ d_t, C_{t} } \Big \{U(d_t, C_t ,\Omega_{t})+\beta \int \big ( p_t    V_{t+1}(\Omega_{t+1}) +(1-p_t)B(\Omega_{t+1})  \big )  d F_t(\Omega_{t+1}|d_t, C_t ,\Omega_{t} ) \Big \}
\end{align}

The continuation value equals to either the value function in next period $V_{t+1}(\Omega_{t+1})$ or the utility from bequest $B(\Omega_{t+1})$.  The bequest utility is a function of  the asset left to the future with the following form:  
\begin{align}
B(\Omega_{t+1})=\iota_1 \frac{(A_{t+1}+\iota_2)^{1-\nu} -1 }{1-\nu}
\end{align}
 Aligned with \cite{de2004wealth} and \cite{de2016medicaid}, I allow bequests to be  luxury goods, which helps explain the little wealth accumulation for poor people. A positive $\iota_2$ reduces the marginal utility of bequest especially for individuals with limited assets, generating extra disincentives of bequest,  as the marginal utility of consumption becomes relatively larger. 

Utility in the future is discounted by the subjective discount factor $\beta$ and the survival rate $p_t$.  The survival rate depends on age and the joint status of physical and cognitive health $h_t$.  $s_t$ is the indicator for survival at age t.  Poor health may shorten the life expectancy and thus change the labor supply and saving decisions under the dynamic setting. 
\begin{align}
p_t=Pr(s_{t+1}=1|s_t=1, h_t )
\end{align}

 To summarize, the union of the state variables $\Omega_t$ includes : age $t$, education $E$, occupation $j$, Average Indexed Monthly Earnings $\text{AIME}_t$, physical health $h_t^p$, cognitive health $h_t^c$, asset $A_t$, Social Security application status in the last period, health insurance type $H_t^{*}$, labor supply status in the last period $d_{t-1}$,  the idiosyncratic components to total income $\zeta_{t}$ and to the non-pecuniary utility $\varepsilon_{t}$. Notice that both the AIME and the assets are included as continuous state variables. In numerical practice, the state space contains 153.12 million state points in total.

\subsection{Channels of Health in Affecting Retirement}
Finally, this subsection concludes the model section by summarizing the channels through which health may affect retirement, and their links to the literature.

As surveyed by \cite{french2017health}, literature typically indicates five channels through which health may affect retirement, which is based on the unitary concept of health  with arguably larger, if not all, weights on physical dimensions.   Firstly, poor health may increase the demand for leisure as the work inflicts extra disutility. This channel could also be important for cognitive health. Cognitive decline may implicitly lead to higher incidence of mistakes, such as bugs for programmers, omitted deadlines for clerks, miscalculation for accountants, typos for editors and so on. While each individual mistake may not be grave, the accumulation could be accompanied by upset and  pressure (\cite{morrow2010predicting}).  Second, poor health could lower the productivity and wage.  Examples in the first channel  may also translate to lower wage, especially under environments where wages are flexibly determined and performance is observed by employers. Third, medical expenditure increases with poor health, which could affect retirement behavior by interacting with the type of health insurance. Previous studies have identified that employer-provided health insurance provides strong incentives for older workers to work until age 65 (\cite{rust1997social}). Fourth, health is directly linked to life-expectancy. Poor cognitive health may well affect mortality rates at advanced ages by developing into Alzheimer, which matters for retirement decisions if individuals are forward-looking (\cite{wolfson2001reevaluation}). Finally, poor health may also affect retirement decision through the application of disability insurance (\cite{maestas2013does}).

Most studies include a subset of these channels and consider the unitary health (e.g. \cite{capatina2015life}, \cite{french2011effects}).  I allow the above channels for both physical and cognitive health, and their quantitative importance will be identified by the data.  The channels of leisure and productivity allow direct occupation-dependent effects via corresponding structural parameters, characterizing the heterogeneous ability requirements across occupations. The importance of other channels also differ between occupations due to their gap in health and socioeconomic status. 

The transition of cognitive health and physical health are interactive. Two individuals with the same physical but different cognitive health status face different probabilities of becoming physically unhealthy in the future. At the same time, these two individuals also face different mortality rates due to the different joint health status.


\section{Solution and Estimation Methods}
\subsection{Model Solution}
I solve the finite-horizon life-cycle model by backward induction. The policy functions of the model, which include the discrete labor supply choice and the continuous consumption decision, has no closed form and are obtained numerically. The solution process follows the steps stated as below.

\begin{enumerate}
 \setlength\itemsep{0.01cm}
  \item Calculate the choice-specific value at age t: $CSV_t(d_t, C_t, \widetilde{\Omega}_t, \zeta_t, \varepsilon_t^d)$, which is the sum of the current payoff:  $ U(d_t, C_t ,\Omega_{t}) $ plus the expected value function at age t+1:  $\beta \int \big ( p_t    V_{t+1}(\Omega_{t+1}) +(1-p_t)B(\Omega_{t+1})  \big )  d F_t(\Omega_{t+1}|d_t, C_t ,\Omega_{t} )$. The latter component requires the solution at age t+1, obtained by solving the model backward.  $ \widetilde{\Omega}_t$ denotes the union of observed state variables, while $\Omega_t$ includes also $\zeta_{t}$ and $\varepsilon_{t}^d$.
   
  \item Given each labor supply choice, search for the optimal consumption that maximizes the choice-specific value function $CSV_t$ evaluated at each possible value of the union of state variables, both observed and unobserved. Notice that the consumption optimization given each labor supply choice is independent of the shock $\varepsilon^d_t$ due to the additivity between pecuniary and non-pecuniary utility.   In this step,  for each labor supply choice, I obtain the optimal consumption $C_t^*(d_t, \widetilde{\Omega}_t, \zeta_t)$ and the corresponding optimal value $CSV_t^{*}( d_t,C_t^*,  \widetilde{\Omega}_t, \zeta_t, \varepsilon_t^d)$. 
  
  \item Compare the choice-specific value  $CSV_t^{*}( d_t,C_t^*,  \widetilde{\Omega}_t, \zeta_t, \varepsilon_t^d)$  across the labor supply choices. The optimal labor supply choice is deterministic conditional on $\zeta_t$ and $ \varepsilon_t^d$. Conditioned on $\zeta_{t}$ only and integrating out the extreme-value type-I distributed $\varepsilon_{t}$, the labor supply decision rule is the conditional choice probability following the logistic closed form. The model solutions are eventually characterized by the labor supply and the optimal consumption decisions. They are deterministic functions of the observed state variables $ \widetilde{\Omega}_t $ and the unobserved state variables $\varepsilon_t^d$ and $\zeta_t$. However, conditional only on $ \widetilde{\Omega}_t $, the decisions are stochastic.
\end{enumerate}

 In the second step, I search for the optimal consumption  for each given labor supply choice. Because the choice-specific value function $CSV_t(d_t, C_t, \widetilde{\Omega}_t, \zeta_t, \varepsilon_t^d)$ may be unsmooth in consumption, due to the consumption floor and the borrowing constraint, it is inappropriate to use derivative-based optimization methods to search for the optimal consumption. Instead, I discretize the consumption into finite grid points and search over these points, after which I implement interpolation to evaluate the value function on the continuous consumption.\footnote{ I use the method by \cite{french2011effects} to facilitate this search process. In particular, I only search over all the grid points in the final stage during the backward induction. For earlier stages, to obtain the optimal consumption for a given set of the state variables $\Omega_t$, I start the search from the initial value which is set as the optimal value of consumption in period t+1 evaluated at the same states as $\Omega_t$. That is, the search in period t starts from the value $C_{t+1}^*(d_{t}, \widetilde{\Omega}_{t}, \zeta_{t})$.  Based on this initial point, I then search over a neighborhood instead of the whole consumption space.  I compared the results with the full search results, and the difference is minimal. More details are provided in Appendix \ref{Appendix:Grid}. } 

\subsection{Estimation}
Upon solving the model, structural parameters are estimated in two steps, aligned with \cite{iskhakov2020effects}, \cite{french2011effects} and \cite{rust1997social}, among others. Parameters of the health transition equations, the survival probability equations, as well as of the functions of medical expenditure, SSDI eligibility, private pension, spousal income and labor income, are estimated in the first step.\footnote{Appendix \ref{Appendix:FirstStage} provides details about the first stage estimation. } Preference parameters and parameters of the type probability functions are estimated jointly in the second step by indirect inference.\footnote{To handle the computational burden, I deploy parallel computing with Fortran. However, while the test programs can be run on HPCs from the lab of my affiliation, the formal program has to be executed on the virtual desktop infrastructure (VDI) system provided by HRS, as the estimation needs to use the restricted data. This constraint indeed brings challenges.} 

Indirect inference is a simulation-based estimation approach, essentially a generalized simulated method of moment (SMM) (\cite{gourieroux1993indirect}, \cite{smith1993estimating}). It searches for structural parameters that simulate the data as close as to the observed data.\footnote{For readers interested in the performance of the simulated method of moments and the maximum likelihood, \cite{eisenhauer2015estimation} provide a comparison based on the dynamic discrete choice model. } SMM adopts a set of moments as the criteria of comparing the simulated  and  actual data, whereas indirect inference is based on the parameters of auxiliary models. Parameters of the auxiliary models are estimated using the actual and simulated data respectively.  Notice that parameters of the auxiliary models estimated on the simulated data are functions of the structural parameters. Indirect inference thus searches for values of the structural parameters that minimize the distance between these two sets of estimated parameters. Whether the auxiliary models are correctly specified does not affect the consistency of the structural estimates, but a set of well-chosen auxiliary models improve the efficiency. Analogous to the hypothesis test of Wald, LR and LM, there are three metrics to construct the objective function. Detailed discussion can be found in \cite{bruins2018generalized}. This paper adopts the metric analogous to Wald and use the diagonal weighting matrix to construct the objective function.\footnote{Elements on the diagonal are the variances of  auxiliary parameters estimated on the actual data.} 

The auxiliary models provide a transparent link between the structural model and the reduced form empirical facts.   The flexibility of choosing auxiliary models allows me to identify the structural model with different variations.  For instance, in different specifications, I use pooled regressions as well as regressions with individual fixed-effects as the auxiliary labor supply models. While the different findings based on different variations are one of the key focus among the reduced work on the retirement effects of health, structural models of retirement have little touch on this issue.   The idea here is in line with rising structural work that exploits more delicate variations to identify structural parameters, such as  \cite{fu2019estimation}  who choose auxiliary models with a regression-discontinuity setup. 

\subsubsection{Auxiliary Models}
Guided by both the reduced form empirical facts  as well as the structural decision rules, I choose the following auxiliary models to help identify the parameters of preference and type probability functions:
\begin{itemize}
 \setlength\itemsep{0.02cm}
  \item The linear probability models that regress the binary labor supply indicator on age dummies from age 51 to 75, separately by occupations. 
  \item The linear probability models that regress the binary labor supply indicator on physical health, cognitive health, age, log assets, and the indicator of negative assets, conditioned on being in the labor force at the last age, separately by occupations.
  \item The regression of asset changes on age dummies from age 51 to 75.
  \item The age-profiles of lower and upper tertiles of assets from age 51 to 75. To reduce variance, observations are grouped into five age groups as [51,55),  [55,60), [60,65) , [65,70) and [70,75).
  \item  The linear probability models that regress the binary labor supply indicator on education, initial physical health, initial cognitive health, initial log assets, initial indicator of negative assets, initial occupations, separately by individual's type about whether he enjoys work.  The regressions are repeated three times, using the labor supply in the first, second and third period after each individual's initial period  respectively. 
  \item The gaps in labor supply from age 51 to 75 between individuals of different types of enjoying work,  different education, different initial physical health, different cognitive health, and  different initial wealth.  
\end{itemize}

\subsubsection{Identification}

For parameters of the pecuniary utility component, the coefficient of risk aversion $\nu$ is mainly identified by the age-profiles of assets. Individual's labor supply also places an upper bound for this coefficient, as suggested by \cite{chetty2006new}.  With larger value of $\nu$, individuals tend to accumulate more assets and stay employed to insure against future risks in health, income and survival. Parameters related to the bequest motives, $\iota_1$ and $\iota_2$, are mainly identified by the assets profiles at older ages, and by the gap between rich and poor individuals. Heuristically, larger assets gap between rich and poor individuals at older ages identify strong bequest motives. \cite{de2004wealth} provides detailed discussions.

For parameters estimated in the first step, the coefficients of physical and cognitive health in wage, medical expenditure and survival equations are identified by the variations of corresponding outcome variables against the variations of health.  For the wage equation, the eligible ages for retirement benefits and Medicare (i.e. the indicator variables of age 62 and 65), conditional on the smooth function of age, are used as exclusion restriction in a Heckman selection model. The underlying assumption is being eligible for retirement benefits only affects participation but not wage. Then the variations of labor supply, net of the ones induced through wage, medical expenditure, life expectancy and all other pay-off variables, against the variations of health pin down the disutility of work due to poor physical and cognitive health $\lambda_{2j}$ and $\lambda_{3j}$. In particular,  the occupation gradients identify the occupation-specific disutility of work due to poor physical and cognitive health respectively. The baseline level of labor supply in each occupation identifies the occupation-dependent disutility of work $\lambda_{1j}$.

In various specifications, I use the pooled regressions as well as the regressions with individual fixed effects as the auxiliary model of labor supply. Parameters of the  auxiliary models based on the pooled regressions capture the correlations between health and labor supply identified by a mix of the within- and between-individual variations. Specifically, the coefficients of physical and cognitive health in the auxiliary model are identified by: (1) an unhealthy individual who is retired and another healthy one who stays in the labor force;  (2)  a given individual who was in the labor force with good health and exited when his health turned bad.  Based on the auxiliary models with individual fixed-effects,  if the second type of variation is rare in the data, the structural parameters related to health will not be identified or identified as having the small contribution to retirement.

Finally, type-specific preference parameters and parameters of the type probability functions are identified by the regressions of labor supply in subsequent waves on the initial state variables, as well as by the gaps in long-term labor supply between individuals of different initial states.  As discussed in Subsection 3.2, initial state variables are linked to individual's unobserved type, which is time-variant and can have impacts on individual's decisions in subsequent periods persistently.  Take the coefficient of physical health of the type probability functions as an example. Ceteris paribus, different types of individuals differ in their disutility of work $\lambda_{1j}^{type}$. If individuals with good initial physical health are persistently more likely to participate in the labor force than those with poor initial physical health, then this cross-sectional variation of the long-term labor supply against the cross-sectional variation of initial physical health will help pin down the link between initial physical health and individual's unobserved type.\footnote{ I have also tried to augment the identification by conditioning on individuals with different work preference, measured by the variable about the degree of enjoying work of each individual obtained from HRS, following \cite{french2011effects}. }


\section{Data and Variables}

The structural model is estimated on the data from the third to eleventh wave of Health and Retirement Study (1996-2011), combined with the administrative data of individual earnings history from the Social Security Administration. HRS is a biennial  longitudinal survey of representative older individuals in the United States. It provides detailed measures of individual's  health, labor market outcomes, as well as financial conditions. I exclude the first and second waves  because the cognitive measure is inconsistent with subsequent waves.\footnote{For the word recall test, the first two waves use a list of 20 words whereas the rest waves use a list with 10 words. I refrain from re-scaling the measure in the first two waves because I found the distribution of the re-scaled measure is nevertheless very different from the subsequent waves .} The  sample consists of male individuals aged 51-61 in their initial waves, covering them until age 75 after which observations are very sparse. Individuals who have never been in the labor force throughout all observed waves are excluded. The estimation sample also drops observations that reported multiple labor supply statuses, as well as that changed occupations or returned to the labor force since the last wave.\footnote{The last two sample restrictions are imposed because the structural model abstracts from the occupation change and assumes out of the labor force as an absorbing state. The change of sample size is modest. Without imposing these two restrictions, we end up with 17,565 observations and 4,091 individuals. } Also, observations with missing values in state variable are dropped. Finally, the sample includes 17,305 observations and 4,085 individuals.

This primary sample is used for the second stage estimation by indirect inference. The first stage estimation is based on an expanded sample.  Due to the sample restriction that individuals should age 51-61 when they entered the sample, the oldest age in the primary sample is only 78. Notice that  agents in  the model form expectations regarding their future health, pension, survival, medical expense and so on until age 90.  For this reason, the first stage estimation is based on the full sample of males from the third to eleventh waves of HRS.

The labor supply is defined as the status of labor force participation, derived from the variable of current job status from the HRS. Working now, unemployed and looking for work, as well as temporarily laid off are considered as in the labor force. Out of the labor force includes disabled, retired and homemaker. One of the main concerns from the literature on the retirement effects of health is the `` justification bias '', about which individuals who retired early tend to report worse health to justify their early retirement  (e.g. \cite{dwyer1999health}). I follow this literature to use more objective measures to instrument the self-reported health, similar to \cite{disney2006ill} and \cite{boundetal10}.  Physical health is thus constructed as a health stock index predicted from an ordered probit of self-reported health regressing on more specific variables regarding individual's physical dimensions of health, which are less likely to suffer from the bias.  Cognitive health is measured by the number of words recalled by the respondent in a standardized words recall test (\cite{wallace1995overview}). This variable reflects the status of individual's episodic memory, which is an important aspect of the fluid cognition. Discussions about health are also presented in the section of empirical facts.
In structural estimation, state variables are discrete and I construct the indicator variables of poor physical and cognitive health.\footnote {Poor physical or cognitive health is defined as when the continuous measure is below the 30 percentile of the distribution. This discretization aims to be consistent with the measure of the self-reported health, of which around 30\% observations report ``poor'' or ``very poor''. The gradients documented in the empirical facts section are robust to this discretization.}

The labor income is measured as the annual labor earnings during the year before the interview of the respondent. Spousal's income includes the overall income from the spouse, including labor income and Social Security benefits, private pension, SSDI and other government transfers. AIME is calculated following the rules of SSA, using the earnings data of the Master Earnings File (MEF) from the SSA.  The variable about assets includes both housing and individual retirement account (IRA). The model explicitly characterizes the defined benefit but not the defined contribution component of the private pension. Therefore, define contribution plan functions equivalently as saving and the IRA is included into the assets variable to reflect the accumulation.  All monetary variables are converted to the dollars in 1999. 


Table \ref{table:desdata} presents the descriptive statistics for the main variables, by  labor supply status and  occupations of the current job. In terms of the economic variables, there is a large disparity across occupations. Individuals from the manual occupation earn less than those from the clerical and professional occupation. These people have much less total assets too,  while those from the clerical occupation fall in the middle. People from the professional occupation have 3.4 times assets as much as those from the manual occupation on average.  Moreover, manual workers tend to have lower AIME, which implies lower Social Security benefits when they retire. Notice that annual workers actually have higher replacement rates whereas they  tend to have shorter life expectancy due to worse health. The statistics  show that individuals from the professional occupations have higher education and better health in both physical and cognitive dimensions. Finally, for observations younger than age 65, professional occupations are more likely to provide  retiree coverage health insurance, whereas the proportion of no insurance is notably lower than the other occupations.

\begin{table}[H]
	\centering
	\footnotesize
	
	\caption{ Descriptive Statistics of Main Variables}
	\begin{minipage}{0.85\textwidth}
		 \tabcolsep=0.45cm\begin{tabular}{p{0.2\textwidth}>{\centering}p{0.1\textwidth}>{\centering}p{0.1\textwidth}>{\centering}p{0.1\textwidth}>{\centering\arraybackslash}p{0.1\textwidth}}
\hline
\hline
		& Manual & Clerical & Professional & Retired \\
		\hline
			Age                              & 59.26         & 59.94         & 59.58    & 66.40  \\			
		                                     &  (4.92)       &  (5.43)       & (5.25)   & (4.74)  \\
\multicolumn{1}{l}{Wage, thousand dollars}	 & 33.67         & 47.72         & 70.74    & -  \\
					                         &  (23.36)      &  (56.58)      & (57.49)  &  -  \\
		\multicolumn{1}{l}{Assets, thousand dollars}	 & 184.85  & 382.41  & 628.55  & 389.56  \\
                                   &    (509.30)     &  (979.25)       &  (1444.42)  &  (866.42)  \\			               
\multicolumn{1}{l}{	AIME, thousand dollars } &    34.27      &    39.40      &   47.18    &		36.45     \\     
			                                 &   (18.26)     &   (19.13)     &    (22.14)   &	  (18.04)  \\
			\multicolumn{1}{l}{Physical Health} &                &       &       &  \\
			\quad Poor & 0.15   & 0.11  &  0.06  & 0.29 \\
			\quad Good &  0.85  & 0.89  &  0.94 &  0.71 \\
		\multicolumn{1}{l}{	Cognitive Health} &        &              &       &  \\
			\quad  Poor &  0.22   & 0.13     & 0.06    & 0.25\\
			\quad Good &  0.78  & 0.87   & 0.94   & 0.75  \\
		\multicolumn{1}{l}{	Education }  &        &            &       &  \\
			\multicolumn{1}{l}{\quad Less High School} &  0.24  & 0.07   & 0.02   & 0.19 \\
		\multicolumn{1}{l}{	\quad High School} &  0.45   & 0.27  & 0.13  & 0.37 \\
		\multicolumn{1}{l}{	\quad Some College} &  0.24   & 0.35  & 0.19  & 0.22 \\
		\multicolumn{1}{l}{	\quad College and Above}&  0.07    & 0.30   & 0.66  & 0.22 \\
		\multicolumn{1}{l}{	Observations }  &   5,532    & 1,819 & 4,150  & 5,806  \\
		\hline
			\multicolumn{1}{l}{	Insurance Type, Age$<$65 }         &       &         &       &  \\
			\multicolumn{1}{l}{	\quad No insurance } & 0.34  & 0.40  & 0.26  & 0.50 \\
			\multicolumn{1}{l}{	\quad Tied insurance}  & 0.29   & 0.24  & 0.30     & 0.03 \\
			\multicolumn{1}{l}{	\quad Retiree covered } & 0.37   & 0.36  & 0.43 & 0.47 \\
			Observations & 4,704      & 1,455    & 3,395  & 1,991 \\
\hline
		\end{tabular} 
		\label{table:desdata}

		\scriptsize{Statistics are presented by occupations and labor supply status. Occupation is defined as the one of  current job. AIME is converted to the annual basis.  Standard deviations are reported in parenthesis.\par }
	\end{minipage}
\end{table}%

\section{Parameter Estimates and Model Fit}
\subsection{First Stage Estimates}

The estimates from the first stage suggest cognitive health affects both  the survival probabilities  as well as the transition of physical health. For two individuals aged 51 and both with good physical health, the one with good cognitive health has life expectancy 32.8 years compared to 28.5 years for the other one with poor cognitive health, if their health remains unchanged throughout the life cycle. If these two individuals have poor physical health instead, their life expectancy is reduced to 21.3 years and 18.6 years respectively. Individuals with poor physical and cognitive health also fact wage penalties which are different across occupations. While professional workers face the largest wage loss under poor cognitive health, they are the least affected under poor physical health.
More results and discussions about the first stage estimates are presented in Appendix \ref{Appendix:FirstStage}. 
The consumption floor and the minimum assets are respectively calibrated to be 4,000\$ and -5,000\$. This minimum assets thus allows for 5,000 \$ maximum debts. 

\subsection{Preference Parameters}

The preference parameters are estimated in the second step by indirect inference. Regardless of extensive explorations, parameters of the type probability functions seem poorly identified. Therefore I am more confident about the results based on the baseline model abstracted from unobserved individual types. Estimates of the model with unobserved individual types and related discussions are presented in Appendix \ref{Appendix:Unobserved}. Parameters estimated in the first step are held fixed in the second step estimation. I estimate the structural preference parameters using different auxiliary models to exploit different variations for identification.
The baseline specification leverages the within-individual variations in health and labor supply. The results are shown in Table \ref{Table:Estiamtes}.

\begin{table}[H]
	\centering
	\footnotesize
	\caption{ Estimates of Preference Parameters for The Baseline Model }
	\begin{minipage}{0.93\textwidth}
		\begin{tabular}{p{0.52\textwidth}>{\centering}p{0.12\textwidth}>{\centering}p{0.12\textwidth}>{\centering\arraybackslash}p{0.12\textwidth}} 
\hline
\hline
			& {Manual}   & {Clerical}   & {Professional}  \\
\hline
\multicolumn{1}{l}{\textbf{Non-pecuniary utility}}   &  & &\\
\quad  Extra disutility of work (poor p.h.) $-\lambda_2$ &0.633 &	0.292 &	0.162 \\
\quad   & (0.147) & (0.168)  & (0.162) \\
\quad  Extra disutility of work (poor c.h.) $-\lambda_3$ & 0.015 &  0.364  & 0.437  \\
\quad   & (0.054) & (0.240)  &  (0.194) \\
\quad  disutility of work   $-\lambda_1$ &  0.410 & -0.085  & 0.101 \\
\quad   & (0.101) & (0.054)  & (0.057) \\
\hline
\multicolumn{1}{l}{\textbf{Pecuniary utility}} &  & &  \\
   \quad  Bequest Motive $\iota$ &  \multicolumn{2}{l}{8.476 (3.98)}       &      \\
      \quad   Coefficient of Risk Aversion $\nu$&  \multicolumn{2}{l}{1.318 (0.110)}       &      \\
\hline
		\end{tabular} \\
		\scriptsize{This table presents the preference parameters estimated in the second step for the baseline model, abstracted from the unobserved types of individuals.  Within-individual variations are exploited to identified these parameters.  Standard errors are in parentheses. The calculation of standard errors is explained in Appendix \ref{Appendix:SE}.  \par }
	\end{minipage}
	\label{Table:Estiamtes}
\end{table}%

The estimate of the coefficient of risk aversion $\nu$ is 1.318, which is in the range of estimates in previous studies, such as   0.960-0.989 by \cite{blaugilleskie08},  1.07 by \cite{rust1997social}, 1.591 and 1.678 by  \cite{vanderklaauwwolpin08}, 2.565 by \cite{haanprowse14}. \cite{french05} and \cite{french2011effects} have larger estimates close to 5. For the parameters related to non-pecuniary utility, the  utility  is normalized as  zero, and  this is  the average utility for individuals with different health status. Notice that we are unable to separately identify the  utility of not working for individuals with different health based on the labor supply and health data.
$-\lambda_1$ characterizes the disutility of work when individual has both good physical and cognitive health. The estimates reveal the largest disutility for the manual occupation.

$-\lambda_2$ and $-\lambda_3$ are the main focuses of this paper, as the literature has emphasized that the disutility of work caused by poor health is an important channel for the retirement effect of health. I find $-\lambda_2$ is positive across all occupations, suggesting that poor physical health leads to extra disutility of work in all occupations. Moreover, this extra disutility is the largest for the manual occupation, with a value of 0.633, followed by the clerical and professional occupations with 0.292 and 0.162  respectively.  To the contrary, poor cognitive health induces extra disutility when work in the clerical and professional occupations with 0.364 and 0.437 respectively. This extra disutility of work is minimal for the manual occupation.

A caveat is that utility is essentially ordinal and incomparable. The subsequent counterfactual experiments help better understand the roles of physical and cognitive health in retirement.

\subsection{Model Fit}   
The key facts focused by this paper are the occupation gradients in the retirement effects of physical and cognitive health. I start by examining the fits of these gradients targeted by the estimation, which are characterized by the coefficients of physical and cognitive health in the auxiliary labor supply models by occupations. Although these facts are targeted, the success of fitting them should not be taken for granted, as we have way more targets than the structural parameters. Moreover, some targeted facts, such as the age profiles of labor supply, are very precisely estimated in the data. They are therefore given heavy weights in the loss function for estimation and dominate the search of optimal structural parameters. In other words, the occupation gradients are targeted, but with much less priority than a number of other targets. The first to the third column of Table \ref{table:fit} show the fits, which are close for both physical and cognitive health and for all occupations.

The above targeted facts are based on the auxiliary models, which need to be relatively simple for the convenience of structural estimation. Therefore, it is better to interpret them as merely correlations. While the gradients estimated with the real data are robust to including a number of control variables, as shown in the empirical facts section, the model simulated gradients need not to be so. For example, the model simulated gradients may be generated by heterogeneity in features other than the ability requirements. To ensure the robustness, I also control for a battery of variables that are not included in the structural model to estimate the simulated relationship between health and retirement.  This check can be interpreted as based on untargeted facts. The results are presented in the fourth to the sixth columns in Table \ref{table:fit}, and the simulated gradients still trace the data well.

\begin{table}[H]
\footnotesize
\centering
	\begin{minipage}{0.99\textwidth}
\caption{Model Fit: Occupation Gradients in Correlation between Health and Retirement}
\begin{tabular}{p{0.11\textwidth}>{\centering}p{0.07\textwidth}>{\centering}p{0.07\textwidth}>{\centering}p{0.07\textwidth}>{\centering}p{0.07\textwidth}>{\centering}p{0.07\textwidth}>{\centering}p{0.07\textwidth}>{\centering}p{0.07\textwidth}>{\centering}p{0.07\textwidth}>{\centering\arraybackslash}p{0.07\textwidth}}
\hline
\hline
                  & \multicolumn{3}{c}{Targeted }        &              \multicolumn{3}{c}{Untargeted: Within} & \multicolumn{3}{c}{Untargeted: Pooled } 
                       \\     \cmidrule(lr){2-4}      \cmidrule(lr){5-7}       \cmidrule(lr){8-10}
                                             & Man.              & Cler.            & Prof.         & Man.                                                                         & Cler.                                     & Prof.                                 & Man.                                                                         & Cler.                                     & Prof.                                \\ \hline
DATA                                             &           &         &   &               &       &   &               &   &  \\     
\hspace{0.1cm} Phy. H.                                    & 0.098***             & 0.061                & 0.026                & 0.106***                                                                       & 0.131***                                     & 0.089***                                     & 0.153***                                                                       & 0.147***                                     & 0.140***                                     \\
                                             & (0.021)              & (0.047)              & (0.032)              & (0.017)                                                                        & (0.040)                                      & (0.026)                                      & (0.015)                                                                        & (0.034)                                      & (0.029)                                      \\
\hspace{0.1cm} Cog. H.                                    & 0.015                & 0.061**              & 0.049**              & 0.014                                                                          & 0.081***                                     & 0.039**                                      & 0.026**                                                                        & 0.046*                                       & 0.046**                                      \\

                                             & (0.013)              & (0.026)              & (0.024)              & (0.011)                                                                        & (0.020)                                      & (0.019)                                      & (0.012)                                                                        & (0.025)                                      & (0.022)                                      \\ \hline
MODEL                            &           &         &   &               &       &   &               &   &  \\
\hspace{0.1cm} Phy. H.                                   & 0.087                & 0.058                & 0.022                & 0.113                                                                          & 0.154                                        & 0.110                                        & 0.154                                                                          & 0.150                                        & 0.143                                        \\
                                             & (0.002)              & (0.005)              & (0.003)              & (0.002)                                                                        & (0.005)                                      & (0.003)                                      & (0.002)                                                                        & (0.004)                                      & (0.003)                                      \\
\hspace{0.1cm} Cog. H.                                  & 0.005                & 0.057                & 0.044                & 0.013                                                                          & 0.068                                        & 0.033                                        & 0.024                                                                          & 0.053                                        & 0.042                                        \\
                                             & (0.001)              & (0.003)              & (0.003)              & (0.001)                                                                        & (0.003)                                      & (0.002)                                      & (0.001)                                                                        & (0.003)                                      & (0.002)                                      \\
                                      \hline
\end{tabular}
\label{table:fit}
\scriptsize{This table presents the occupation gradients in the correlation between retirement and physical and cognitive health, estimated with real or simulated data.  Standard errors in parentheses. *** p$<$0.01, ** p$<$0.05, * p$<$0.1. As the parameters estimated on the simulated data can have arbitrarily small standard errors by increasing the number of simulations, the stars of significance levels are not labelled.}
	\end{minipage}  
\end{table}

The above gradients are estimated based on within-individual variations. While the cross-sectional relationship between health and labor supply is not the main focus of our model, we are curious about how well the model fits these untargeted facts. Based on the pooled regression of labor supply on health, the last three columns show the fits are also good.

\begin{figure}[H]
\centering
 	\caption{Model Fit: Age Profiles of Labor Supply by Occupations }
 	\begin{minipage}{0.85\textwidth}
 		\includegraphics[height=8.6cm]{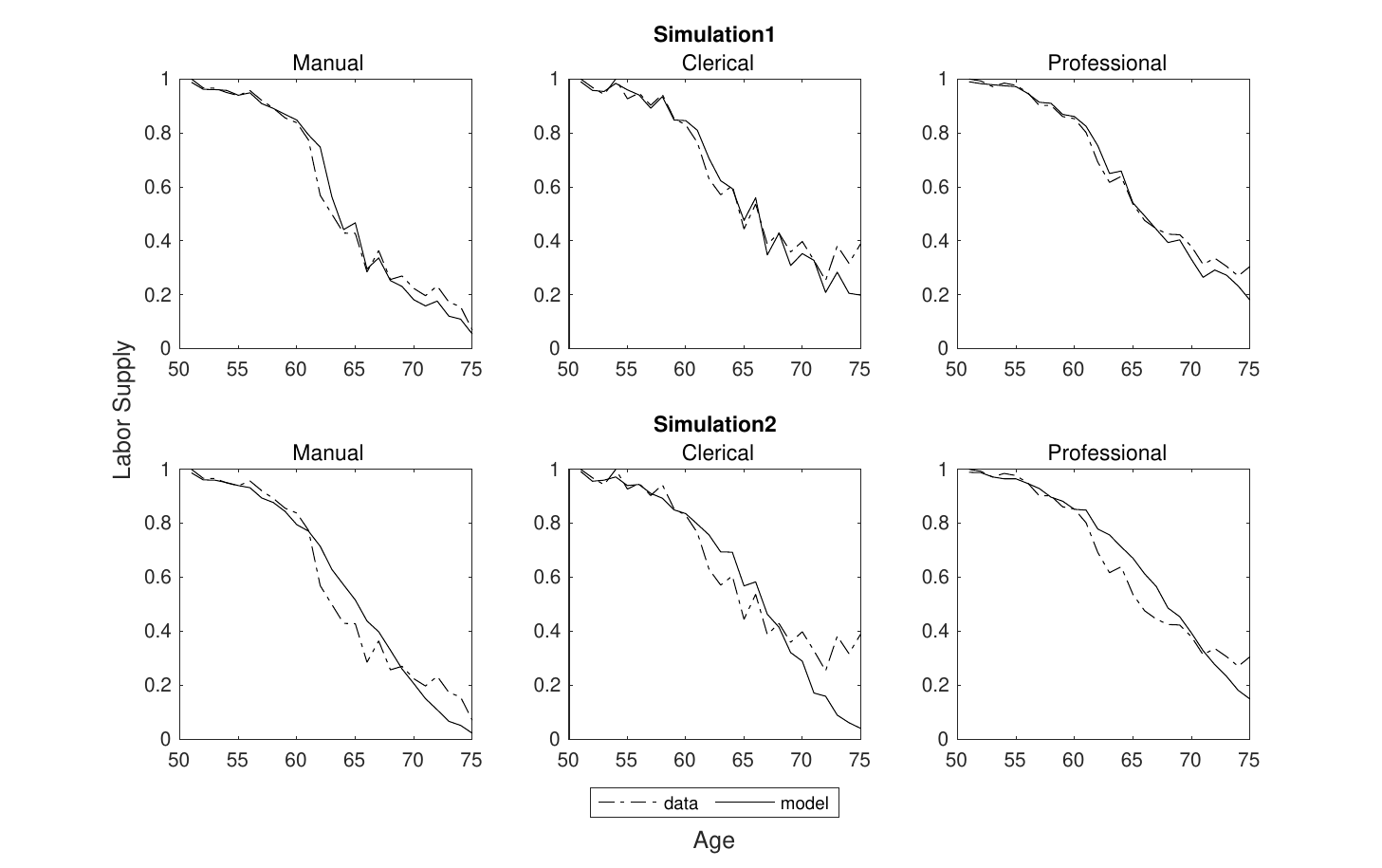} \\	    
 		\scriptsize{These figures present the model fits about the age profiles of labor supply, measured by the LFP rates. Simulation1 refers to labor supply simulated by one-period-ahead approach. Simulation2 refers to simulations starting from the initial values.   \par}	
 	\end{minipage}
 	\label{figure:modelfit}
 \end{figure}

Besides the relationship between health and retirement, how the model predicts labor supply across occupations is also important. The average retirement ages are 64.2, 66.1 and 66.4 for the manual, clerical and professional occupations in the data. To compare, the simulated ages are 64.5, 65.4 and 66.9 respectively. There are two  challenges to trace the age profiles of labor supply by occupations.   First,  I refrain from including leisure effects of ages that are directly dependent on occupations into the model. Indeed, much of the existing work typically includes  direct leisure effects of ages to help keep track of the life cycle employment decline. Studies that target heterogeneous age profiles of labor supply are rare.  Second, the  simulations implemented during the structural estimation are based on the one-period-ahead approach instead of starting from the initial period. That is, for the subsequent periods, only labor supply is simulated, and the rest state variables are taken the ones in the data instead of being simulated. Therefore the structural parameters are chosen to fit the labor supply one period ahead instead of to fit the whole life cycle. Figure \ref{figure:modelfit} shows the simulated labor supply both by one-period-ahead and from the initial period. While the age profiles of labor supply simulated by one-period-ahead are highly close to the reality, the ones simulated from the initial states also trace the data reasonably close.

\section{Results}

Upon estimating the structural model, I first quantify the importance of different health dimensions across occupations under existing pension rules. To see how poor health constrains labor supply responses to pension reforms and the welfare implications,  I then simulate retirement behaviors and welfare changes under counterfactual pensions rules about the Full Retirement Age.

\subsection{Importance of Health under Existing Pension Rules}

The role of health in retirement is obscure by simply comparing two otherwise the same individuals differing only in their health status. For instance,  in a reduced form regression of labor supply on health, it is unclear whether we should control for wealth. On the one hand, wealth is an informative indicator of individual's socioeconomic status, which captures omitted variables correlated with both health and labor supply. On the other hand, wealth is an important channel through which poor health affects labor supply. Including it into the regression makes it a bad control.
Moreover, the full picture of health in retirement may be missing with contemporaneous variations. A regression of labor supply with exogenous health shock at the age of 65 may nevertheless fail to deliver a complete answer about the retirement effect of health given its static nature.

To understand the role of health in retirement,  an ideal experiment would be to compare two otherwise the same individuals such that one neither experiences nor expects health shocks through out the life cycle, while the other faces realistic health process. This experiment can hardly be implemented in labs or in fields. Alternatively, the structural model provides a clear framework to understand the role of health in retirement from a life cycle perspective. In particular, we are able to turn off the channels through which poor health affects retirement, and examine the changes in labor supply at each age.

I start with two experiments in the baseline to explore the role of overall health across occupation, as well as the roles of physical and cognitive health regardless of occupations. These baseline experiments shed light on the heterogeneity by occupations and by health dimensions separately. Then I move to quantify the importance of the interaction between these two factors.

\subsubsection{Role of Overall Health by Occupation}
In the first experiment, I quantify the employment decline across occupations driven by the overall health, aggregating both physical and cognitive dimensions.  Specifically, I turn off all the channels of poor health and simulate the labor supply at each ages. I quantify the declines of employment between age 51 and 70 under this counterfactual setting and compare them with the ones in reality. \footnote{The reality refers to the benchmark simulation from the model, which closely tracks the real data. Poor health affects the calculation of employment rates also by reducing the number of survived individuals, generating misleading interpretation on the channels through which health affects retirement. Therefore, in Section 7.1 and 7.2, the analyses are based on the simulated sample without attrition (but poor health still affects the decision making by changing the life expectancy). }

Under this counterfactual scenario, the employment decline from age 51 to 70 is 61.2 percentage points. Comparing to the 73.4 percentage points under the reality, poor overall health explains 16.6\% of the employment decline between age 51 and 70. By occupation, the overall health explains 19.8\%, 19.7\% and 11.2\% of the employment decline for the manual, clerical and professional occupation respectively. To compare, \cite{blundell2021impact} find the employment decline between age 50 and 70 is explained by the cognition and instrumented subjective health by 15.8\% for high school dropouts, 12.2\% for high school graduates, and 14.2 \% for college graduates.  

Another way to see the role of health is through the average retirement ages under the counterfactual setting with no poor health. In reality, the average retirement ages for manual, clerical and professional workers are 64.4, 65.4 and 66.8 respectively. When the effects of poor overall health are muted, these retirement ages rise to 66.7, 67.4 and 67.9. In particular, the gap in average retirement ages between manual and professional workers reduces from 2.5 years to 1.2 years.

\subsubsection{Dynamic Roles of Physical and Cognitive Health}
I then examine the different roles of physical and cognitive health, pooling all the occupations together. By shutting down underlying channels, I find physical and cognitive health respectively explain 10.0\% and 6.2\% of the employment declines between age 51 to 70. At the same time,  the average retirement age rises from 65.1 to 66.4 and 65.8 respectively.

It is more interesting to examine how physical and cognitive health dynamically affect retirement at different ages, as the incidence of poor physical and cognitive health is different over age.  I quantify the  impact of cognitive health relative to physical health at each age t as follows:
\begin{align*}
\text{\textit{Relative Impact at t}}= \frac{Y_t^{c*}- Y_t}{Y_t^{p*}- Y_t}
\end{align*}
The numerator represents the increase in labor supply at each age $t$ after muting cognitive health channels while the denominator represents the increase after muting physical channels.

\begin{figure}[H]
\footnotesize
\centering
	\begin{minipage}{0.78\textwidth}
\caption{Labor Supply Effect of Cognitive Relative to Physical Health}
   \hspace{1cm}    \includegraphics[height=6.2cm]{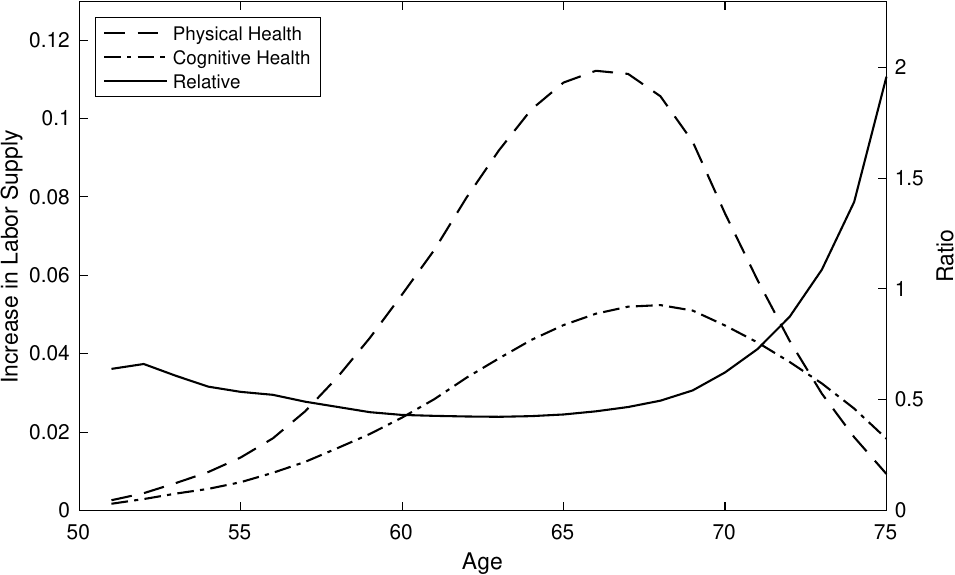}  
       		\label{figure:chvsph}

		\scriptsize{This figure presents the labor supply effect of cognitive health relative to physical health at each age. The left axis shows the increased labor supply by muting the channels of poor physical or cognitive health. The right axis shows their ratio.  \par }
	\end{minipage}          
\end{figure}

Figure \ref{figure:chvsph} shows how this relative impact varies with age.  The effect of cognitive health is half that of physical health, which remains stable until late 60s and has since grown exponentially. In particular, the effect of cognitive health overtakes physical health at around age 73 and has almost doubles in three years. This finding emphasizes the necessity to take into account the cognitive dimension of health, especially for evaluating retirement policies targeting at advanced ages.

\subsubsection{Interaction between Health and Occupation}
The previous subsection explores the retirement effect of health by occupation and by dimension. The key insight of this paper is that different health dimensions may interact with the heterogeneous job requirements in affecting retirement behaviors. To highlight this interplay,  I quantify the share of employment decline driven by physical and cognitive health for each occupation separately.  While the heterogeneous effects of health via other channels are also interesting, this section focuses on the channels of disutility of work and of productivity, which characterize the heterogeneous job requirements across occupations.

\begin{figure}[H]
\footnotesize
\centering
	\begin{minipage}{0.78\textwidth}
\caption{Employment Decline Explained by Physical and Cognitive Health and by Occupation}
     \hspace{2.3cm}     \includegraphics[height=6.2cm]{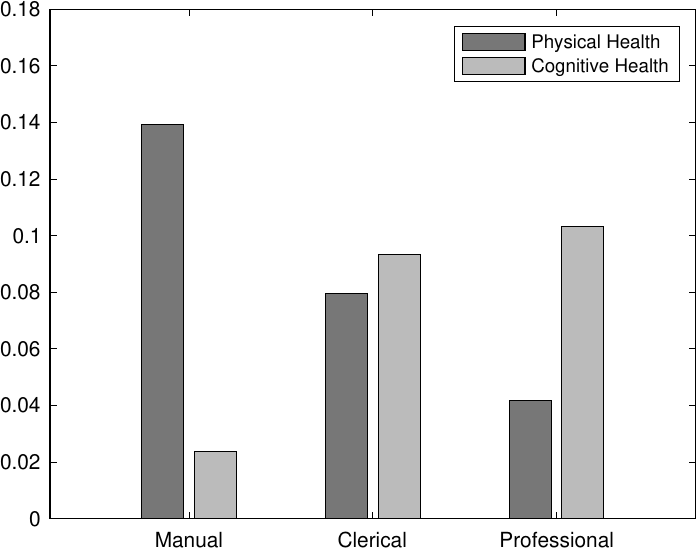}
                    \label{figure:CHPHbyOCC}

          \scriptsize{This figure presents the shares of employment decline between age 51 and 70 that can be explained by the disutility of work and productivity channels of physical and cognitive health. }
\end{minipage}  
\end{figure}

 Figure \ref{figure:CHPHbyOCC} reveals significant heterogeneity in the retirement effects of health once we take into account the interaction with different occupations.   For the manual occupation, poor physical health explains 13.9\% of the employment decline from age 51 to 70, whereas poor cognitive health accounts for 2.4\%. To the contrary, while 4.2\% of the employment decline for the professional occupation is attributed to poor physical health, 10.3\% is explained by poor cognitive health. For the clerical occupation, the effects of physical and cognitive health are on par with each other.

Regarding average retirement age, muting the effect of poor physical health would increase it from 64.5, 65.5, and 66.9 to 66.2, 66.4, and 67.4 for manual, clerical, and professional occupations respectively. The resulting retirement age gap between manual and professional workers would reduce from 2.4 to 1.1 years.  Conversely, muting the effect of poor cognitive health would result in a counterfactual retirement age of 64.8, 66.4, and 67.8, widening the retirement age gap to 3.1 years.

\subsubsection{Counterfactual Ability Requirements}
The previous subsection reveals heterogeneous retirement effects of health across occupations. It is also interesting to know how does the difference in ability requirements between occupations matter. To assess its quantitative importance, I simulate the average retirement age under counterfactual ability requirements. In the first two sets of experiments, I explore how would labor supply change if workers from the occupations with more demanding health requirements faced the mildest requirements.  
 Professional workers on average retire 2.4 years later than manual workers in reality. Specification (1) in Table \ref{table:ctability} shows that this gap would shrink to 1.3 years by 46\%, if the requirement for physical health faced by manual workers is as mild as professional workers.
To the contrary, Specification (2) shows that when professional workers face milder cognitive requirement, the gap would further widen to 3.1 years by 29\%.

\begin{table}[H]
\footnotesize
\centering
	\begin{minipage}{0.9\textwidth}
\caption{Average Retirement Age under Counterfactual Ability Requirements}
 \tabcolsep=0.29cm \begin{tabular}{p{0.22\textwidth}>{\centering}p{0.12\textwidth}>{\centering}p{0.07\textwidth}>{\centering}p{0.07\textwidth}>{\centering}p{0.07\textwidth}>{\centering}p{0.07\textwidth}>{\centering}p{0.07\textwidth}>{\centering}p{0.07\textwidth}>{\centering}p{0.07\textwidth}>{\centering\arraybackslash}p{0.07\textwidth}}
\hline
\hline
 Ave. Retirement Age  &  Baseline  & \multicolumn{2}{c}{(1)} & \multicolumn{2}{c}{(2)} & \multicolumn{2}{c}{(3)} & \multicolumn{2}{c}{(4)} \\
 \hline
     \hspace{0.5cm}    Manual        & 64.5     &  \multicolumn{2}{c}{  65.7  (1.2) }         &   \multicolumn{2}{c}{  64.6 (0.1) }           &  \multicolumn{2}{c}{  63.3  (-1.2)  }   &   \multicolumn{2}{c}{ 64.4  (-0.1) }   \\
       \hspace{0.5cm}    Clerical              & 65.4     &     \multicolumn{2}{c}{  65.9 (0.5)  }                  &  \multicolumn{2}{c}{  66.3  (0.9)  }                  &   \multicolumn{2}{c}{  65.2   (-0.3)  }                 &  \multicolumn{2}{c}{ 64.6        (-0.8)   }                \\
       \hspace{0.5cm}     Professional             & 66.9     &   \multicolumn{2}{c}{ 67.0   (0.1)  }                  &  \multicolumn{2}{c}{  67.7  (0.8)   }                 &    \multicolumn{2}{c}{  66.9     (-0.1) }                  &   \multicolumn{2}{c}{  66.2    (-0.8) }        \\
                \multicolumn{1}{l}{  \hspace{0.5cm} Pro. - Man. }   & 2.4      &       \multicolumn{2}{c}{  1.3   (-1.1) }                   &    \multicolumn{2}{c}{  3.1  (0.7)   }                 &  \multicolumn{2}{c}{  3.6  (1.2)   }                 &   \multicolumn{2}{c}{  1.8   (-0.7)}  \\
\hline
\end{tabular}   
\label{table:ctability}

\scriptsize{This table presents the average retirement ages under the counterfactual ability requirements. The changes relative to the baseline are reported in parenthesis. }
\end{minipage}
\end{table}

What if workers faced the  most demanding requirement? According to Specifications (3) and (4), the retirement age gap would rise from 2.4 to 3.6 years if manual workers faced the same cognitive requirements as professional workers. On the other hand, the gap would narrow to 1.8 years if the professional occupation is as physically demanding as the manual occupation. Overall, these results indicate that heterogeneous ability requirements explain a great share of the labor supply gap across occupations.

\subsection{Policy Experiments}

Previous results reveal the heterogeneous roles of physical and cognitive health  in retirement across occupations. Notice that these results are obtained under the current Social Security rules. What's the implication of this interaction for Social Security reforms in the future? Facing the reduced pension benefits under the policy proposals that incentivize late retirement, older workers typically use their labor supply as a buffer to partially insure this negative income shock. However, poor health tends to limit this behavioral response. Policymakers indeed concern about workers in those physically-demanding occupations, whereas the above results suggest cognitive health appears to be important for clerical and professional workers. Ignoring the cognitive dimension of health, we may systematically overestimate the labor supply responses of a rising share of workers from cognitively demanding occupations and underestimate their welfare losses.

The following policy experiment simulates workers' labor supply responses to pension reforms and evaluates the welfare implications.
The considered policy change is to increase the full retirement age (FRA) to 70.\footnote{This proposal has been proposed in various times. For instance, Social Security Policy Options 2015 by Congressional Budget Office examines the options to increase FRA to age 68, to age 70, and to increase it by one month per birth year. In April 2017, Martin Feldstein, chairman of the Council of Economic Advisers under President Reagan, suggested in Wall Street Journal that increasing the FRA to age 70 can offset revenue loss from the tax cut under the government of Donald Trump.} 
The current rule stipulates 100\% benefits for retirement starting at age 66 of the sample cohort. Benefits are reduced by 6.67\% every one year earlier than the FRA, up to 3 years. For every additional earlier year, the benefits are reduced by 5\% instead. Benefits are higher  for retirement starting after the FRA, by 8\% each year up to age 70.  Under the current rule, benefits are 75\% and 132\% of the full benefits respectively for individuals retire at age 62 and 70. Under the counterfactual experiment, individuals will receive full benefits at age 70 but no higher benefits for  further delayed retirement. Benefits reduction for early retirement is still 6.67\% each year up to 3 years and 5\% each year for the rest. The retirement at age 62 thus leads to 55\% of the full benefits under the counterfactual rule.

\subsubsection{Framework of Heterogeneous Potential Outcomes}

To understand workers' behavioral response to the reform, I augment the analysis with a framework of heterogeneous potential outcomes. Languages are borrowed from \cite{angrist1996identification}, while the following discussion focuses on how does the exogenous policy change $z$ affect the choice $d$ ( the ``first stage''), instead of the welfare consequence of retirement (the ``treatment effect'').

Notice that the choice-specific utility is given as follows:
\begin{align}
    U(d_t,C_t,\Omega_t)= &  \frac{1}{1-\nu} (C_{t}^{1-\nu}-1) + \lambda_{1j}  d_{t} + \big (\lambda_{2j} h_{t}^p+\lambda_{3j} h_{t}^c \big)  \cdot d_{t} +\varepsilon_{t}^{d}   \notag  \\  & + \beta \int \big ( p_t    V_{t+1}(\Omega_{t+1}) +(1-p_t)B(\Omega_{t+1})  \big )  d F_t(\Omega_{t+1}|d_t, C_t ,\Omega_{t} )
\end{align}
An individual decides to work if the utility is greater than the one of retirement:
\begin{align*}
    U(d_t=0,C_t(d_t=0),\Omega_t) <  U(d_t=1,C_t(d_t=1),\Omega_t) 
\end{align*}
Subsequent discussions will focus on the heterogeneous behaviors across individuals at each given age $t$.  To save notations, I thus abstract from the subscript of time $t$ and introduce the subscript of individual $i$ instead.\footnote{Meanwhile, variables with superscript $'$ denote ones in the next period.}   Inserting  Equation (11) into the above inequality yields: 
\begin{align}
    \nu_i < \text{PS}_i+ \text{NPS}_i + \text{ES}_i
\end{align}
\begin{align*}
    \text{where \hspace{0.3cm }PS}_i\hspace{0.1cm}=& \hspace{0.2cm} \frac{1}{1-\nu}  \big [( W_i + \widetilde{Y}_i (d_i=1) )^{1-\nu}  -  ( SS_i + P_i + \widetilde{Y}_i (d_i=0) )^{1-\nu}  \big ]  \\
    \text{NPS}_i\hspace{0.1cm}=& \hspace{0.2cm} \lambda_{1j}  + \big (\lambda_{2j} h_{i}^p+\lambda_{3j} h_{i}^c \big) \\
    \text{ES}_i\hspace{0.1cm}=& \hspace{0.2cm}  \beta \big [ \text{Emax }  (\Omega_{i}^{'}  | d_t=1, C_{i}(d_i=1), \Omega_i  ) -  \text{ Emax }  (\Omega_{i}^{'} | d_t=0, C_{i}(d_i=0), \Omega_i  ) \big ]
\end{align*}
\hspace{0.3cm}  and  \hspace{0.3cm} $ \widetilde{Y}_i (d_i) = (1+r) A_i-A^{'}_i(d_i)+ Y_i^s + SSDI_i + \zeta_i - ME_i    $  \\

The above inequality indicates that the individual decides to work if the resistance $\nu_i=\varepsilon_i^0-\varepsilon_i^1$ is smaller than the work surplus, which consists of three components. The pecuniary surplus $\text{PS}_i$ mainly weighs on the labor income of work $W_i$ and the pension income of retirement $SS_i+P_i$.\footnote{There is a second order effect due to inter-temporal decisions of consumption and savings, as saving $A_{'}(d_i)$ is conditional on labor supply $d_i$. } Nonpecuniary utility $\text{NPS}_i$ captures the disutility of work, whereby poor health has an extra effect. Finally, through the expected future surplus $\text{ES}_i$, the model captures dynamic features such as pension accruals contributed by work.

Denote the threshold of work as $\mu=\text{PS}+ \text{NPS} + \text{ES}$, and $\mu_0$ and $\mu_1$ as the ones before and after the reform respectively.\footnote{ The threshold is heterogeneous across individuals. To facilitate exposition, following discussions focus on the heterogeneity aroused from the idiosyncratic leisure preference (i.e. the resistance $\nu_i$), conditional on individuals having the same threshold.}   Then we consider how does the reform of raising the FRA  affect individual's heterogeneous decisions to work. 
First of all, higher FRA leads to lower Social Security benefits received by individuals. For instance, an individual retired at age 66 can have 100\% benefits under the current rules but 75\% after the reform. This effect can be captured by a discounted Social Security benefits $\rho \cdot SS_i$, where $0<\rho<1$.   The reduction of benefits tends to increase the pecuniary surplus of work $\text{PS}$, leading to $\mu_1>\mu_0$. Secondly, the higher FRA also  implies different accruals of Social Security. For instance, an additional year of work from age 66 to 67 corresponds with an increase by 8\% of the benefits under the existing rules. Under new rules, it leads to an increase by 5\%. This effect is captured by a decrease in $\text{ES}$, lowering the work threshold $\mu$ so that $\mu_1<\mu_0$.

Because of the heterogeneity in both the work resistance $\nu_i$ and the work threshold $\mu_i$, there will be rich heterogeneity in labor supply responses to the policy change. Considering the potential work decisions $d_i(z_i)$ under the current policy $z_i=0$ and the reform $z_i=1$, We can classify individuals at each age into the following four types:
\begin{itemize}
	 \setlength\itemsep{0.02cm}
\item Always-takers:  $d_{i}(z_i=1)=d_{i}(z_i=0)=1 $
\item Never-takers:  $d_{i}(z_i=1)=d_{i}(z_i=0)=0 $
\item Compliers: $d_{i}(z_i=1)>d_{z_i=i}(0) $
\item Defiers: $d_{i}(z_i=1)<d_{i}(z_i=0)$
\end{itemize}
We are curious about the share of these four types of individuals at each age, especially the comliers who delay their retirement in response to the reform and the never-takers who fail to do so. While this framework is static in a sense that the shares of different types are defined at each given age, their evolution with age sheds light on the dynamic effects of pension reform.\footnote{Alternatively, we can focus on the behavior of each individual through out his life-cycle before and after the reform. We can readily define those responsive to the reform as compliers. However, this alternative fails to disentangle never-takers and always-takers, as both of them do not respond to the policy change.}

\subsubsection{Role of Health in Pension Reform}
A primary concern in policy design is the potential constraint that poor health may impose on workers' labor supply responses, thereby reducing the effectiveness of reforms. Let us consider an extreme scenario in which all workers are unable to postpone retirement. From the government's perspective, while the policy change generates a budget surplus by reducing benefit payments, it loses the contribution that workers could have made to the pension system through extended years of work and the potential for delayed benefit collection. On the worker side, they fail to use labor supply to insure against the negative income shock induced by the reform and suffer greater welfare losses.

To examine the role of poor health under pension reforms, the analysis aims to understand how workers would respond to the policy change in a scenario where poor health does not result in increased work burden or reduced productivity. For simplicity, the following analysis temporarily assumes away defiers to the policy change.

\begin{figure}[H]
	\footnotesize
	\centering
	\begin{minipage}{0.92\textwidth}
		\begin{center}
			\caption{Health and Heterogeneous Effects of Pension Reform}
			\hspace{0.2cm}\includegraphics[height=9.2cm]{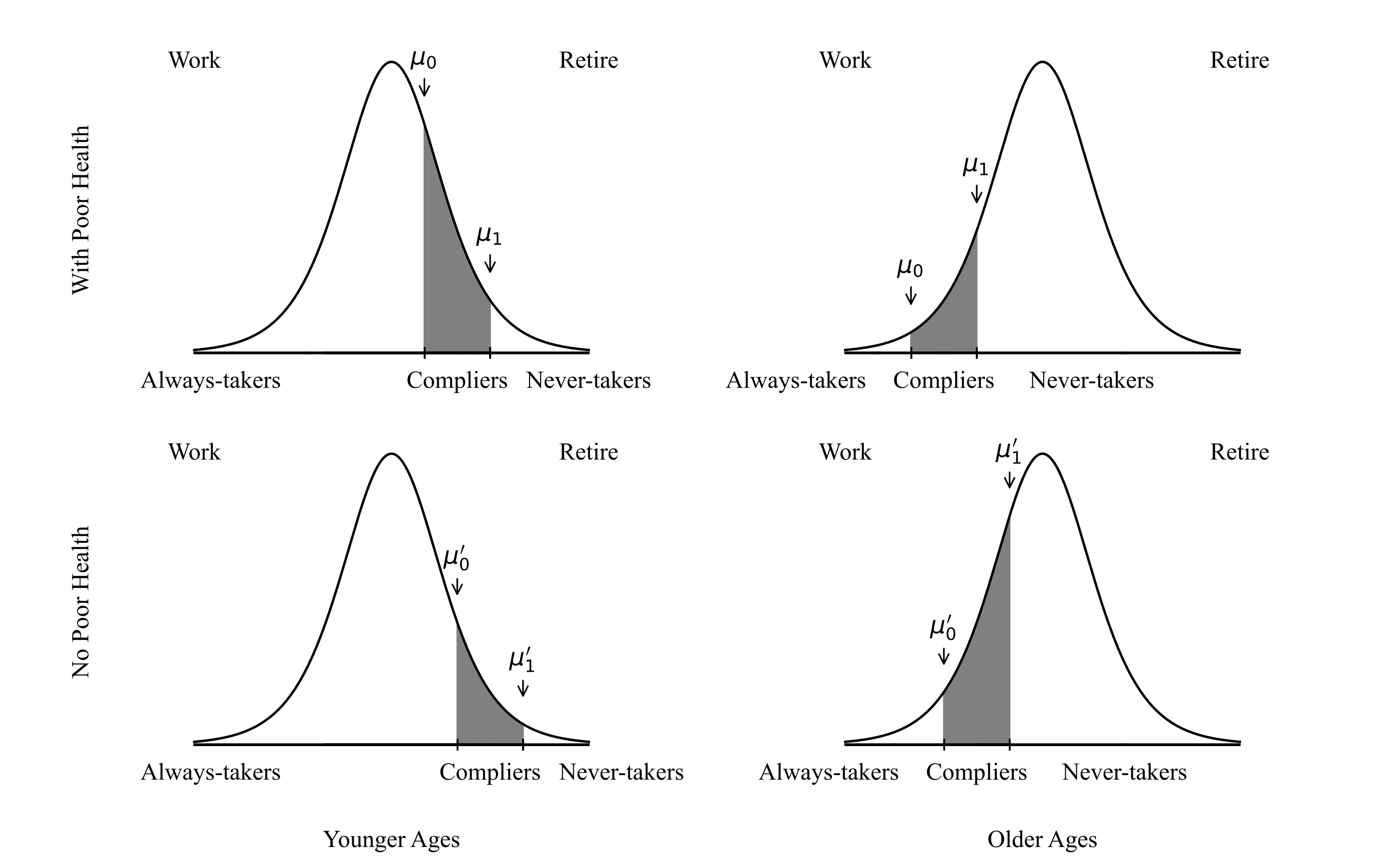}
			\label{figure:Comliers-Theory}   
		\end{center}
		\scriptsize{This figure shows how the sub-populaiton of always-takers, compliers, and never-takers depends on health and age. A: Always-takers; C: Compliers; N: Never-takers}
	\end{minipage}  
\end{figure}

Under the counterfactual setting that poor health does not incur disutility of work and wage penalty,  both the nonpecuniary surplus $\text{\text{NPS}}$ and the pecuniary surplus $\text{\text{PS}}$  increase, provided that $\lambda_{2j}<0$, $\lambda_{3j}<0$, $\kappa_{5j}<0$ and $\kappa_{6j}<0$. The resulting retirement threshold $\mu$ would increase to $\mu^{'}$, the one under no health constraint. For simplicity, we abstract from the second order effects resulted from nonlinearity, so that $\mu_0^{'}-\mu_0=\mu_1^{'}-\mu_1$.     Figure  \ref{figure:Comliers-Theory}  demonstrates that without the constraint of poor health, there will be less never-takers and more always-takers. However, the change of the compliant sub-population turns out to be indeterminate. For work thresholds with dominating always-takers, such as those younger ages, there will be less compliers to the policy change under no health constraint, as most workers already work and they have fewer leeway to further increase the labor supply. On the other hand, for retirement thresholds where never-takers dominate, eliminating health constraint indeed leads to more compilers to the policy change.

\subsubsection{Experiment Results}

The above analysis shows that while there will be more workers unable to adapt to the reform with health limitations, the share of compliant workers is indeterminate. It crucially depends on the retirement threshold $\mu$, particularly its evolution with age, indicating the necessity of a dynamic analysis. Theoretical analysis also  suggests that the shift of work threshold $\mu$ due to the reform needs not to be monotone and uniform. Increasing the FRA may lead to both compliers and defiers at the same time. Without the monotonicity assumption, we cannot follow the framework of \cite{angrist2009mostly} in counting the size of the complier group.\footnote{For instance, it is of particular interest to measure the proportion of policy compliers among those who do not work under the current rules $P( d_i(1)>d_i(0) | d_i=0)$. The calculation $P( d_i(1)>d_i(0) | d_i=0) = \frac{P( d_i=0 | d_i(1)>d_i(0))    P(d_i(1)>d_i(0) )      }{ P(d_i=0 )} =  \frac{P( Z_i=0) E(d_i|Z_i=1 ) - E(d_i|Z_i=0 )    )      }{ P(d_i )=0}  $   requires the monotonicity assumption to deliver     $ P(d_i(1)>d_i(0)= E(d_i|Z_i=1 ) - E(d_i|Z_i=0 ) $.    } 

However, the structural model brings us benefits to clearly identify complier, defier, always-taker, and never-taker at the individual level. In particular, both $d_i(z_i=0)$ and $d_i(z_i=1)$ are observed for the same person. The size of compliers and never-takers can thus be directly calculated.

\begin{figure}[H]
\footnotesize
\centering
	\begin{minipage}{0.95\textwidth}
\caption{Share of Compliers and Never-takers to the Reform}
          \includegraphics[height=10cm]{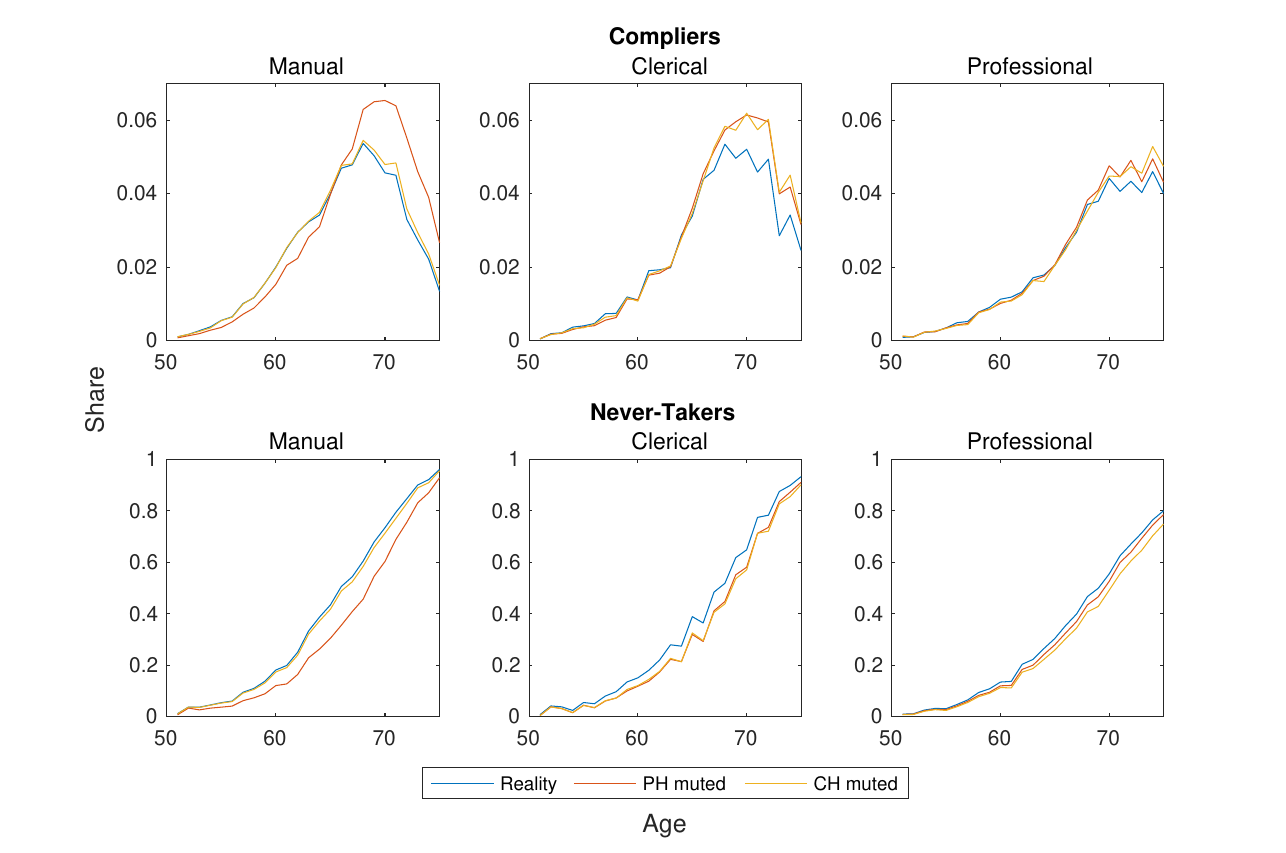}
                    \label{figure:Compliers1}   
          \scriptsize{This figure presents the shares of compliers and never-takers to the reform among all individuals at each age, by occupation respectively. Three sets of results are reported: the reality, the counterfactual setting with no constraint of physical health (PH), and the counterfactual setting with no constraint of cognitive health (CH). }
\end{minipage}  
\end{figure}

Figure \ref{figure:Compliers1} presents the share of compliers who would have retired but induced to work  $d_i(z_i=1)>d_i(z_i=0)$, as well as the proportion of never-takers who are unresponsive   $d_i(z_i=1)=d_i(z_i=0)=0$ at each age.  There are few compliant workers at younger age as most of them choose to work regardless of the reform. Meanwhile, the share of workers that are unresponsive to the reform increases with age. The policy change of increasing FRA to 70 induces workers to work mainly at around age 70, since which manual workers become less responsive but professional workers remain compliant.

\begin{figure}[H]
\footnotesize
\centering
	\begin{minipage}{0.95\textwidth}
\caption{Share of Compliers and Never-takers to the Reform, \\Conditional on Retirement under Current Policy}
          \includegraphics[height=10cm]{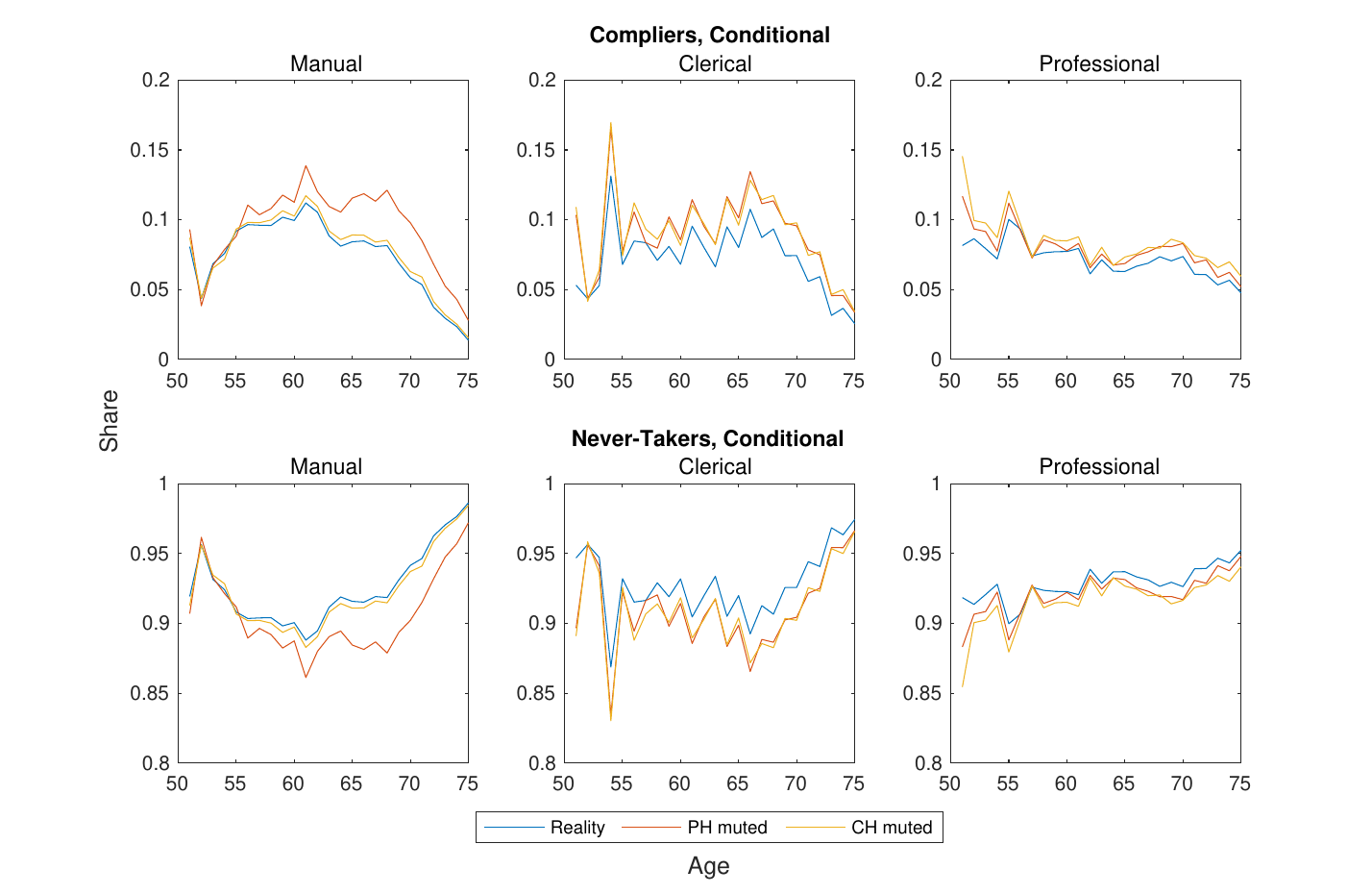}
                    \label{figure:Compliers2}   
          \scriptsize{This figure presents the shares of compliers and never-takers among individuals who retire under the current policy at each age, by occupation respectively. Three sets of results are reported: the reality, the counterfactual setting with no constraint of physical health (PH), and the counterfactual setting with no constraint of cognitive health (CH). }
\end{minipage}  
\end{figure}

It is particularly interesting to see how does the interaction between poor health and job requirements dampen the labor supply responsiveness. Figure \ref{figure:Compliers1} reveals that poor physical health impedes workers' capacity to delay their retirement in response to the policy change among all occupations, while this impact is largest for manual workers. To the contrary, the effect of poor cognitive health appears to be minimal for manual workers but at least as large as physical health for clerical and professional workers. Similar findings are revealed in Figure \ref{figure:Compliers2}, which presents the proportion of compliers and never-takers among retired workers under the current Social Security rules, the group of workers mostly relevant to the policy change.

To obtain an overall assessment on impacts of the health-occupation interplay on policy effectiveness, I calculate the change in the average retirement age across occupations, taking into account different behavioral responses all together. As Table \ref{table:policy_laborsupply} shows, when the FRA increases to 70, manual, clerical, and professional workers would delay their retirement by 0.65, 0.64 and 0.51 years respectively.  Moreover, the health-occupation interaction shows nontrivial impacts on labor supply responsiveness. Muting the constraint of poor physical health, workers in these occupations would further delay their retirement by 13.7\%, 4.8\% and 3.3\% in response to the policy change. Conversely, manual workers show no extra capacity to postpone retirement when the channels of cognitive health are shut down, while clerical and professional workers further extend their work horizon by 7.7\% and 5.0\% respectively.\footnote{The results show that, without poor cognitive constraint, the additional years of work by manual workers are less.  This is consistent with the previous analysis that no health constraint may also imply fewer compliers and thus smaller responsiveness. }

\begin{table}[H]
	\footnotesize
	\centering
	\begin{minipage}{0.9\textwidth}
		\caption{Labor Supply Responsiveness to the Policy Change}
		\tabcolsep=0.08cm \begin{tabular}{p{0.32\textwidth}>{\centering}p{0.10\textwidth}>{\centering}p{0.10\textwidth}>{\centering}p{0.10\textwidth}>{\centering}p{0.10\textwidth}>{\centering}p{0.10\textwidth}>{\centering\arraybackslash}p{0.10\textwidth}}
			\hline
			\hline
			&  Manual  &   &  Clerical&   &    Professional    &   \\
			\hline
			FRA=66  &   &   &   &   &     &    \\
			\hspace{0.3cm} - Reality  & 64.50  &  - & 65.44  & -  &  66.94   &  -  \\
			\hspace{0.3cm} - Physical Shut-Down  &  66.23 & (1.72)  & 66.37  & (0.93)  &  67.34   & (0.39)    \\		
			\hspace{0.3cm} - Cognitive Shut-Down  &  64.75 &  (0.25) & 66.43  &  (0.98) & 67.81    &  (0.87)  \\								
			\hline
			FRA=70  &   &   &   &   &     &    \\
			\hspace{0.3cm} - Reality  & 65.15  & -  &66.08   & -  & 67.45    & -   \\
			\hspace{0.3cm} - Physical Shut-Down  &66.97   & (1.81)   &67.04   & (0.96)  &67.86     & (0.41)   \\		
			\hspace{0.3cm} - Cognitive Shut-Down  &65.40   & (0.24)   & 67.11  & (1.03)  & 68.34    & (0.89)    \\		
			\hline		
			Changes  &   &   &   &   &     &    \\
			\hspace{0.3cm} - Reality  & 0.65  & -  & 0.64  & -  & 0.51    & -   \\
			\hspace{0.3cm} - Physical Shut-Down  &0.74   &  (13.7\%) &  0.67 & (4.8\%)  &  0.52   &  (3.3\%)    \\		
			\hspace{0.3cm} - Cognitive Shut-Down  &0.64   & (-1.7\%)  & 0.69  & (7.7\%)  & 0.53    & (5.0\%)   \\		
			\hline			
		\end{tabular}   
		\label{table:policy_laborsupply}

		\scriptsize{This table presents the average retirement age by occupation under the current FRA and the counterfactual FRA. It presents results under the reality and when channels of poor health are shut down.
	Differences relative to the reality are reported in parenthesis. }
	\end{minipage}
\end{table}

What is the welfare implication of the constrained labor supply responsiveness? Labor supply is an important mechanism for workers to insure against the income loss caused by the reform. As poor health impairs the capacity to work, workers with poor health, particularly from occupations that have demanding health requirements, may fail to adapt to the policy change and their  welfare losses can be exacerbated. Table \ref{table:policy_welfare} shows that as of age 56, the present discounted value (PDV) of individual utility decreases under the reform. When workers are able to better adapt to the reform under no physical health constraint, manual, clerical, and professional workers would experience 13.0\%, 6.7\% and 2.5\% less welfare losses respectively. If no constraint of cognitive health, an individual would experience minimal reduction in welfare losses if he is from the manual occupation. However, the welfare losses would be significantly mitigated by 11.4\% and 10.6\% respectively,  if he  comes from the clerical and professional occupations.

To compensate for the welfare losses exacerbated by poor health, the subsidy for a manual and a professional worker respectively costs 5,005 and 598 dollars when only the physical dimension is considered, resulting in a gap of 4,408 dollars. However, if the welfare losses due to cognitive constraint is also taken into account, these subsidies increase to 5,198 and 2,247 respectively, and the gap narrows significantly by 33 \% to 2,951 dollars.\footnote{ The calculation is explained in the next-subsection. }

\begin{table}[H]
	\footnotesize
	\centering
	\begin{minipage}{0.90\textwidth}
		\caption{Welfare Consequences of Constrained Labor Supply Responsiveness}
		\tabcolsep=0.08cm \begin{tabular}{p{0.32\textwidth}>{\centering}p{0.10\textwidth}>{\centering}p{0.10\textwidth}>{\centering}p{0.10\textwidth}>{\centering}p{0.10\textwidth}>{\centering}p{0.10\textwidth}>{\centering\arraybackslash}p{0.10\textwidth}}
			\hline
			\hline
			&  Manual  &   &  Clerical&   &    Professional    &   \\
			\hline
			FRA=66  &   &   &   &   &     &    \\
			\hspace{0.3cm} - Reality  & 27.16  &  - & 27.54  & -  &  32.20   &  -  \\
			\hspace{0.3cm} - Physical Shut-Down  &  28.10 & (0.94)  & 28.00  & (0.46)  &  32.34   & (0.14)    \\		
			\hspace{0.3cm} - Cognitive Shut-Down  &  27.23 &  (0.08) & 28.00 &  (0.46) & 32.54    &  (0.33)  \\								
			\hline
			FRA=70  &   &   &   &   &     &    \\
			\hspace{0.3cm} - Reality  & 26.65  & -  & 27.05  & -  &  31.98    & -   \\
			\hspace{0.3cm} - Physical Shut-Down  & 27.66   & (1.01)   & 27.54  & (0.49)  & 32.13    & (0.14)   \\		
			\hspace{0.3cm} - Cognitive Shut-Down  & 26.73   & (0.08)   &  27.56  & (0.52)  & 32.34  & (0.36)    \\		
			\hline		
			Changes  &   &   &   &   &     &    \\
			\hspace{0.3cm} - Reality  & -0.51 & -  & -0.49  & -  & -0.22   & -   \\
			\hspace{0.3cm} - Physical Shut-Down  & -0.44  &  (-13.0\%) &  -0.46 & (-6.7\%)  & -0.22   &  (-2.5\%)    \\		
			\hspace{0.3cm} - Cognitive Shut-Down  & -0.50   & (-0.5\%)  & -0.44  & (-11.4\%)  & -0.20    & (-10.6\%)   \\		
			\hline					
		\end{tabular}   
		\label{table:policy_welfare}

		\scriptsize{This table presents the present discounted value (PDV) of individual utility as of age 56 by occupation.
			It presents results under the reality and when channels of poor health are shut down.
			Differences relative to the reality are reported in parenthesis.}
	\end{minipage}
\end{table}

\subsubsection{Welfare Implication and Shifting Job Requirements}

Finally, in order to shed light on the evolving impacts of poor health over time, a back-of-the-envelope calculation is conducted to compare the subsidies required to compensate for the welfare losses of a typical worker in the present day with those in the 1960s.

 First, the compensation variation for welfare losses caused by the reform in reality is calculated. The results indicate that lump-sum transfers needed to compensate for the losses are 38.5, 20.0 and 23.9 thousand dollars respectively for a typical manual, clerical and professional worker aged 56. By combining these numbers with the estimates presented in the last panel of Table \ref{table:policy_welfare}, the required subsidy amounts for the welfare losses exacerbated by health constraints are obtained. Subsequently, based on the occupation composition in 1968 and 2015, the subsidy for an average worker can be determined.\footnote{The shares of manual, clerical and professional employments are 50.2\%,  23.0\%, and 25.8\% in 1968, and 38.7\%, 22.6\% and 38.4\% in 2015. They do not sum up to one because armed forces are left out. }

 Considering the physical dimension alone, the results show that the subsidy for a typical worker in 1968 would amount to 2,975  dollars. As jobs become less physically demanding over time, this subsidy declines to 2,469 dollars in 2015 by 506 dollars (17\%). However, the welfare implications of poor cognitive health increase concurrently, leading to an additional subsidy of 288 dollars for cognitive constraints. Taking both dimensions into account, the subsidy would be 325 rather than 506 dollars less than 1968, indicating an overestimation of 55\%. These findings highlight the nontrivial implications of considering shifting job requirements in the design of future pension reforms.





\section{Conclusion}

Since the 1960s, significant changes in labor markets have led to increased cognitive requirements, while jobs on average are becoming less physically demanding. Previous studies on the retirement effects of health have primarily focused on physical health dimensions and factors related to the supply side of the labor market, overlooking the rising cognitive requirements from the other side of the market.
In this paper, I present new insights into the relationship between health and retirement across different occupations. Through the estimation of a dynamic structural model, I examine the retirement effects of different health dimensions and their interaction with the heterogeneous ability requirements across occupations. Additionally, I analyze the heterogeneous response in labor supply and the unequal welfare consequences in the context of retirement policy reforms that are actively discussed by a number of governments facing population aging.

These findings provide new perspectives for understanding both retirements across occupations and trends of retirement over time. The interaction between health and occupation may also offer fresh insights for understanding retirement behaviors across countries, where the occupation composition differs substantially, and across gender, as females and males engaged in notably different work. One reminder is that the current analysis is conducted under a partial equilibrium framework, with specific focus on individual behavioral responses to policy changes. In order to fully assess the consequences of occupation and job changes on older workers, a model with general equilibrium is needed, though there is a trade-off between capturing the intricacies of individual decision making and the general equilibrium effects.

\newpage
\begin{appendices}

\section{\Large Physical Health Index}
\label{Appendix:PH}
This appendix section presents the age profiles and variations for the measure of physical health used in this paper, i.e. the health index constructed as the self-reported health instrumented by a battery of objective measures. Specifically, I estimate the ordered probit regressions with independent variables about  the functional limitations, diseases and basic demographic and socioeconomic status. Variables related to functional limitations include: the summary of Activities of Daily Living (ADL), the summary of Instrumental Activities of Daily Living (IADLs), the mobility index, the large muscle index, the gross motor index and the fine motor index. Variables related to diseases include whether the individual ever had: (1)high blood pressure; (2)diabetes; (3)cancer; (4)lung disease; (5)heart problems; (6)stroke; (7)arthritis.\footnote{Details about these variables are referred to Rand HRS documentations.} Demographic and socioeconomic variables consist of the age dummies, years of schooling, education degrees, birth year dummies and assets.

 \begin{figure}[H]
  \begin{center}
 	\caption{Age Profiles of Physical Health }
 	\begin{minipage}{0.9\textwidth}
 	\begin{center}
 		\includegraphics[height=6.6cm]{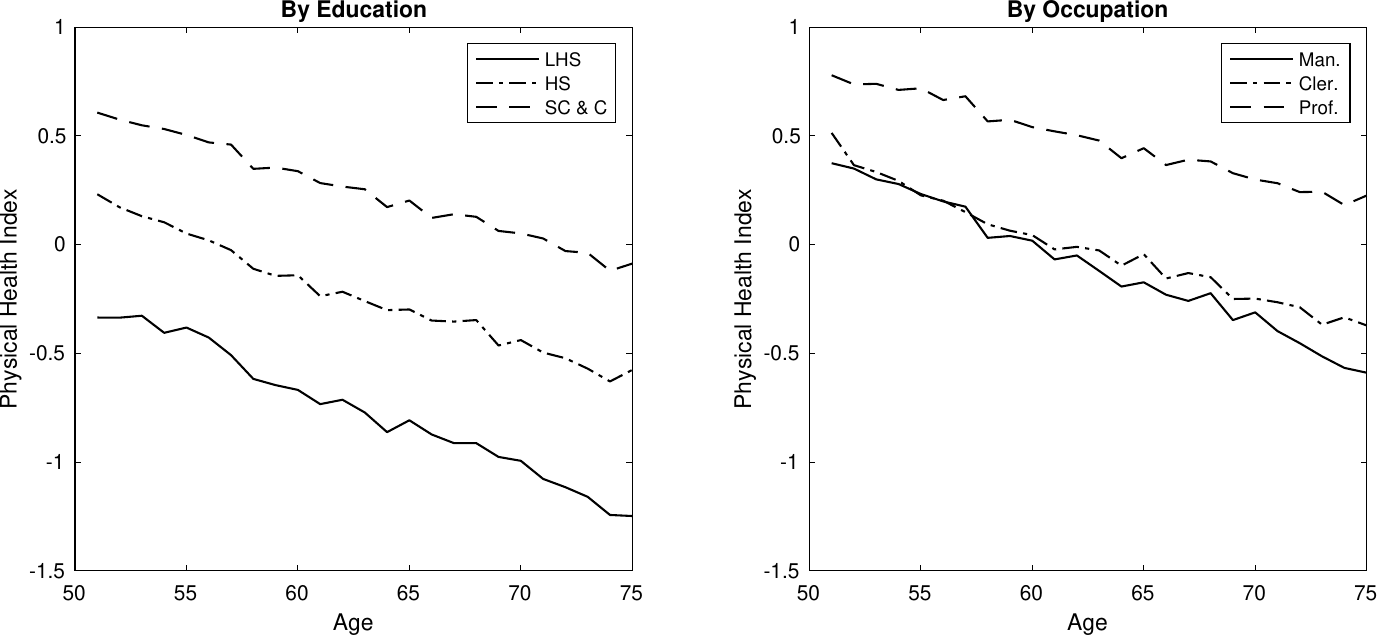} 	    
 	\end{center}
 		\scriptsize{The figures present the variations of physical health index from age 51 to 75. Results are obtained by regressing physical health index on age dummies, controlling for individual fixed effects.  LHS: less than high school; HS: high school; SC\&C: some college and college. Man.: manual; Cler.: clerical; Prof.:professional. \par}	
 	\end{minipage}
 	\label{figure:phgraph}
 	\end{center}  
 \end{figure}

\begin{table}[H]
 	\centering
 	 		\footnotesize
 	\begin{minipage}{0.825\textwidth}
 	\caption{Variations of Physical Health by Education and Occupation}
		 \tabcolsep=0.12cm
		 \begin{tabular}{p{0.28\textwidth}>{\centering}p{0.10\textwidth}>{\centering}p{0.10\textwidth}>{\centering}p{0.10\textwidth}>{\centering}p{0.10\textwidth}>{\centering}p{0.10\textwidth}>{\centering\arraybackslash}p{0.10\textwidth}}
 			\hline
    \hline
                     & \multicolumn{3}{c}{Education} &   \multicolumn{3}{c}{Occupation}  \\
 \cmidrule(lr){2-4}
 \cmidrule(lr){5-7}                     
\multicolumn{1}{c}{} & LHS                  & HS                            & SC\&C              & Man.               & Cler.                      & Prof.         \\ \hline
Age 51         & 0.23                 & 0.56                         & 0.86                & 0.56                & 0.76                          & 0.86                \\
Age 61             & -0.09                 & 0.20                          & 0.56                & 0.16                 & 0.39                          & 0.61                \\

Age 71         & -0.52                 & -0.12                          & 0.25                 & -0.23                 & 0.09                           & 0.34                \\
Standard Deviation   & 0.91                 & 0.78                          & 0.39                 & 0.73                 & 0.73                           & 0.65                 \\ 
\hline
 Drop from 51 to 61        & -0.33                & -0.36                         & -0.29                & -0.40                & -0.37                          & -0.25                \\
 Drop from 61  to 71        & -0.43                & -0.32                         & -0.31                & -0.39                & -0.30                          & -0.27                \\
 			\hline
\end{tabular}\\
 		\label{table:phtable}%
 		\scriptsize{This table presents variations of physical health index by education and occupation. The values are predicted from the regressions with cubic ages to reduce noise. Individual fixed effects are controlled to obtain within-individual variations. LHS: less than high school; HS: high school; SC\&C: some college and college.  Man.: Manual; Cler.: Clerical; Prof.: Professional. I include observations that have already retired in order to keep track of the transition at all ages. Their occupations are defined as the last ones while working. 
 			 \par}				
 	\end{minipage}
 	
\end{table}

For comparison, I also present the results of self-reported health, the most common measure used by previous studies.

 \begin{figure}[H]
  \begin{center}
 	\caption{Age Profiles of Self-Reported Health }
 	\begin{minipage}{0.99\textwidth}
 	\begin{center}
 		\includegraphics[height=6.6cm]{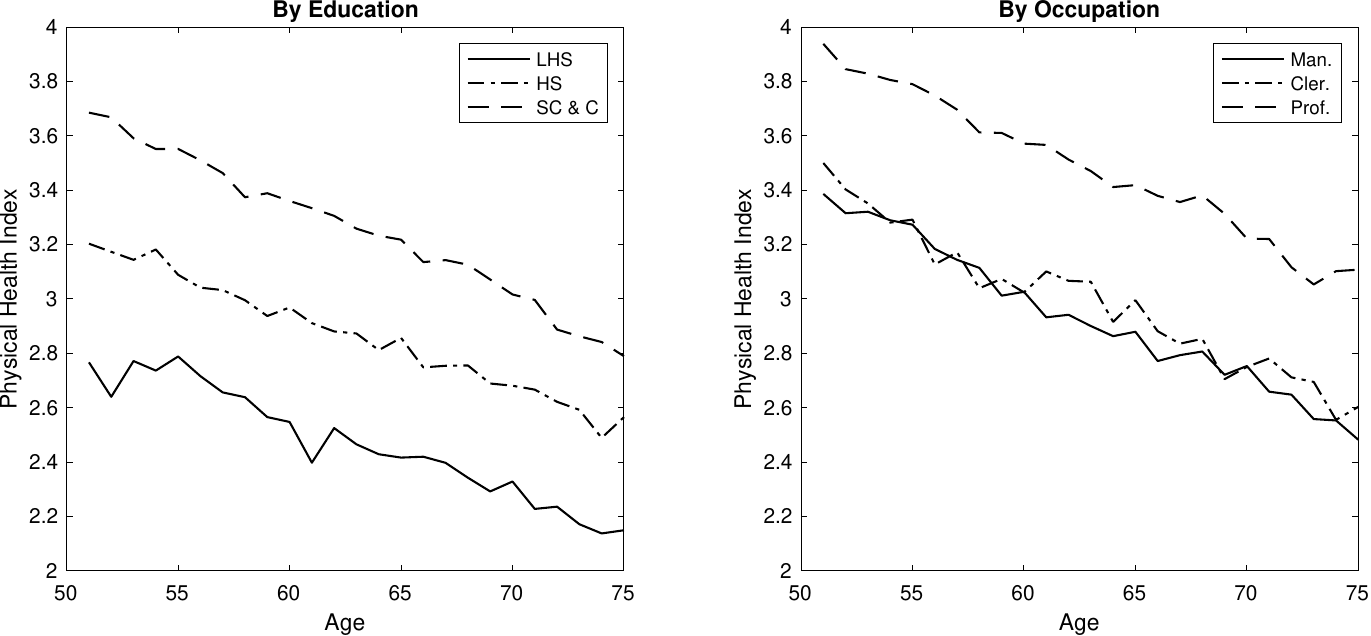} 	    
 	\end{center}
 		\scriptsize{The figures present the variations of self-reported health from age 51 to 75. Results are obtained by regressing self-reported health on age dummies, controlling for individual fixed effects.  LHS: less than high school; HS: high school; SC\&C: some college and college. Man.: manual; Cler.: clerical; Prof.:professional. \par}	
 	\end{minipage}
 	\label{figure:shgraph}
 	\end{center}  
 \end{figure}

\begin{table}[H]
 	\centering
 	 		\footnotesize
 	\begin{minipage}{0.825\textwidth}
 	\caption{Variations of Self-Reported Health by Education and Occupation}
		 \tabcolsep=0.12cm
		 \begin{tabular}{p{0.28\textwidth}>{\centering}p{0.10\textwidth}>{\centering}p{0.10\textwidth}>{\centering}p{0.10\textwidth}>{\centering}p{0.10\textwidth}>{\centering}p{0.10\textwidth}>{\centering\arraybackslash}p{0.10\textwidth}}
 	\hline
  \hline
                     & \multicolumn{3}{c}{Education} &   \multicolumn{3}{c}{Occupation}  \\
 \cmidrule(lr){2-4}
 \cmidrule(lr){5-7}                     
\multicolumn{1}{c}{} & LHS                  & HS                            & SC\&C              & Man.               & Cler.                      & Prof.         \\ \hline
Age 51         & 3.22                & 3.56                        & 3.98               & 3.60               &3.82                        & 4.07                \\
Age 61             & 2.99                &3.30                         & 3.65               & 3.24              &3.50                         & 3.72                \\

Age 71         & 2.68                 &3.01                         & 3.27                & 2.89                 & 3.16                          & 3.35                \\
Standard Deviation   & 1.13                 & 1.09                         & 1.08                 & 1.06                & 1.05                          &1.00                 \\ 
\hline
 Drop from 51 to 61        & -0.23                & -0.26                         & -0.33                & -0.37                & -0.32                          & -0.35                \\
 Drop from 61  to 71        & -0.31                & -0.29                         & -0.38                & -0.35                & -0.34                          & -0.37                \\
 		\hline
\end{tabular}
 		\label{table:shtable}%

 		\scriptsize{This table presents variations of self-reported health by education and occupation. The values are predicted from the regressions with cubic ages to reduce noise. Individual fixed effects are controlled to obtain within-individual variations. LHS: less than high school; HS: high school; SC\&C: some college and college.  Man.: Manual; Cler.: Clerical; Prof.: Professional. I include observations that have already retired in order to keep track of the transition at all ages. Their occupations are defined as the last ones while working. 
 			 \par}				
 	\end{minipage}
 	
\end{table}

\section{Occupation Classification and Ability Requirements}
\label{Appendix:OCC}

In HRS, occupations are reported as 4-digit codes consistent with the Census. The occupations from wave 1 to wave 7 are consistent with Census 1980, while the codes based on Census 2000 are applied since wave 8. For confidentiality, the 4-digit codes are masked and classified into 17 groups in wave 1-7 and 25 groups in wage 8-12.  Table \ref{table:OccClass} lists the mapping between the three categories defined in this paper and HRS 2-digit masked occupations.

Table \ref{table:onet1} and \ref{table:onet2} show how much each detailed ability is required by the three occupation groups. The last column of each table reports the ratio of the professional relative to the manual occupation. Physical abilities are significantly less required by the professional occupation, whereas cognitive abilities are in the contrast.

\begin{table}[H]
  \centering
  \begin{minipage}{0.95\textwidth}
\footnotesize
\caption{Occupation Classifications}
    \begin{tabular}{p{0.12\linewidth} | p{0.40\linewidth} | p{0.40\linewidth}}
\hline
\hline
  Categories   &  Waves 1-7  & Waves 8-12 \\ \hline
Manual& (10)Farming, forestry, fishing; (11)Mechanics and repair; (12)Construction trade and extractors; (13)Precision production; (14)Operators: machine; (15)Operators: transport, etc.; (16)Operators: handlers, etc.; (5)Service: private household, cleaning and building services; (6)Service: protection; (7)Service: food preparation; (8)Health services; (9)Personal services & (19)Farming, Fishing, and Forestry; (20)Construction Trades (21)Extraction Workers; (22)Installation, Maintenance, and Repair; (23)Production; (24)Transportation and Material Moving; (12)Healthcare Support; (13)Protective Service; (14)Food Preparation and Serving Related; (15)Building and Grounds Cleaning and Maintenance; (16)Personal Care and Service \\
     \hline
 Clerical   &   (3)Sales; (4)Clerical, administrative support   & (17)Sales and Related; (18)Office and Administrative Support \\
  \hline
Professional & (1)Managerial specialty operation; (2)Professional specialty operation and technical support  &   (1)Management; (2)Business and Financial; (3)Financial Specialists; (4)Computer and mathematical; (5)Architecture and Engineering; (6)Life, Physical, and Social Science; (7)Community and Social Service; (8)Legal; (9)Education, Training, and Library; (10)Arts, Design, Entertainment, Sports, and Media; (11)Healthcare Practitioners and Technical\\
\hline
    \end{tabular}  
    \footnotesize{Member of Armed Forces is excluded.}
    \label{table:OccClass}
      \end{minipage}
\end{table}


\begin{table}[H]
	\centering
		\footnotesize
	\begin{minipage}{0.87\textwidth}
		\caption{Physical Ability Requirements across Occupations}
		
		\tabcolsep=0.12cm
\begin{tabular}{p{0.35\textwidth}>{\centering}p{0.14\textwidth}>{\centering}p{0.14\textwidth}>{\centering}p{0.14\textwidth}>{\centering\arraybackslash}p{0.14\textwidth}}		
		
			\hline
			\hline
			& \multicolumn{1}{c}{Manual} & \multicolumn{1}{c}{Clerical} & \multicolumn{1}{c}{Professional} & \multicolumn{1}{c}{Prof./Man.} \\
			\hline
			
			\hspace{0.5cm}  Dynamic Flexibility & 7.05  & 0.88  & 0.65  &\nprounddigits{2}$\numprint{0.093}$  \\
			\hspace{0.5cm}   Dynamic Strength & 33.65 & 10.54 & 9.39  & \nprounddigits{2}$\numprint{0.279}$ \\
			\hspace{0.5cm}    Explosive Strength & 10.49 & 2.21  & 4.53  & \nprounddigits{2}$\numprint{0.431}$ \\
			\hspace{0.5cm}    Extent Flexibility & 42.42 & 13.77 & 11.07 & \nprounddigits{2}$\numprint{0.261}$ \\
			\hspace{0.5cm}   Gross Body Coordination & 35.95 & 12.40 & 11.69 & \nprounddigits{2}$\numprint{0.325}$ \\
			\hspace{0.5cm}    Gross Body Equilibrium & 27.52 & 9.27  & 9.25  & \nprounddigits{2}$\numprint{0.336}$\\
			\hspace{0.5cm}   Stamina & 40.79 & 14.43 & 13.56 & \nprounddigits{2}$\numprint{0.332}$\\
			\hspace{0.5cm}    Static Strength & 44.69 & 17.23 & 13.54 & \nprounddigits{2}$\numprint{0.303}$\\
			\hspace{0.5cm}    Trunk Strength & 48.98 & 22.95 & 21.57 & \nprounddigits{2}$\numprint{0.440}$\\
			\hline
		\end{tabular} \\
		\scriptsize{The table presents the  physical ability requirement scores for the occupation groups defined in this paper. For each ability, the score for a given occupation group is the weighted-sum of the scores of six-digit occupations under this group. The scores of six-digit occupations are obtained from O*NET and the weights are the employment shares in 2014 obtained from the CPS data .  \par }
		\label{table:onet1}%
	\end{minipage}
\end{table}%

\begin{table}[H]
	\centering
		\footnotesize
	\begin{minipage}{0.88\textwidth}
		\caption{Cognitive Ability Requirements across Occupations}
		
\tabcolsep=0.12cm
\begin{tabular}{p{0.35\textwidth}>{\centering}p{0.14\textwidth}>{\centering}p{0.14\textwidth}>{\centering}p{0.14\textwidth}>{\centering\arraybackslash}p{0.14\textwidth}}				
			\hline
		\hline
			& \multicolumn{1}{c}{Manual} & \multicolumn{1}{c}{Clerical} & \multicolumn{1}{c}{Professional} & \multicolumn{1}{c}{Prof./Man.} \\
			\hline
			\hspace{0.5cm}      Oral Expression & 58.28 & 71.19 & 76.01 & \nprounddigits{2}$\numprint{1.304}$\\
			\hspace{0.5cm}      Oral Comprehension & 60.75 & 71.90 & 75.49 & \nprounddigits{2}$\numprint{1.243}$\\
			\hspace{0.5cm}      Number Facility & 29.76 & 41.45 & 44.46 & \nprounddigits{2}$\numprint{1.494}$\\
			\hspace{0.5cm}      Mathematical Reasoning & 28.81 & 41.66 & 47.18 & \nprounddigits{2}$\numprint{1.638}$\\
			\hspace{0.5cm}      Information Ordering & 53.15 & 55.63 & 63.62 & \nprounddigits{2}$\numprint{1.197}$\\
			\hspace{0.5cm}      Inductive Reasoning & 49.61 & 52.51 & 66.53 & \nprounddigits{2}$\numprint{1.341}$\\
			\hspace{0.5cm}      Fluency of Ideas & 34.76 & 40.34 & 55.79 & \nprounddigits{2}$\numprint{1.605}$\\
			\hspace{0.5cm}      Flexibility of Closure & 40.77 & 38.46 & 48.59 & \nprounddigits{2}$\numprint{1.192}$\\
			\hspace{0.5cm}      Deductive Reasoning & 52.94 & 54.67 & 69.11 & \nprounddigits{2}$\numprint{1.305}$\\
			\hspace{0.5cm}      Category Flexibility & 45.73 & 50.35 & 56.76 & \nprounddigits{2}$\numprint{1.241}$\\
			\hspace{0.5cm}      Memorization & 30.24 & 35.49 & 40.82 & \nprounddigits{2}$\numprint{1.350}$\\
			\hspace{0.5cm}      Written Expression & 41.96 & 58.29 & 68.25 & \nprounddigits{2}$\numprint{1.627}$\\
			\hspace{0.5cm}      Written Comprehension & 48.43 & 62.69 & 72.76 & \nprounddigits{2}$\numprint{1.502}$\\
			\hspace{0.5cm}      Visualization & 41.79 & 30.28 & 41.53 & \nprounddigits{2}$\numprint{0.994}$\\
			\hspace{0.5cm}      Time Sharing & 40.08 & 39.54 & 43.29 & \nprounddigits{2}$\numprint{1.080}$\\
			\hspace{0.5cm}      Speed of Closure & 32.34 & 32.08 & 40.08 & \nprounddigits{2}$\numprint{1.239}$\\
			\hspace{0.5cm}      Spatial Orientation & 21.94 & 3.36  & 5.14  & \nprounddigits{2}$\numprint{0.234}$\\
			\hspace{0.5cm}      Selective Attention & 48.75 & 49.00 & 52.41 & \nprounddigits{2}$\numprint{1.075}$\\
			\hspace{0.5cm}      Problem Sensitivity & 58.88 & 57.59 & 70.51 & \nprounddigits{2}$\numprint{1.198}$\\
			\hspace{0.5cm}      Perceptual Speed & 41.66 & 39.47 & 44.24 & \nprounddigits{2}$\numprint{1.062}$\\
			\hspace{0.5cm}      Originality & 33.29 & 38.79 & 53.50 & \nprounddigits{2}$\numprint{1.607}$  \\
			\hline
		\end{tabular}

		\scriptsize{The table presents the cognitive ability requirement scores for the occupation groups defined in this paper. For each ability, the score for a given occupation group is the weighted-sum of the scores of six-digit occupations under this group. The scores of six-digit occupations are obtained from O*NET and the weights are the employment shares in 2014 obtained from the CPS data .  \par }
		\label{table:onet2}%
	\end{minipage}
\end{table}%

\section{Calculation of Social Security Benefits }
\label{Appendix:SS}

Given the complexity of the Social Security scheme, there are several modelling issues and simplifications to be discussed:

\subsection{Eligibility for Social Security benefits}
Individuals are entitled to Social Security retirement benefits only after they earned 40 credits. The credits are linked to the annual earnings and each year a maximum 4 credits can be earned. For example, in 2016 one credit is received for each \$1260. For most people, this requires them to work at least 10 years to be qualified for the Social Security retirement benefits. Because of the curse of dimensionality, I do not maintain the credits that individual has earned as a state variable in the model. Instead, all individuals are assumed to be qualified as long as they reach the early retirement age. Given that the average work experience at age 62 is very long, this should not be a very strict assumption.\footnote{As far as I know, the only exception which keeps the earned credit as a state variable is \cite{vanderklaauwwolpin08}. }

\subsection{State Variables}
As described before, AIME serves as a state variable in the model and I calculate PIA and Social Security benefits based on it. The dependence of benefits on the age at which individual begins drawing benefits requires adding this age as another state variable. Given the multiple values this variable can take, adding it as a state variable will significantly expand the state space of current model, which is already tremendously large. Instead, I follow \cite{french2011effects} and reflect the adjustment from PIA to real benefits in the transition process of AIME to exclude the starting age of benefit-taking as a state variable. The cost of doing it is to add another binary values state variable: whether the individual is the first or subsequent year taking benefits.

To be specific, take the individual starts to draw benefits at age 67 as an example. By Social Security rule, the benefits individual takes is 1.08 times of his PIA, not only for benefits collected at the age of 67 but also all the subsequent ages. To convert the PIA to real benefits amounts, say, at the age of 69, a variable records that individual began collecting benefits at age 67 is necessary to obtain the adjustment coefficient 1.08. To avoid doing this, at the age of 67 when individual collects the benefits for the first time, in the transition process of AIME, I multiply the AIME by the adjustment coefficient. Notice that the adjustment coefficient is only known at age 67 but not subsequent ages without keeping the age of first-time-benefit-drawing as a state variable. In the all following ages, the adjustment from PIA to real benefits is not needed because it is already reflected in the AIME. However, without modelling the Social Security application as a choice, I assume the first year of not working after age 62 as the beginning time of drawing benefits. In the case that individual reenters working after receiving Social Security benefits, I cannot distinguish whether it is his first time or not if he stops working without the assistance of extra variables. Therefore I add a binary state variable to record whether it is the first time of benefit-drawing. This variable, together with AIME, will determine individual's Social Security retirement benefits eventually.\footnote{\cite{french2011effects} models the Social Security application. Thus whether it is the first time of benefits taking can be determined directly from the choice variable. \cite{vanderklaauwwolpin08} does not have Social Security application as a choice variable. Nevertheless they add the age at which individual begins drawing Social Security benefits as a state variable. }

\section{Private Pension}
\label{Appendix:PrivatePension}
Private pension is an important supplement to Social Security, particularly for people with high income. \cite{coile2007future} reveals that private pension has equivalent importance in incentivizing older people to retire.
There are two main plans of the private pensions: the defined-benefit pension and the defined-contribution pension (hereinafter DB and DC plans). The DB plan was previously prevalent whereas the DC plan has become popular recently for the sake of alleviating the increasing burden upon employers. The private pension plans are employer-specific and very heterogeneous. Completely modelling this essentially requires solving the model with respect to every individual respectively.\footnote{Examples include \cite{blaugilleskie08} and \cite{boundetal10}}. Alternatively, \cite{vanderklaauwwolpin08} abstracts from modelling the DC plan because it requires adding an extra decision variable similar to saving \footnote{The contribution to DC pension is similar to savings through pension wealth.} \cite{french2011effects} further abstract from modelling the private pension based on the detailed employer-specific plans. Instead, they construct a complex while reduced-form model to predict the private pension without specifically distinguishing the DB and DC plans. This paper follows the approach adopted by \cite{french2011effects}.

\section{Optimal Consumption}
\label{Appendix:Grid}
In a few cases, the choice-specific value function is not a concave function of consumption. For example, when the consumption in period t is over a threshold so that the assets left for next period is too low in a certain range such that the individual is hitting the consumption floor in period t+1 in this assets range. This will lead to the situation that expected value function in period t is declining in current consumption and then becomes flat after the assets left for next period is lower than a threshold (i.e. after the current period consumption is higher than a threshold). By widening the searching frame, say set \emph{Cnear} to 10 instead of 5, this issue can be addressed. I also tried another strategy to avoid increasing the search frame which leads to slower searching speed. If the non-concave case happens, in which $U(Cbest+Cnear)>U(Cbest)<U(Cbest+Cnear)$, I reset \emph{Cbest} to \emph{Cbest-Cnear} or \emph{Cbest+Cnear} depending on which provides the higher utility. This leads to slightly higher bias in the optimal consumption if \emph{Cnear} is small, but the bias is trivially small.

\section{First-Stage Estimation}
\label{Appendix:FirstStage}
\subsection{Mortality Rates}


Mortality rates are assumed to depend on age and the joint status of physical and cognitive health. Following \cite{french05}, I use the Bayesian rule to calculate the health-dependent mortality rates. As the physical and cognitive health have been discretized into two states, there are four joint health status: $\{h_t^p=0,h_t^c=0\}$, $\{h_t^p=1,h_t^c=0\}$, $\{h_t^p=0,h_t^c=1\}$ and $\{h_t^p=1,h_t^c=1\}$, where $h_t^{p}$ and $h_t^{c}$ are the indicators of poor physical and cognitive health respectively. As shown by the formula below, the health-dependent mortality rate can be decomposed into the unconditional mortality rate $P_t( s_{t+1}=0 | s_{t}=1 )$ and a health-specific shifter $\frac{P_t(h_t^p, h_t^c | s_{t+1}=0, s_{t}=1 )}{P_t(h_t^p, h_t^c | s_{t}=1)}$. 

\begin{align*}
  P_t(s_{t+1}=0 | s_{t}=1, h_t^p, h_t^c) & = \frac{P_t(h_t^p, h_t^c | s_{t+1}=0, s_{t}=1 )}{P_t(h_t^p, h_t^c | s_{t}=1)} \times P_t( s_{t+1}=0 | s_{t}=1 )
\end{align*}

To see how this shifter works, consider two examples. First, there should be little share of individuals with poor health at age 51, whereas those died between 51 and 52 are very likely to suffer poor health. These two facts imply a large poor health shifter to the unconditional mortality rate at 51: having poor health at younger ages is rare but it raises the individual mortality rate significantly from the average level. To compare, many  individuals alive  at 81 may have poor health, regardless of their survivals to age 82 or not. In this case, having poor health leads to only modest shift to the mortality rate at age 81. Since most individuals have poor health at age 81, being unhealthy does not change the individual mortality rate substantially.

The unconditional mortality rate $P_t( s_{t+1}=0 | s_{t}=1 )$ is obtained from Social Security Administration actuarial life tables. I use the HRS data to estimate the health shifter $P_t(h_t^p, h_t^c | s_{t+1}=0, s_{t}=1)/ P_t(h_t^p, h_t^c | s_{t}=1)$. It is estimated on the expanded sample instead of the primary sample, because the primary sample has very few observations at  old ages. To obtain smooth functions, I use quadratic polynomials of age to predict $P_t(h_t^p, h_t^c | s_{t+1}=0, s_{t}=1)$ and $P_t(h_t^p, h_t^c | s_{t}=1)$. Estimates of the mortality rates by health status are presented in the figure below. The mortality rates are the lowest if the individual is both physically and cognitively healthy, followed by the ones under only poor cognitive health. The mortality rates are further higher when individuals have poor physical health only, and the rates end up the highest if both physical and cognitive health are poor. 

\begin{figure}[H]
\caption{ Estimates of Mortality Rates by Joint Health Status}
\centering
          \includegraphics[height=8.6cm]{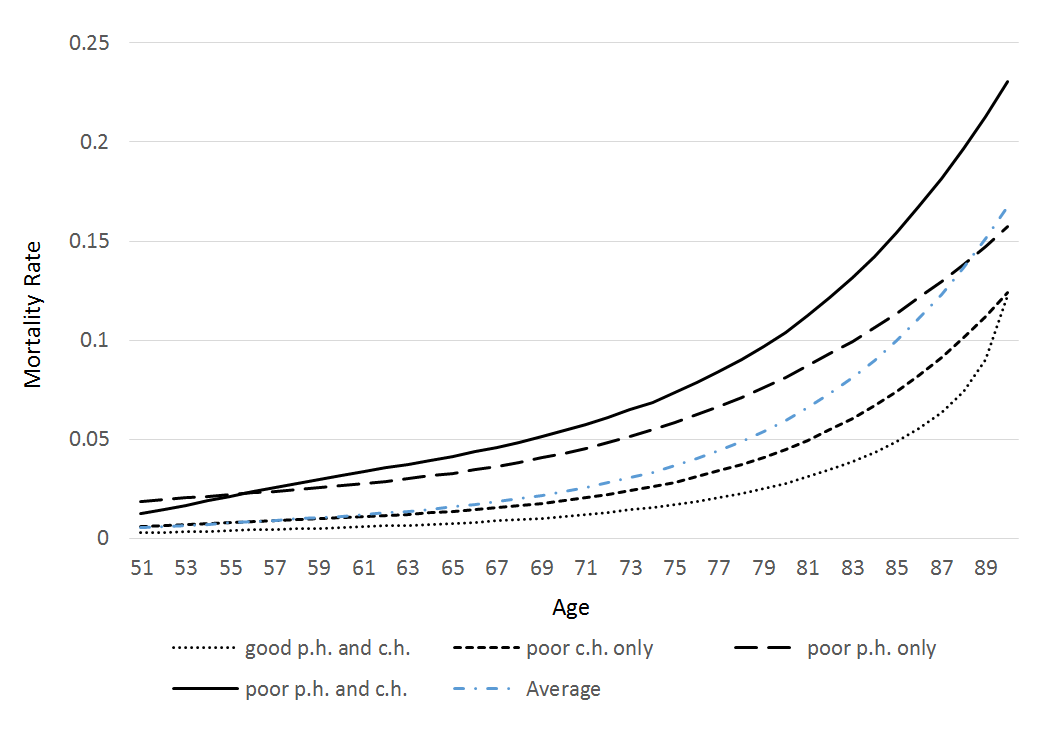}
\end{figure}


\subsection{Health Transition}

 The health transition is also based on the joint status of physical and cognitive health with four states: $\{h_t^p=0,h_t^c=0\}$, $\{h_t^p=1,h_t^c=0\}$, $\{h_t^p=0,h_t^c=1\}$ and $\{h_t^p=1,h_t^c=1\}$, where value 1 denotes poor health status. The joint health state in period t+1 is assumed to depend on the current joint health state, age and occupation in period t, as well as education. Specifically, I estimate multinomial logit regressions of the joint health state in period t+1 on quadratic ages, the education indicators and the occupation indicators in period t, separately conditioning on each joint health state in period t.\footnote{I do not estimate the regressions separately by the interaction of education, occupation and joint health states, which leads to 4*4*4=64 regressions, because sample sizes are very small under these finer classifications. Education and occupation thus affect health transition probabilities mainly by shifting the constant term, though they also interact with age via the nonlinear logit functional form. According to my exploratory estimates, difference between these two specifications is very small. } The identification assumption is that, given the current status of health, the future health  is affected by the current LFP and the corresponding occupations, but not vice versa.

 The estimates reveal the age-specific transition probabilities between 4*4=16 joint health states, with heterogeneity by education and occupation. Results are provided as below. In general, probabilities of getting into worse health states become higher as age increases, while the probabilities of recovering from worse health states decline with age. People with lower education are more likely to transit into  worse health states and  less likely to recover from those states.

The age profiles by occupations are presented as individuals exited their occupations and become out of the LF at different ages.  In general, working, regardless of the occupations, reduces the likelihood of transiting into poor health in the next period.

\begin{figure}[H]
	\caption{Transition Probabilities of Health conditional on $\{h_t^p=0,h_t^c=0\}$}
	\begin{minipage}{0.5\textwidth}
		\centering
		\includegraphics[height=4.1cm]{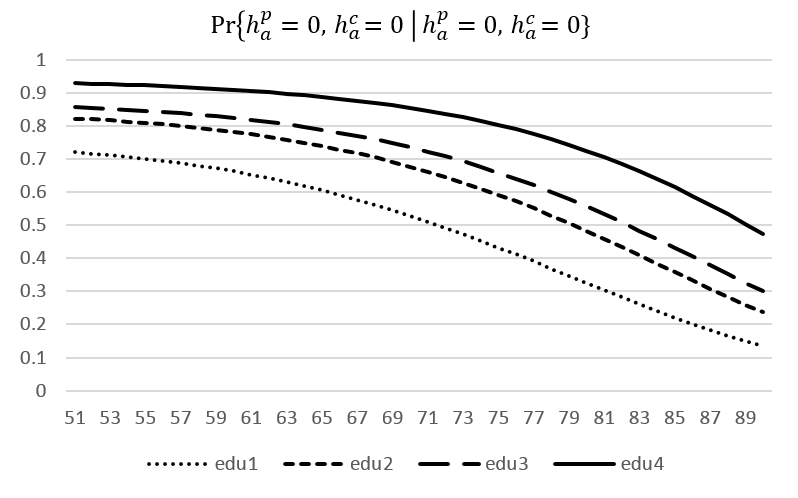}
	\end{minipage}
	\begin{minipage}{0.5\textwidth}
		\centering
		\includegraphics[height=4.1cm]{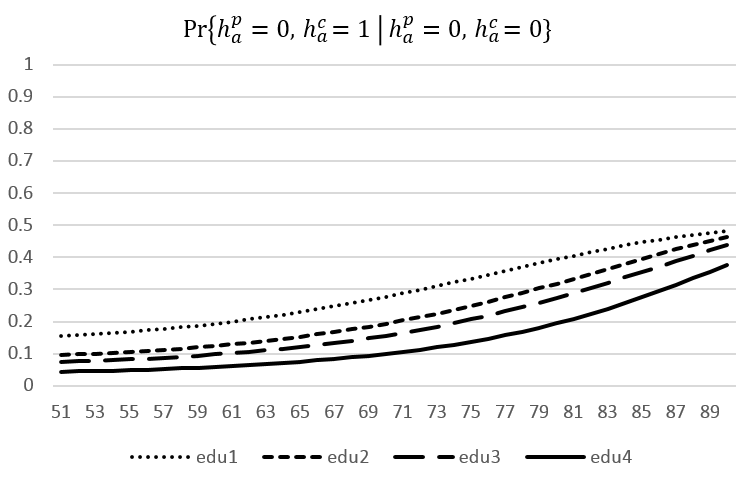}
	\end{minipage}
	\begin{minipage}{0.5\textwidth}
		\centering
		\includegraphics[height=4.1cm]{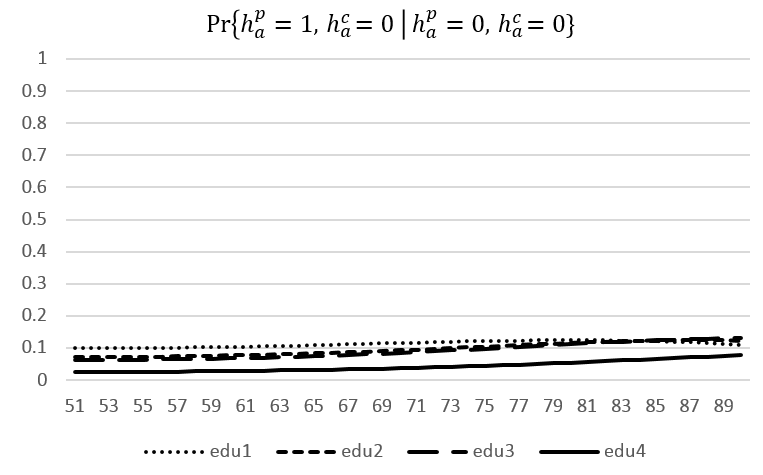}
	\end{minipage}
	\begin{minipage}{0.5\textwidth}
		\centering
		\includegraphics[height=4.1cm]{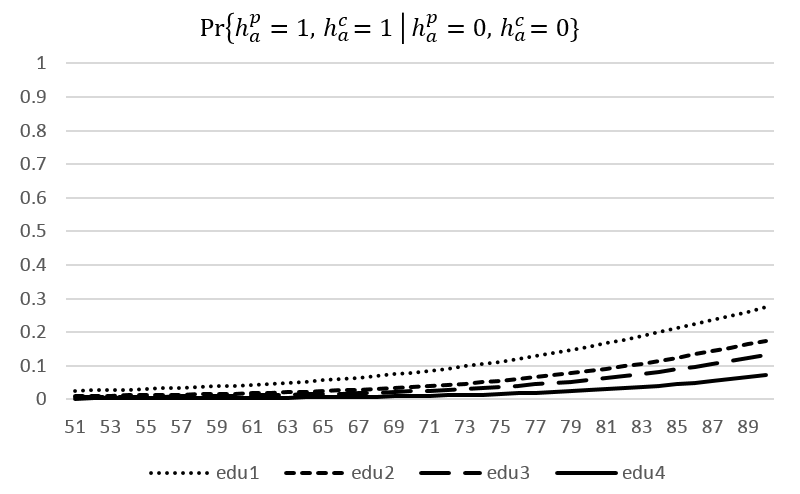}
	\end{minipage}
\end{figure}

\begin{figure}[H]
\caption{Transition Probabilities of Health conditional on $\{h_t^p=1,h_t^c=1\}$}
\begin{minipage}{0.5\textwidth}
\centering
          \includegraphics[height=4.1cm]{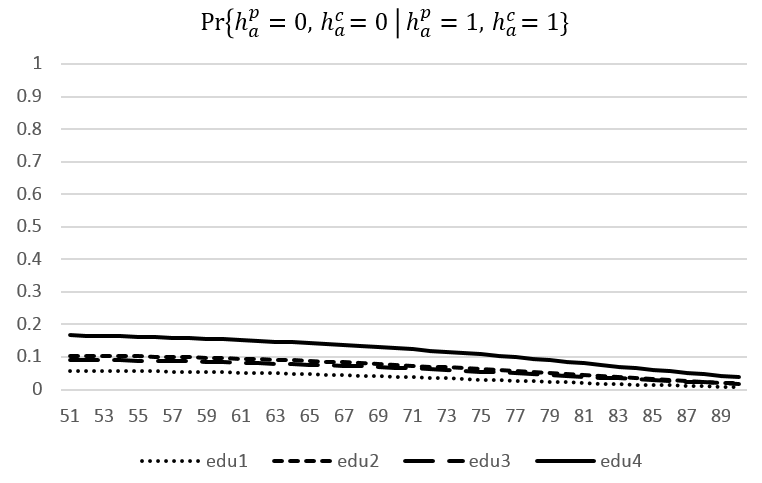}
  \end{minipage}
  \begin{minipage}{0.5\textwidth}
\centering
          \includegraphics[height=4.1cm]{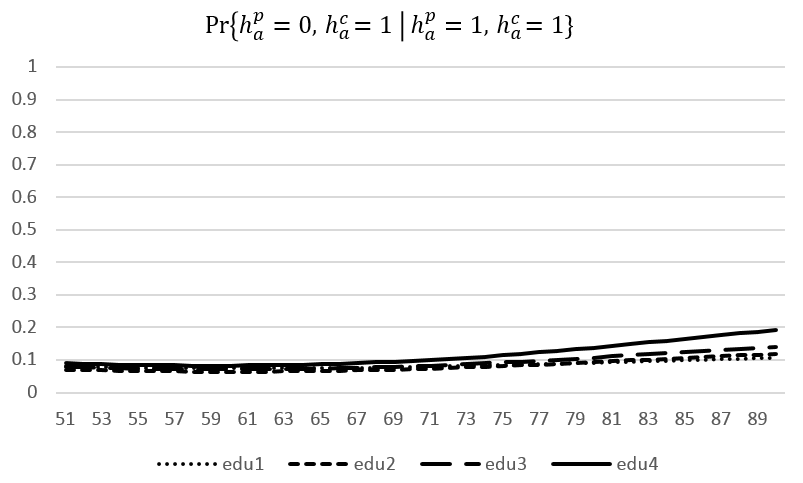}
  \end{minipage}
\begin{minipage}{0.5\textwidth}
\centering
          \includegraphics[height=4.1cm]{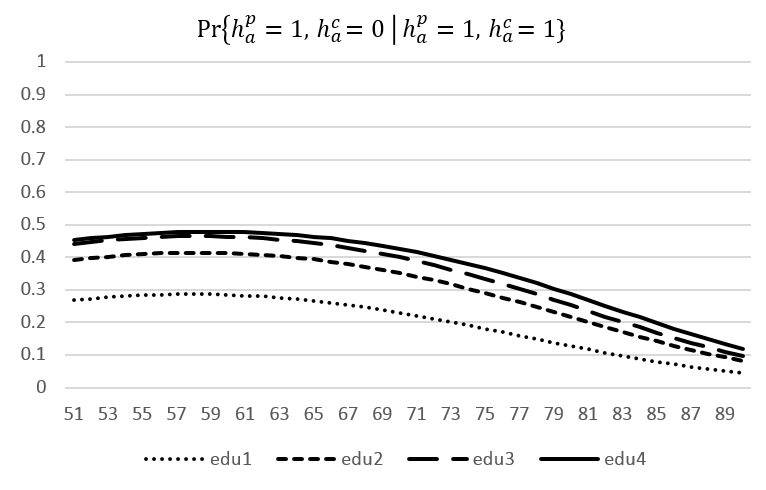}
  \end{minipage}
  \begin{minipage}{0.5\textwidth}
\centering
          \includegraphics[height=4.1cm]{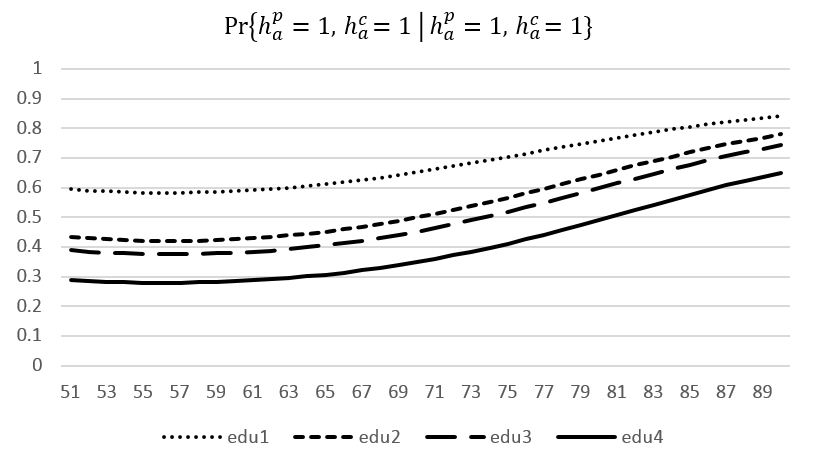}
  \end{minipage}
\end{figure}

\begin{figure}[H]
\caption{Transition Probabilities of Health conditional on $\{h_t^p=0,h_t^c=1\}$}
\begin{minipage}{0.5\textwidth}
\centering
          \includegraphics[height=4.1cm]{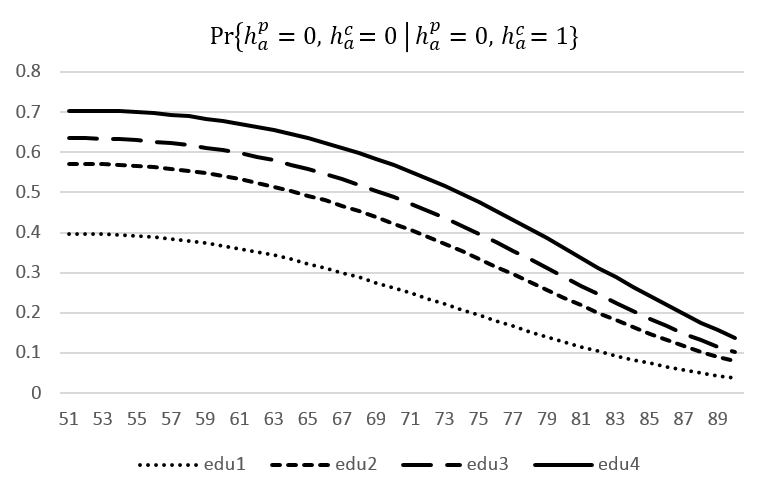}
  \end{minipage}
  \begin{minipage}{0.5\textwidth}
\centering
          \includegraphics[height=4.1cm]{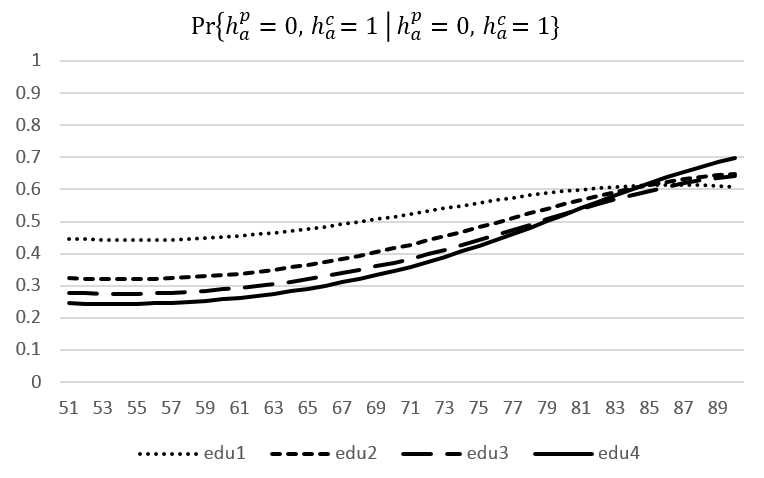}
  \end{minipage}
\begin{minipage}{0.5\textwidth}
\centering
          \includegraphics[height=4.1cm]{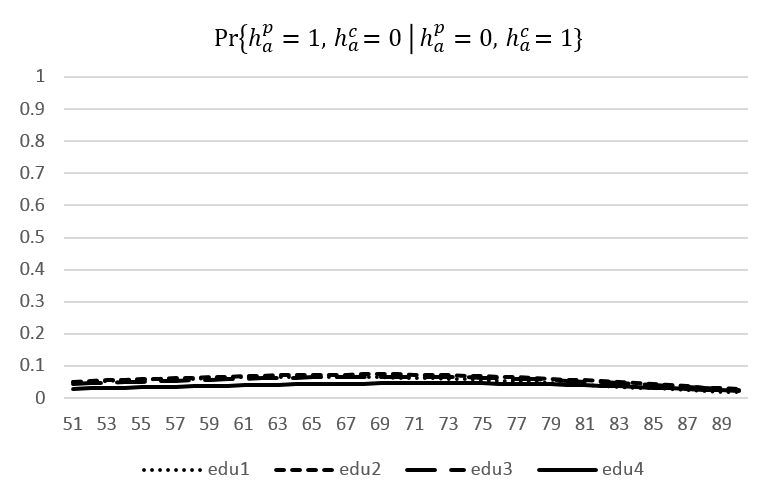}
  \end{minipage}
  \begin{minipage}{0.5\textwidth}
\centering
          \includegraphics[height=4.1cm]{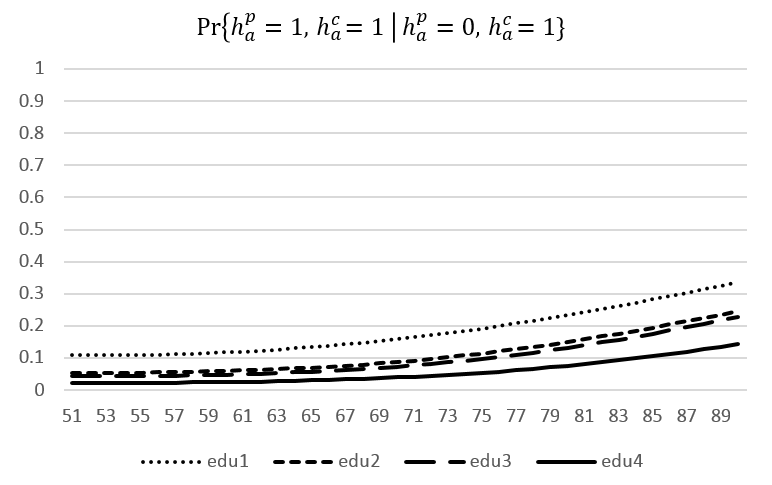}
  \end{minipage}
\end{figure}

\begin{figure}[H]
\caption{Transition Probabilities of Health conditional on $\{h_t^p=1,h_t^c=0\}$}
\begin{minipage}{0.5\textwidth}
\centering
          \includegraphics[height=4.1cm]{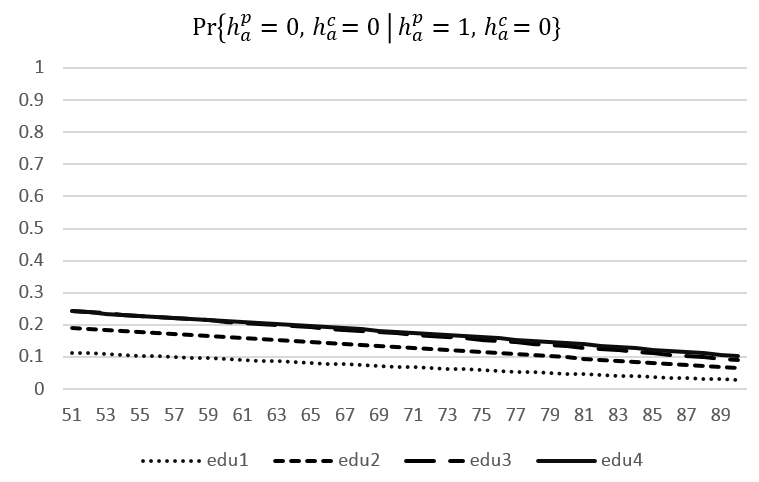}
  \end{minipage}
  \begin{minipage}{0.5\textwidth}
\centering
          \includegraphics[height=4.1cm]{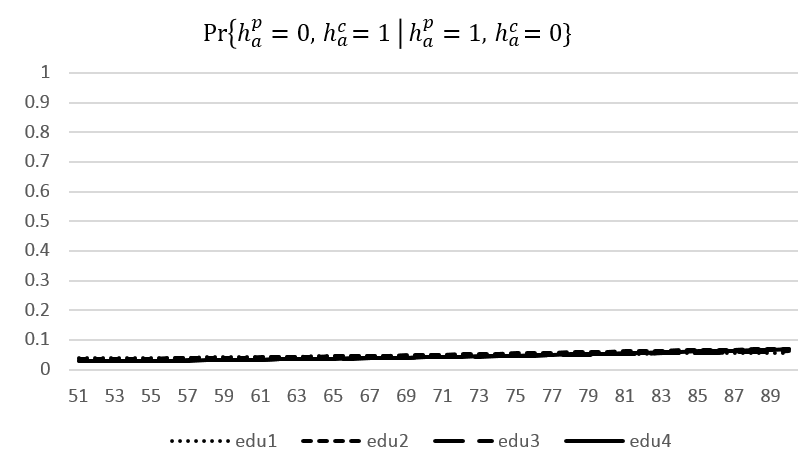}
  \end{minipage}
\begin{minipage}{0.5\textwidth}
\centering
          \includegraphics[height=4.1cm]{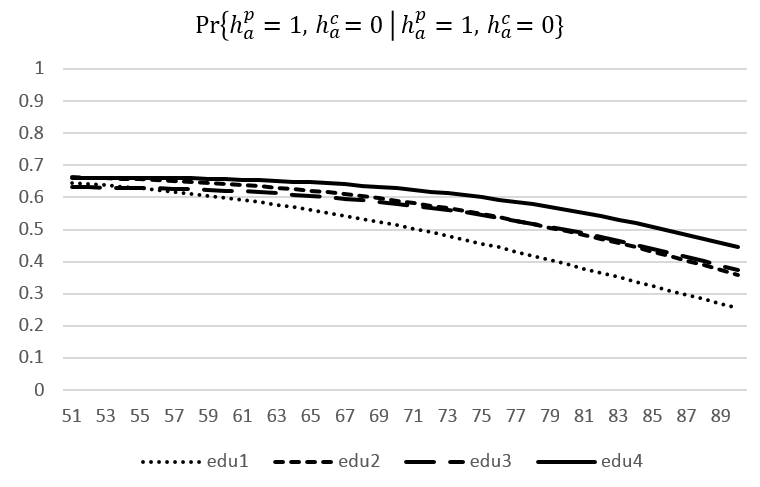}
  \end{minipage}
  \begin{minipage}{0.5\textwidth}
\centering
          \includegraphics[height=4.1cm]{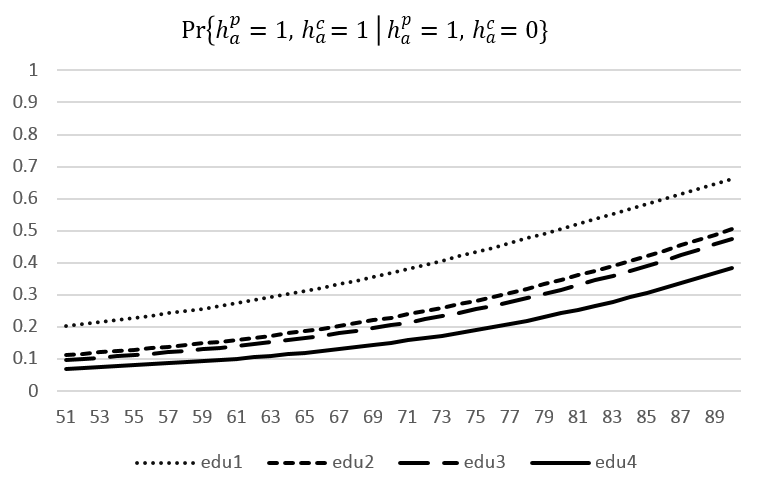}
  \end{minipage}
\end{figure}

\subsection{Wage Equations}
The model includes idiosyncratic shocks to the total income instead of the labor earnings. The reason is if the idiosyncratic shocks are attached to labor earnings, the consumption after retirement will be homogeneous conditional on the observed state variables. With idiosyncratic income shocks independent of labor supply,  the wage equations end up with no selection and it is possible to estimate them in the first stage, separately from the preference parameters. Nevertheless, I estimate the wage equations by Heckman selection model, using the indicators of age 62 and 65 as the exclusion restriction to augment the identification. The underlying assumption is that, once controlling for the smooth effects of age on wage, reaching age 62 and 65 only affects individuals labor supply but not wage. Individuals are eligible for the retirement benefits by reaching age 62, while they are eligible for Medicare at age 65. While individuals' productivity is definitely affected by age, this effect should be smooth.  Their productivity is unlikely to suffer a sudden change at these specific ages. 

The estimates of wage equations are reported below. The occupation-dependent effects of physical and cognitive health still exist though seem less clear. The reduction in wage due to poor physical health is slightly smaller in the professional occupation than other occupations.  The effects of poor cognitive health is much larger in the clerical and professional occupation than the manual occupation. For the other estimates, 

\begin{table}[H]
	\centering
	\footnotesize
	\caption{ Estimates of Wage Equations}
	\begin{minipage}{0.88\textwidth}
\tabcolsep=0.15cm
\begin{tabular}{p{0.42\textwidth}>{\centering}p{0.16\textwidth}>{\centering}p{0.16\textwidth}>{\centering\arraybackslash}p{0.16\textwidth}}
			\hline
   			\hline
			& {Manual }   & {Clerical}   & {Professional}  \\
			\hline
			\quad  Constant $\ln r_j + \kappa_{1j}$                    & 0.818                       & 0.944             & 4.334  \\
			&   (\npnoround{0.567})       &   (0.354)         &   (0.904)     \\
			\quad   Age $\kappa_{2j}$                                  & 0.156                        & 0.043            &  0.010 \\
			&   (0.023)                    &   (0.042)         &   (0.0026)     \\
			\quad   Age-Squared $\kappa_{3j}$                         & -0.0030                     & -0.0011           & -0.0003 \\
			&   (\npnoround{0.0004})       &   (0.0007)         &   (\npnoround{0.0004})     \\                                                            
			\quad  Education: high school $\kappa_{42j}$               & 0.151                      & 0.232     & -0.007    \\
			&   (\npnoround{0.027})       &   (0.088)         &   (\npnoround{0.129})     \\
			\quad  Education: some college $\kappa_{43j}$              & 0.186                      & 0.356  & 0.155     \\
			&   (\npnoround{0.032})       &   (\npnoround{0.089})         &   (\npnoround{0.128})     \\
			\quad  Education: college and above $\kappa_{44j}$        & 0.259   & 0.459    & 0.289     \\
			&   (\npnoround{0.045})       &   (0.093)         &   (\npnoround{0.127})     \\
			\quad  Poor physical health  $\kappa_{5j}$ & -0.074 & -0.109 & -0.042  \\
			&   (\npnoround{0.028})       &   (\npnoround{0.068})         &   (\npnoround{0.056})     \\
			\quad  Poor cognitive health $\kappa_{6j}$ & -0.073 & -0.154 & -0.189  \\
			&   (\npnoround{0.025})       &   (\npnoround{0.061})         &   (\npnoround{0.053})     \\                                         
			\quad Observations                    &   5,509        &    1,819      &  4,144     \\
			\hline
		\end{tabular} 
		\label{table:wageestimates}

		\scriptsize{This table reports the estimates of wage equations. The skill price and the constant term in the human capital $r_j$ and $\kappa_{1j}$ cannot be separately identified. Education takes four discrete values in the model, so $\kappa_{4xj}$ correspond with the premium of three discrete higher education levels (less than high school is omitted as baseline). Standard errors in parentheses. \par }
	\end{minipage}
\end{table}%

\subsection{Others}
I estimate the equation of out-of-pocket medical expenditure by Tobit, taking into account the censoring. Out-of-pocket medical expenditure is computed as sum of the one from the respondent and from his wife, if any. In the structural model, medical expenditure depends on the joint status of physical and cognitive health, the health insurance type, as well as the age. In estimating the equation, I also control for assets, education, income, as well as mental health status. In general, the results suggest poor physical health has significant effect on medical expenditures, with the effect of cognitive is small and ambiguous. One caveat is that the medical expenditure is deterministic in the model, due to the complexity of the current model with two types of health as well as three occupations. The channel of medical expense could be underestimated. Readers who are interested in the role of risk of medical expenditure are referred to \cite{french2011effects}, who focus on the modelling about the stochastic medical expenditure.

The benefits of SSDI is calculated based on the rule of Social Security Administration. I approximate the probability of having SSDI for those younger than 65 by their joint health status, the nonparametric age function, as well as their labor supply status and occupation. Those older than age 65 convert their SSDI to the retirement benefits automatically. 

The spousal income is predicted by individual's own education and a cubic function of age. Specifically, I allow two separate functions of age and education, each for the probability of having positive spousal income and for the amount of spousal income conditional on having.

\section{Model with Unobserved Types}
\label{Appendix:Unobserved}
In my extensive exploration, I find the parameters of the type probability functions are hard to precisely identified, although the heterogeneous disutility of work across types are well identified. This finding is consistent with \cite{heckman1984method}, who also find while estimates of the underlying mixing distribution are not reliable, the structural parameters are estimated quite well. While there are vast theoretical discussions on the identification of finite mixture model, such as \cite{kasahara2009nonparametric}, two features in the current paper complicate the identification: First, the current estimation is based on moments instead of likelihood functions that most theoretical work relies on. Moments-based methods such as the method of simulated moments and the indirect inference leverage only partial information in the data. To the best of my knowledge, there is little discussion on which moments are necessary for the identification of the finite mixture model, especially among empirical studies. \cite{iskhakov2020effects} also attribute the weak identification of parameters related to their mixture model to this potential reason. \footnote{See their discussion in the 30th footnote.} Second, the outcome variable is discrete instead of continuous. Intuitively, we have to rely on variations in the long-term labor force participation rates across individuals to identify the mixture structure, which imposing higher data requirements.

In various exercises, I have tried different identification strategies following previous empirical work: (1) Regressing the labor supply in subsequent waves on the initial state variables,  \`{a} la \cite{dix2014trade}. The idea is that unobserved types differing in the disutility of work have persistent effects on the labor supply, which can be captured by the regression of subsequent labor supply on initial state variables; (2) Including extra variables as the exclusion restriction, \`{a} la \cite{french2011effects} and \cite{vanderklaauwwolpin08}. I include the variable about individual's work preference, which measures how much the individual enjoys working. I also apply the practice in (1) separately by individual's work preference. While there are only seven parameters to be estimated and more than a hundred moments, I find only the heterogeneous disutility of work across types is identified. The coefficients in the type probability functions share the same signs but their magnitudes are highly sensitive to the initial values. \footnote{ Parameters related to the mixture model includes the parameter capturing the difference in disutility of work between types, as well as the parameters of the type probability functions: the constant term, the coefficient of work preference, education, initial physical health, initial  cognitive health, and of the initial assets. Notice that the coefficients of occupations in the type probability functions are not separately identified from the disutility of good health $\lambda_1$, which is also constant and occupation-specific.
}

Only when I target the gaps in life cycle labor supply between different initial states, which are described in the Section of Solution and Estimation Methods, parameters of the type probability functions are reasonably identified. The rationale is as follows. Different initial values, say physical health, associate with individuals of different types with different disutility of work, and this unobserved heterogeneity will translate to the gap in life cycle labor supply between individuals with different initial values. However, as so many moments about the life cycle labor supply conditional on different initial state variables are included, the model is distracted from targeting the more central moments, such as the labor supply by occupations, and it ends up with worse fit for these important facts which are important to our counterfactual experiments. 

For all the above reasons, I am less confident about the estimates based on the model with heterogeneous types. The counterfactual experiments are thus implemented based on the model without unobserved types, of which estimates are stable and transparently identified. Results based on the model with unobserved types are presented in the following table. Individual unobserved types mainly affect the estimation of $\lambda_1$. The central parameters, such as $\lambda_2$, $\lambda_3$ and $\nu$, are quite robust to different specifications.

\begin{table}[H]
	\centering
	\footnotesize
	\caption{ Estimates of Preference Parameters for The Model with Unobserved Types }
	\begin{minipage}{0.96\textwidth}
		\begin{tabular}{p{0.52\textwidth}>{\centering}p{0.12\textwidth}>{\centering}p{0.12\textwidth}>{\centering\arraybackslash}p{0.12\textwidth}} 
			\hline
   \hline
			& {Manual}   & {Clerical}   & {Professional}  \\
			\hline
\multicolumn{1}{l}{\textbf{Non-pecuniary utility}}   &  & &\\
\quad  Extra disutility of work (poor p.h.) $-\lambda_2$ &0.831 &	0.565 &	0.202 \\
\quad   & (0.015) & (0.021)  & (0.017) \\
\quad  Extra disutility of work (poor c.h.) $-\lambda_3$ & 0.167 &  0.517  & 0.317  \\
\quad   & (0.013) & (0.025)  &  (0.026) \\
\quad  disutility of work  (Type1) $-\lambda_1^{type1}$ &  0.420 & -0.083  & 0.152 \\
\quad   & (0.008) & (0.004)  & (0.006) \\
\quad  disutility of work (Type2) $-\lambda_1^{type2}$ &  0.189  &	-0.315   &	-0.080  \\
\hline
\multicolumn{1}{l}{\textbf{Pecuniary utility}} &  & &  \\
   \quad  Bequest Motive $\iota$ &  \multicolumn{2}{l}{12.05 (0.388)}       &      \\
      \quad   Coefficient of Risk Aversion $\nu$&  \multicolumn{2}{l}{1.269 (0.009)}       &      \\
\hline
		\end{tabular} \\
		\label{table:estimates2}
		\scriptsize{This table presents the preference parameters estimated in the second step for the baseline model.  Within-individual variations are exploited to identified these parameters. $\lambda_1^{type2}=\lambda_1^{type1}+\delta_\lambda$, where $\lambda_1^{type1}$ and $\delta_\lambda$ are directly estimated. The $\lambda_1^{type2}$ are induced. Standard errors are in parentheses.  \par }
	\end{minipage}
\end{table}%

\section{Standard Error}
\label{Appendix:SE}

The asymptotic variance of the structural estimates is given the following formula:
\begin{align*}
(G_0^{'} \Omega_0 G_0)^{-1}  ( G_0^{'} \Omega_0 \Lambda_0 \Omega_0 G_0 ) (G_0^{'} \Omega_0 G_0)^{-1}       \qquad \qquad \qquad \qquad \qquad \qquad \qquad   & \\
 where  \qquad \qquad \qquad   \qquad \qquad \qquad \qquad \qquad \qquad \qquad \qquad \qquad \qquad \qquad \qquad \qquad \qquad \qquad \qquad \qquad \qquad  &\\
 \Lambda_0=Var[s_i(\varphi_0)]=E[s_i(\varphi_0)s_i^{'}(\varphi_0)]  \quad \qquad \qquad  \qquad \qquad \qquad \qquad \qquad &\\
 and  \quad \qquad \qquad \qquad   \qquad \qquad \qquad \qquad \qquad \qquad \qquad \qquad \qquad \qquad \qquad \qquad \qquad \qquad \qquad \qquad \qquad \qquad  &\\
 G_0=E[\bigtriangledown_{\varphi} s_i(\varphi_0)] \quad  \qquad \qquad \qquad \qquad  \qquad \qquad \qquad \qquad \qquad &
\end{align*}

$\varphi$ is the vector of structural parameters estimated in the second step. $s_i (\varphi)$ is the the simplified notation for the score $s(x_i(\varphi), \widehat{\theta})$, which is evaluated at the parameters of auxiliary models estimated on the real data and at the simulated data based on the structural parameters $\varphi$. $G$ is the Jacobian matrix of the derivative of scores with respect to the structural parameters. $\Omega$ is the weighting matrix.

Based on the above formula, I obtain the consistent estimator as:
\begin{align*}
\hspace{5cm} (\widehat{G}^{'}  \widehat{\Omega}  \widehat{G})^{-1}  ( \widehat{G}^{'} \widehat{\Omega}  \widehat{\Lambda}  \widehat{\Omega} \widehat{G} ) (\widehat{G}^{'} \widehat{\Omega} \widehat{G})^{-1}       \qquad \qquad \qquad \qquad \qquad \qquad \qquad   
\end{align*}

\end{appendices}

\newpage
\bibliographystyle{te}

\bibliography{Citation}

\end{document}